\documentclass[useAMS,usenatbib,usegraphicx]{mn2e}

\usepackage{aas_macros}
\usepackage{amssymb}
\usepackage{fixltx2e}
\usepackage{multirow}
\usepackage{color}
\usepackage{array}

\title[SILCC: Zooming-in on molecular clouds]{SILCC-Zoom: The dynamical and chemical evolution of molecular clouds}
  \author[D. Seifried et al.]
  {D.~Seifried,$^1$\thanks{seifried@ph1.uni-koeln.de} S.~Walch,$^1$ P.~Girichidis,$^{2,3}$ T.~Naab,$^2$ R.~W\"unsch,$^4$ R.~S.~Klessen,$^{5,6}$ 
   \newauthor
   S.~C.~O.~Glover,$^5$  T. Peters,$^2$ P. Clark$^7$ \\
  $^1$I. Physikalisches Institut, Universit\"at zu K\"oln, Z\"ulpicher Str. 77, 50937 K\"oln, Germany\\
  $^2$Max-Planck-Institut f\"{u}r Astrophysik, Karl-Schwarzschild-Str. 1, 85741 Garching, Germany \\
  $^3$Heidelberg Institute for Theoretical Studies, Schloss-Wolfsbrunnenweg 35, 69118 Heidelberg, Germany\\
  $^4$Astronomical Institute, Academy of Sciences of the Czech Republic, Bocni II 1401, 141 31 Prague, Czech Republic\\
  $^5$Universit\"{a}t Heidelberg, Zentrum f\"{u}r Astronomie, Institut f\"{u}r Theoretische Astrophysik, Albert-Ueberle-Str. 2, 69120 Heidelberg, Germany\\
  $^6$Universit\"{a}t Heidelberg, Interdisziplin\"{a}res Zentrum f\"{u}r Wissenschaftliches Rechnen, Im Neuenheimer Feld 205, 69120 Heidelberg, Germany\\
  $^7$School of Physics \& Astronomy, Cardiff University, 5 The Parade, Cardiff CF24 3AA, Wales, UK\\
  }

\date{Released 2017}

\pagerange{\pageref{firstpage}--\pageref{lastpage}} \pubyear{2015}

\begin{document}

\label{firstpage}

\maketitle

\begin{abstract}
We present 3D "zoom-in" simulations of the formation of two molecular clouds out of the galactic interstellar medium. We model the clouds -- identified from the SILCC \mbox{simulations --} with a resolution of up to 0.06 pc using adaptive mesh refinement in combination with a chemical network to follow heating, cooling, and the formation of H$_2$ and CO including (self-) shielding. The two clouds are assembled within a few million years with mass growth rates of up to \mbox{$\sim$ 10$^{-2}$ M$_{\sun}$ yr$^{-1}$} and final masses of \mbox{$\sim$ 50\,000 M$_{\sun}$}. A spatial resolution of \mbox{$\lesssim$ 0.1 pc} is required for convergence with respect to the mass, velocity dispersion, and chemical abundances of the clouds, although these properties also depend on the cloud definition such as based on density thresholds, H$_2$ or CO mass fraction. To avoid grid artefacts, the progressive increase of resolution has to occur within the free-fall time of the densest structures (1 -- 1.5 Myr) and \mbox{$\gtrsim$ 200} time steps should be spent on each refinement level before the resolution is progressively increased further. This avoids the formation of spurious, large-scale, rotating clumps from unresolved turbulent flows. While CO is a good tracer for the evolution of dense gas with number densities $n \geq$ 300 cm$^{-3}$, H$_2$ is also found for $n \lesssim 30$ cm$^{-3}$ due to turbulent mixing and becomes dominant at column densities around 30 -- 50 M$_{\sun}$ pc$^{-2}$. The CO-to-H$_2$ ratio steadily increases within the first 2 Myr whereas $X_\rmn{CO}\simeq$ \mbox{1 -- 4 $\times$ 10$^{20}$ cm$^{-2}$ (K km s$^{-1}$)$^{-1}$} is approximately constant since the CO(1-0) line quickly becomes optically thick.
\end{abstract}


\section{Introduction}

Molecular clouds (MCs) condense out of the warm interstellar medium (ISM) on scales of several 100 pc and host filamentary substructures on sub-pc scales \citep[see e.g. the reviews by][]{Andre14,Dobbs14,Klessen16}. They consist of molecular hydrogen H$_2$, which can only be traced indirectly with observations, mostly by means of CO line emission, dust emission, or dust extinction. Since they are observed to have large non-thermal line widths \citep[e.g.][]{Larson81,Solomon87}, the cloud sub-structure appears to be shaped by supersonic turbulence. The crux, however, is that supersonic turbulence is expected to decay in a crossing time \citep[e.g.][]{Stone98,MacLow98,Elmegreen04} unless it can be sustained by some physical process, e.g. stellar feedback \citep[see e.g.][for reviews]{Elmegreen04,MacLow04,Ballesteros07}, gravity-driven turbulence \citep[e.g.][]{Vazquez08,Klessen10,Goldbaum11,Ballesteros11,Krumholz16}, or magnetorotational instabilities \citep{Kim03}. With a turbulent dynamical state being observed, the idea of short-lived, dynamically evolving MCs is widely accepted \citep{MacLow04}. It is thus likely that the turbulent, internal sub-structure of MCs is imprinted already in the earliest phase of their formation process.

This conception has several important implications. First, the formation and evolution of MCs cannot be modelled in isolation but should be simulated \textit{together with the surrounding, multi-phase ISM.} Since the turbulent motions within the cloud seem to be inherited from large scale flows \citep[e.g.][]{Brunt09}, the physical processes driving these flows need to be modelled self-consistently. This includes e.g. thermal instability to seed cold clouds \citep{Vazquez00}, self-gravity leading to the collapse of cold clouds, gas accretion onto the forming MCs, as well as nearby supernova explosions, which affect the ISM on scales of several 100 pc and larger \citep{Klessen16}.

Second, the chemical abundances of the gas evolve on timescales comparable to the dynamical time of cloud formation and therefore, cloud formation and molecule formation need to be \textit{modelled simultaneously and coupled together}. The formation timescale of molecular hydrogen roughly scales as $\tau_{\rm H_2} \approx 1 \;{\rm Gyr} \left( n/ 1\;{\rm cm}^{-3}\right)^{-1}$ \citep{Hollenbach79,Glover07a,Glover07b,Glover10} leading to $\tau_{\rm H_2} \sim$ 10 Myr at a mean number density of $n =100 \;{\rm cm}^{-3}$. The free fall timescale at this density is $\tau_{\rm ff} = \sqrt{3 \pi /(32 G \rho)} \approx 5$ Myr, and thus the chemical state of the gas cannot be assumed to be in equilibrium. \citet{Glover07b} have shown that in turbulent environments H$_2$ can form more rapidly than $\tau_{\rm H_2}$ at the respective mean gas density \citep[see also][]{Micic12}, possibly also enhanced to due turbulent mixing \citep{Valdivia16}.

Simulating the formation and evolution of a MC in its large-scale environment with a non-equilibrium chemical network is a numerically challenging task due to the large dynamic range of the problem. Moreover, an on-the-fly treatment of molecule formation and of approximate radiative transfer to compute (self-)shielding are challenging and time-consuming, but are becoming feasible \citep{Dobbs08b,Clark12,Inoue12,Pettitt14,Smith14b,Seifried16, Walch15,Richings16a,Richings16b,Hu16,Szucs16,Valdivia16,Hu17}. It is also evident that the detailed modelling of the dense, molecular phase is important for the understanding of galaxy formation on larger scales \citep{Naab16}.

\citet{Smith14b} study MC formation from kpc scales down to 0.3 pc scales. They investigate the amount of CO-dark molecular gas in a Milky Way-like galaxy and report extremely  elongated filamentary clouds with length $>$ 100 pc as observed by, e.g., \citet{Goodman14} and \citet{Ragan14}. However, these simulations do not account for the effects of gas self-gravity or stellar feedback. When studying synthetic observations of H$_2$, CO, and HI of a Milky Way-like disc simulated by \citet{Dobbs13}, \citet{Duarte15} show that models without feedback and self-gravity fail to reproduce the observed properties of the Milky Way. Furthermore, by re-simulating a MC formed in a full galactic disc simulation including chemistry, \citet{Dobbs15} demonstrate that clouds tend to only approximately follow the linewidth - size relation found by \citet{Larson81} with a significant scatter. Moreover, synthetic emission maps of this cloud show that CO picks up only small sub-complexes within giant MCs, \citet[][but see also \citealt{Smith14b}]{Duarte16}. CO emission estimates thus miss a significant fraction of the molecular H$_2$ content, which agrees with observational results by e.g. \citet{Tielens85,Dishoeck88,Xu16}.

\citet{Richings16a,Richings16b} study the chemical evolution of low-mass disc galaxies and their embedded MCs showing that the CO-to-H$_2$ conversion factor can have a significant scatter and that up to 86\% of the H$_2$ might be not traceable by CO observations. However, with a mass resolution of 750 M$_{\sun}$, they are unable to resolve the detailed substructure of the clouds as well as the accurate formation of H$_2$ and CO. At significantly higher resolution, \citet{Hu16,Hu17} have investigated the formation of H$_2$ in dwarf galaxies and the impact of interstellar radiation.

On somewhat smaller scales ($\sim$ 500 pc), \citet{Ibanez15} present a zoom-in study of MCs (down to scales of $\sim$ 0.5 pc) that form out of the multi-phase ISM in supernova-driven stratified galactic discs, but do not include a chemical network. They find that with self-gravity turned off initially, the velocity dispersions of the clouds are too low and move onto the Larson relation only once self-gravity is switched on. 

Furthermore, using a setup with periodic boundary conditions, \citet{Padoan16b,Padoan16} investigate MC and star formation in supernova-driven periodic boxes \citep[see also][]{Gatto15,Li15}. Like \citet{Dobbs15} the authors argue that the MCs tend to follow the Larson relation with a significant scatter of about one order of magnitude. However, as pointed out by \citet{Rey15}, starting simulations with an already pre-existing cloud can be problematic since it does not match the full complexity of galactic clouds and the potential impact of star formation in the earliest stage.

This long-standing problem of realistic initial conditions for simulations of MC formation and evolution is the motivation for the present study. Although, as discussed before, a wealth of simulations have tackled this problem, they usually lack one or more important aspects: the larger-scale environment, the inclusion of a chemical network, self-gravity, or high ($\sim$ 0.1 pc) spatial resolution. In particular for the latter aspect, so far a detailed resolution study related to the formation of MCs \textit{as well as} their chemical evolution is lacking in the literature.

In this work we try to remedy the aforementioned shortcomings by making use of the SILCC\footnote{SImulating the LifeCycle of molecular Clouds; www.astro.uni-koeln.de/$\sim$silcc} simulations \citep{Walch15,Girichidis16}, which model the evolution of the multi-phase ISM in a part of a stratified galactic disc (500 pc $\times$ 500 pc $\times$ $\pm$ 5 kpc) including a chemistry network describing the formation of H$_2$ and CO. The simulations thus provide detailed thermodynamical and chemical conditions, which allow a self-consistent modelling of the formation of MCs out of the diffuse ISM. We make use of a zoom-in strategy
which allows us -- for the first time --
to simultaneously model MCs with
\begin{enumerate}
 \item a high, effective spatial resolution of up to 0.06 pc,
 \item their large-scale environment at lower resolution, and
 \item their detailed chemical evolution.
\end{enumerate}

Zoom-in simulations are becoming more prominent in the context of MC formation \citep[e.g.][]{Renaud13,Dobbs15,Ibanez15,Ibanez17,Kuffmeier16} and we address important questions related to the zoom-in strategy in this paper. 
We will also investigate which resolution is required to accurately model the dynamical \textit{and} chemical evolution of clouds. Furthermore, we will show that similar to ISM simulations on larger scales \citep{Walch15,Girichidis16}, a simple density threshold criterion does not recover the same cloud structure that we find when we examine the chemical state of the gas. Moreover, we demonstrate that the 
H$_2$ abundance is enhanced in low-density gas ($n$ $<$ 30 cm$^{-3}$) due to turbulent mixing as previously suggested by \citet{Glover07b} and \citet{Valdivia16}.

The structure of the paper is as follows: We first present the zoom-in strategy used in this work (Section~\ref{sec:simulations}). In Section~\ref{sec:fiducial} we discuss MC properties obtained by different selection criteria and the evolution of the H$_2$ content. Next, we investigate which resolution is required to properly model the chemical and dynamical evolution of the MCs in Section~\ref{sec:resolution}. In Section~\ref{sec:comparison}, we further show that the time, over which the resolution is increased, can change the structure of the resulting cloud. We show a first application in the context of synthetic observations in Section~\ref{sec:XCO} and conclude in Section~\ref{sec:conclusion}.

\section{Numerical methods and simulation setup}
\label{sec:simulations}

The zoom-in simulations discussed in this work are based on the SILCC simulations presented in detail in \citet{Walch15} and \citet{Girichidis16}. In the following we briefly describe the numerical methods (Section~\ref{sec:numerics}) and initial conditions (Section~\ref{sec:ICs}). 
The zoom-in technique is introduced in Section~\ref{sec:zoom}.

\subsection{Numerical methods}
\label{sec:numerics}

We use the adaptive mesh refinement (AMR), finite-volume code FLASH version 4 \citep{Fryxell00,Dubey08} to solve the (magneto-)hydrodynamical equations. The 'Bouchut 5-wave solver' guarantees positive entropy and density \citep{Bouchut07,Bouchut10,Waagan09,Waagan11}. The resolution on the base-grid is 3.9 pc.

We model the chemistry of the ISM using a simplified chemical network for hydrogen and carbon chemistry \citep{Nelson97,Glover07a,Glover07b,Glover10,Glover12}. The network is designed to follow the chemical abundances of H$^+$, H, H$_2$, C$^+$, and CO as well as free electrons and atomic oxygen. We assume solar metallicity, with  fixed elemental abundances of carbon and oxygen given by $x_\rmn{C} = 1.41 \times 10^{-4}$ and $x_\rmn{O} = 3.16 \times 10^{-4}$, respectively \citep{Sembach00}. Initially, all carbon is in the form of C$^+$ and all hydrogen is in the form of atomic hydrogen.

The chemical network also describes the thermal evolution of the gas including detailed heating and cooling processes using the chemical abundances provided by the network \citep{Glover10,Glover12,Walch15}. For the heating via the photoelectric effect as well as photo-dissociation reactions, the simulations include a uniform interstellar radiation field (ISRF) with a strength of G$_0$ = 1.7 \citep{Draine78} in units of the Habing field \citep{Habing68}. The ISRF is attenuated due to shielding by the surrounding gas and dust. The necessary column densities of H$_2$, CO, and the total gas are calculated via the TreeCol algorithm \citep{Clark12b}. The cosmic ray ionisation rate is set to a constant value of 1.3 $\times$ 10$^{-17}$ s$^{-1}$ in the entire simulation domain. More details of the implementation of the chemical network in FLASH and its coupling to the TreeCol algorithm is explained in \citet{Walch15} and \citet{Wunsch17}.

Both self-gravity of the gas as well as a background potential, which represents the old stellar component of the galactic disc, are taken into account. For self-gravity of the gas we solve the Poisson equation with a tree based method \citep{Wunsch17}, which is part of the official FLASH release. The external gravitational acceleration due to the pre-existing stellar component in the galactic disc is modelled with an isothermal sheet with $\Sigma_\rmn{star}$ = 30 M$_{\sun}$ pc$^{-2}$ and a scale height of 100 pc, originally proposed by \citet{Spitzer42}.

Turbulence is generated in the simulations by supernova (SN) feedback. For a single SN explosion, an energy of 10$^{51}$ erg is injected in the form of thermal energy if the mean density in the surrounding spherical region (i.e. the injection region with a minimum radius of 4 cells) is low enough to resolve the Sedov-Taylor phase of the SN. If the Sedov-Taylor phase is unresolved, we heat the gas within the injection region to $10^4$ K and inject the momentum, which the swept-up shell has gained at the beginning of the momentum-driven snowplough phase \citep[see][for details, but also \citealt{Walch15b,Haid16}]{Gatto15}. 

The rate at which SNe are injected is based on the Kennicutt-Schmidt relation \citep{Schmidt59,Kennicutt98}, which relates the disc's gas surface density $\Sigma_\rmn{gas} = 10\;{\rm M}_{\odot} \;{\rm pc}^{-2}$ (see Section~\ref{sec:ICs}) to a typical star formation rate surface density $\Sigma_\rmn{SFR}$, i.e. $\Sigma_\rmn{SFR} \propto \Sigma_\rmn{gas}^{1.4}$. In a next step,  $\Sigma_\rmn{SFR}$ is translated into a SN rate by assuming a standard initial mass function \citep{Chabrier01}. This defines an SN rate with which the gas is heated and stirred. 

In the original SILCC simulations \citep{Walch15, Girichidis16}, we investigate how a different positioning of the SNe relative to the dense gas influences the structure of the ISM. We find that {\it mixed driving} results in an ISM with solar neighbourhood properties. Mixed driving means that 50\% of the SNe are injected at random positions (modulo a weighting of the vertical positions with a Gaussian distribution with a scale height of 50 pc) and 50\% are injected at local density peaks. In this way we mimic the explosions of massive stars in either their parental cloud or diffuse regions due to runaway O/B-stars. The simulation with mixed driving is the starting point of the work presented here. 

We switch off the SN driving when we start the zoom-in simulations and discuss the impact of nearby SN explosions on the evolution of the forming MC in a subsequent paper.

\subsection{Initial conditions of the SILCC simulation}
\label{sec:ICs}

Our initial conditions represent a small section of a galactic disc at low redshift with solar neighbourhood properties and solar metallicity. The simulation box has a size of 500 pc $\times$ 500 pc $\times$ \mbox{$\pm$ 5 kpc} with periodic boundary conditions applied along the $x$- and $y$-direction and outflow conditions along the $z$-direction. The gas surface density is $\Sigma_\rmn{gas}$ = 10 M$_{\sun}$ pc$^{-2}$ and the initial vertical distribution of the gas is modelled with a Gaussian profile
\begin{equation}
 \rho(z) = \rho_0 \times \textrm{exp}\left[ - \frac{1}{2} \left( \frac{z}{h_z} \right)^2 \right]
\end{equation}
with $h_z$ = 30 pc and $\rho_0 = 9 \times 10^{-24}$ g cm$^{-3}$. The resulting total gas mass is \mbox{$M_\rmn{disc} = 2.5 \times 10^6$ M$_{\sun}$}. The gas near the disc midplane has an initial temperature of 4500 K and consists of atomic hydrogen and C$^+$. At large scale heights the density is set to a constant value of $\rho=10^{-28}$ g cm$^{-3}$, representing hot, ionised halo gas with T = 4 $\times$ 10$^8$ K.

The gas is initially at rest. From the start of the simulation until the time when we start to zoom-in on the forming molecular clouds, we drive SNe at a constant rate of 15 Myr$^{-1}$ using mixed driving (see Section~\ref{sec:numerics}). In addition to the SN driving, the heating and cooling of the gas plus the gravitational forces self-consistently generate structures and motions within the disc gas.

\subsection{Initial conditions of the zoom-in simulations}
\label{sec:zoom}

A number of dense and cold MCs develop from the diffuse ISM in this SILCC simulation \citep[see][for a discussion]{Walch15}. These MCs have different histories. They continuously accrete gas and sometimes merge with each other. In addition, they are pushed from different sides by nearby SN explosions. We select some of these clouds by eye and re-simulate them at a high spatial and temporal resolution to study the evolution of their internal sub-structure and chemistry. We call this a "zoom-in" simulation.

To properly follow the evolution of the clouds' internal structure, we rewind the simulation and start to zoom-in at a time when the typical number density within the zoom-in region does not exceed a few times 10 cm$^{-3}$. Such low densities are a crucial requirement to properly model the initial formation process of the cloud. This limits the number of possible cloud candidates since in particular at later times the clouds will already be significantly denser. Moreover, in order to allow for a clean convergence study, we choose MCs which form in (relative) isolation and are not affected by any significant merger. In particular at late times the MCs start to merge, which leaves us with very few compact clouds after $\sim$ 100 Myr \citep[see e.g. Figs.~6 and~7 in][]{Walch15}.

In this paper, we choose two molecular clouds, MC1 and MC2, which form at the same time but at different positions in the galactic midplane. Initially, the precursors of the two MCs are formed from gas swept-up by SN explosions \citep{Koyama00,Inoue09,Inutsuka15}. As discussed later on, the cloud properties are representative for typical Galactic MCs. While the surrounding multi-phase ISM is modelled at a fixed resolution of 3.9 pc, we gradually increase the resolution within the zoom-in region (see below, Section~\ref{sec:instantaneous}). The starting time for both zoom-in runs is $t_0 = 11.9$ Myr and the extent of the two zoom-in regions are specified in Table \ref{tab:overview}. Since the MCs do not have any significant bulk motion with respect to the grid, we do not have to shift the zoom-in region with time.

\begin{table}
 \caption{Centre (second column) and extent (third column) of the zoom-in regions for runs MC1 and MC2. The starting time $t_0$ of the zoom-in elapsed from the start of the SILCC simulation is given in column 4.}
 \label{tab:overview}
 \centering
 \begin{tabular}{cccc}
 \hline
 run & centre & volume & $t_0$\\
 & [pc] & [pc$^3$]  & [Myr] \\
 \hline
 MC1 & (146,126,0) & 88 $\times$ 78 $\times$ 71 &  11.9 \\
 MC2 & (42,188,0) & 97 $\times$ 81 $\times$ 58 &  11.9 \\
 \hline
 \end{tabular}
\end{table}

\subsection{Refinement strategy}
\label{sec:instantaneous}
{\sc Flash} is an adaptive mesh refinement (AMR) code based on the PARAMESH library \citep{MacNeice00} with which the resolution can be increased in a spatially and temporally adaptive way \citep{Fryxell00}. The user has to define refinement criteria upon which a decision whether the resolution should be increased or decreased is made. Further, in {\sc Flash} the spatial resolution can only differ by a factor of 2 for neighbouring cells.

The minimum refinement level, with which the SILCC simulation is run before $t_0$, corresponds to a base grid resolution of 3.9 pc. To increase the resolution further, we use two refinement criteria. Up to a spatial resolution of 0.5 pc (3 levels above the base grid), we refine on variations in the gas density. This estimate is based on the second derivative of the density field normalised by the average of the gradient \citep[following][]{Lohner87} and allows to capture low-density fluctuations. This criterion is quite sensitive such that the entire region is refined to the maximum possible resolution. Furthermore, we also refine on the local Jeans length, $L_{\rm Jeans}$, which is computed for each cell. We require that $L_{\rm Jeans}$ is resolved with at least 16 cells in each dimension \citep{Truelove97,Federrath11}.

For refinement above 0.5 pc, we only use the Jeans refinement criterion. To carry out a computationally feasible simulation\footnote{In collapsing cold cores the minimum Jeans length decreases with increasing density up to $\rho \approx 10^{-13}\;{\rm g\; cm}^{-3}$, where the gas becomes optically thick to infrared radiation. It is unfortunately impossible to resolve such a high density in our simulations.}, we have to set a {\it global maximum refinement level}, $l_{\rm max, tot}$ with a corresponding effective spatial resolution d$x$. In our fiducial simulation,  $l_{\rm max, tot}$ corresponds to \mbox{d$x$ = 0.12 pc} (or 5 levels above the base grid). Density structures that form on these scales will not be resolved and we cannot make a prediction on their further fragmentation. However, we show that an effective resolution of d$x$ = 0.12 pc is sufficient to reach convergence in terms of H$_2$ and CO mass fractions within the forming molecular clouds (see Section~\ref{sec:resolution}).

A stepwise refinement is applied within the zoom-in region. Therefore, we define a refinement time scale, $\tau$, over which the {\it current maximum refinement level}, $l_{\rm max, curr}$ (with the current maximum resolution d$x_\rmn{curr}$) is increased from the base grid level at $t_0$ to $l_{\rm max, tot}$ at time $t_0 + \tau$. We test different values of $\tau=$4.5, 2.25, 1.5, 1.0, and 0.0 Myr, where $\tau=0.0$ Myr implies that the code is enabled to instantaneously increase the resolution to $l_{\rm max, tot}$. For $\tau >0$ we spend a certain time on each intermediate maximum refinement level $l_{\rm max, curr}$ and thus, slowly increase $l_{\rm max, curr}$ to $l_{\rm max, tot}$. In Table~\ref{tab:levels} we list the time spent on each refinement level for different $\tau$. The time spent corresponds to a typical number of time steps for which the code runs with an intermediate $l_{\rm max, curr}$. For $\tau=1.0$ Myr, we typically spend 200 time steps on each level before advancing to the next higher level. According to the CFL condition, a typical time step decreases with decreasing d$x_\rmn{curr}$ and therefore the time spent on an intermediate $l_{\rm max, curr}$ decreases with increasing $l_{\rm max, curr}$. For other values of $\tau$, the times on each intermediate level are scaled up.

We note that so far the choice of $\tau$ is somewhat heuristic. However, in Section~\ref{sec:comparison}, we show that a refinement time of $\tau$ = 1.5 Myr yields good results, whereas for $\tau$ $<$ 1 Myr -- due to the low number of time steps per intermediate $l_{\rm max}$ -- numerical artefacts appear. We also perform runs with different maximum refinement levels and determine the resolution needed to reach convergence in cloud mass and chemical abundances in Section~\ref{sec:resolution}. All runs performed in this work are listed in Table~\ref{tab:runs}.

\begin{table}
\begin{minipage}{\linewidth}
 \caption{Overview of the progressive increase of the spatial resolution d$x_\rmn{curr}$ for four different total refinement times $\tau$. The times given below refer to the start of the zoom-in simulation at $t_0$.}
 \label{tab:levels}
 \centering
\begin{tabular}{ccccc}
 \hline
 $\tau$ = & 4.5 Myr & 2.25 Myr & 1.5 Myr & 1.0 Myr\\
 \hline
 \hline
  d$x_\rmn{curr}$ & time & time & time & time \\
  {[p}c] & [Myr] & [Myr] & [Myr] & [Myr] \\
 \hline
 2.0  & 0 - 1.25 & 0 - 0.625 & 0 - 0.5 & 0 - 0.4  \\
 1.0  & 1.25 - 2.5 & 0.625 - 1.25 & 0.5 - 1.0 & 0.4 - 0.65   \\
 0.5  & 2.5 - 3.5 & 1.25 - 1.75 & 1.0 - 1.25 & 0.65 - 0.85 \\
 0.24  & 3.5 - 4.5 & 1.75 - 2.25 & 1.25 - 1.5 & 0.85 - 1.0 \\
 0.12 & $>$ 4.5 & $>$ 2.25 & $>$ 1.5 & $>$ 1.0 \\
 \hline
 \vspace{0.1cm}
\end{tabular}
\end{minipage}
\begin{minipage}{\linewidth}
\caption{Overview of the simulations performed in this work, giving the run name, the maximum spatial resolution d$x$, and the refinement time $\tau$, at which the maximum spatial resolution is reached. The fiducial runs are MC1\_tau-1.5\_dx-0.12 and MC2\_tau-1.5\_dx-0.12. Then we have a set of runs used for the spatial resolution study (see Section \ref{sec:resolution}), and a set of runs used to show the impact of the refinement time (see section \ref{sec:comparison}) for both clouds, MC1 and MC2.}
\label{tab:runs}
\centering
\begin{tabular}{lll}
 \hline
 run name & d$x$ [pc] & $\tau$ [Myr]  \\
 \hline
 \hline
 MC1\_tau-1.5\_dx-0.12 & 0.12 & 1.5\\
 \hline
 MC1\_tau-1.65\_dx-0.06 & 0.06 & 1.65\\
 MC1\_tau-1.25\_dx-0.24 & 0.24 & 1.25\\
 MC1\_tau-1.0\_dx-0.5 & 0.5 & 1.0\\
 MC1\_tau-0.5\_dx-1.0 & 1.0 & 0.5\\
 MC1\_dx-3.9 & 3.9 & -- \\
\hline
 MC1\_tau-4.5\_dx-0.12 & 0.12 & 4.5\\
 MC1\_tau-2.25\_dx-0.12 & 0.12 & 2.25\\
 MC1\_tau-1.0\_dx-0.12 & 0.12 & 1.0\\
 MC1\_tau-0\_dx-0.12 & 0.12 & 0.0\\
  \hline
  \hline
 MC2\_tau-1.5\_dx-0.12 & 0.12 & 1.5\\
 \hline
 MC2\_tau-1.65\_dx-0.06 & 0.06 & 1.65\\
 MC2\_tau-1.25\_dx-0.24 & 0.24 & 1.25\\
 MC2\_tau-1.0\_dx-0.5 & 0.5 & 1.0\\
 MC2\_tau-0.5\_dx-1.0 & 1.0 & 0.5\\
 MC2\_dx-3.9 & 3.9 & -- \\
 \hline
 MC2\_tau-4.5\_dx-0.12 & 0.12 & 4.5\\
 MC2\_tau-2.25\_dx-0.12 & 0.12 & 2.25\\
 MC2\_tau-1.0\_dx-0.12 & 0.12 & 1.0\\
 MC2\_tau-0\_dx-0.12 & 0.12 & 0.0\\
 \hline
\end{tabular}
\end{minipage}
\end{table}

\section{Fiducial runs}
\label{sec:fiducial}

In order to study the impact of resolution and refinement time on the simulation results (see Section~\ref{sec:convergence}), we have to define certain quantities that are used to compare the simulations. In this section, we therefore discuss the results for runs MC1 and MC2 (see Section~\ref{sec:zoom}) with the fiducial parameters $\tau$ = 1.5 Myr and d$x$ = 0.12 pc (see Section~\ref{sec:instantaneous}). These are called MC1\_tau-1.5\_dx-0.12 and MC2\_tau-1.5\_dx-0.12 (see Table~\ref{tab:runs}). Starting from $t_0$, every simulation is evolved for 5 Myr, so the final time is $t_{\rm end} = t_0 + 5$ Myr. In all figures shown in this paper, the time refers to the time elapsed since $t_0$.

\begin{figure*}
 \includegraphics[width=\textwidth]{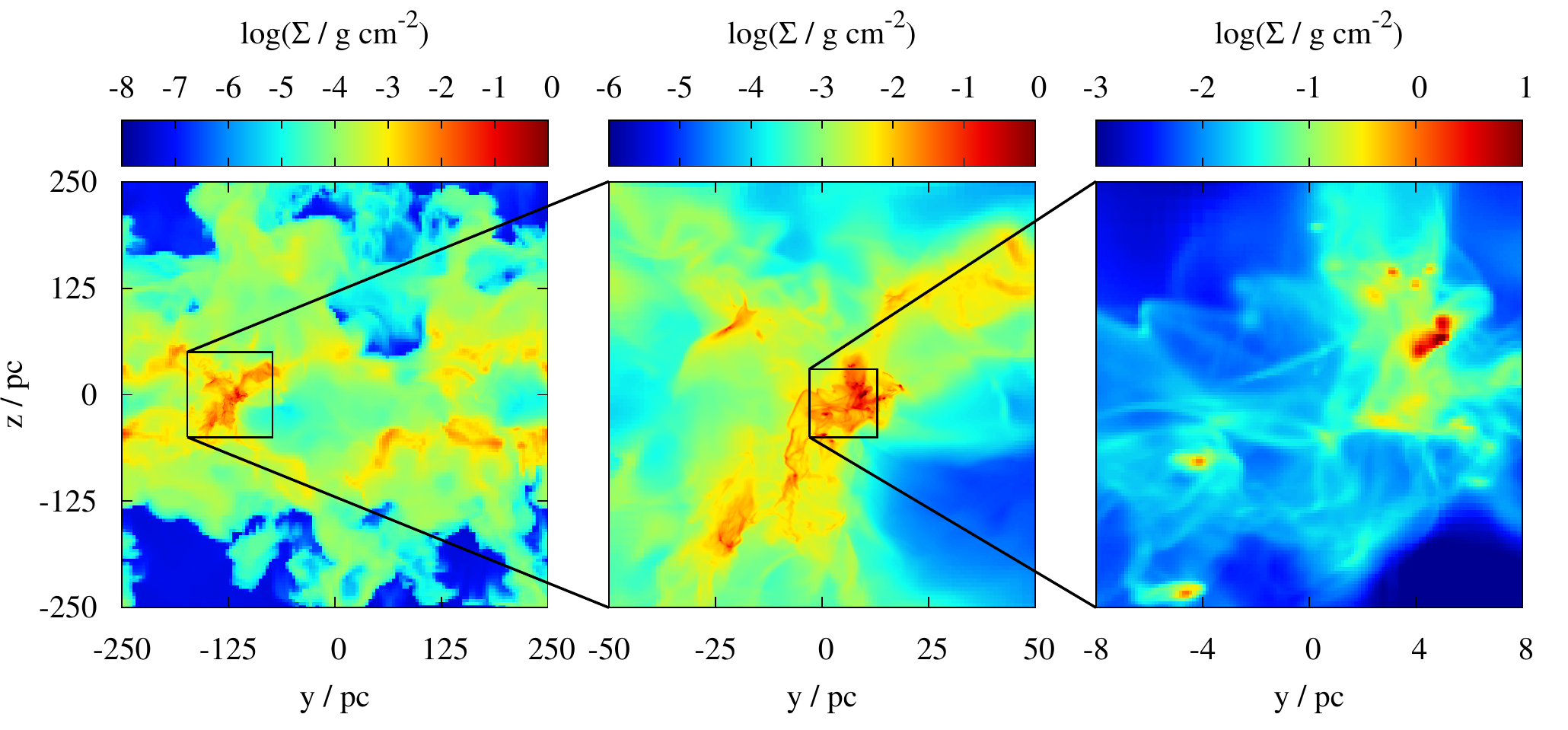}
 \caption{Zoom-in from the galactic disc (the full extent of the SILCC simulation in y-direction) onto the cloud of run MC1\_tau-1.5\_dx-0.12 and finally on the central part of the cloud at $t_{\rm end} = t_0+5$ Myr (from left to right). Shown is the total gas column density. We emphasize that the right panel represents only a small part of the full zoom-in region (middle panel). The full simulation domain extends beyond the plotted vertical range up to $\pm$ 5 kpc. Also note that the colour scale as well as the coordinate system is adapted in each panel. The figure demonstrates the power of the applied zoom-in technique to resolve the complex morphology of molecular clouds, which are evolving within the large-scale ISM. }
 \label{fig:zoom}
\end{figure*}

\begin{figure*}
\includegraphics[width=\textwidth]{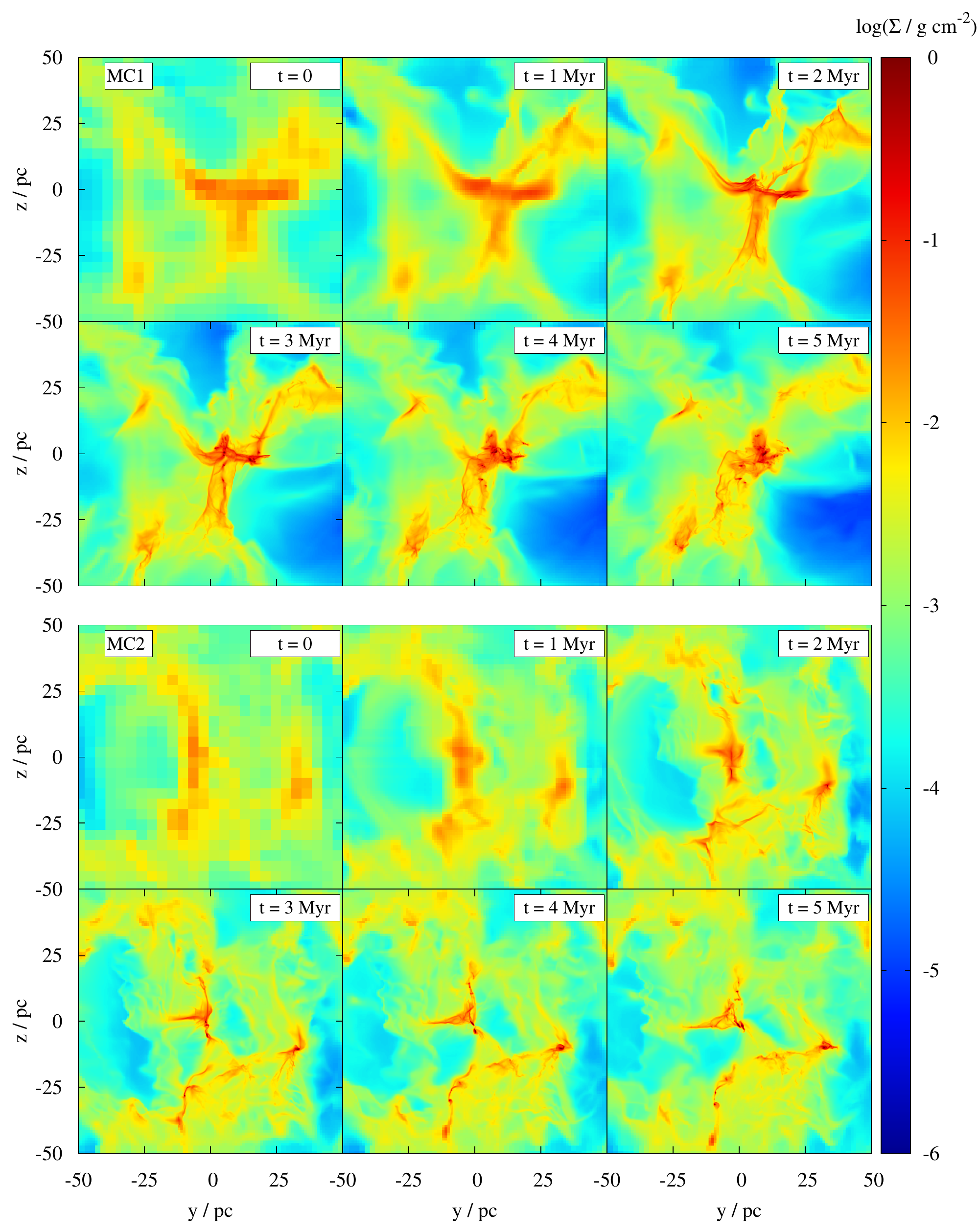}
\caption{Time evolution of the total gas column density in the zoom-in region of run MC1\_tau-1.5\_dx-0.12 (top) and MC2\_tau-1.5\_dx-0.12 (bottom), where the cloud is forming (from top left to bottom right).}
\label{fig:coldens}
\end{figure*}

\begin{figure*}
\includegraphics[width=\textwidth]{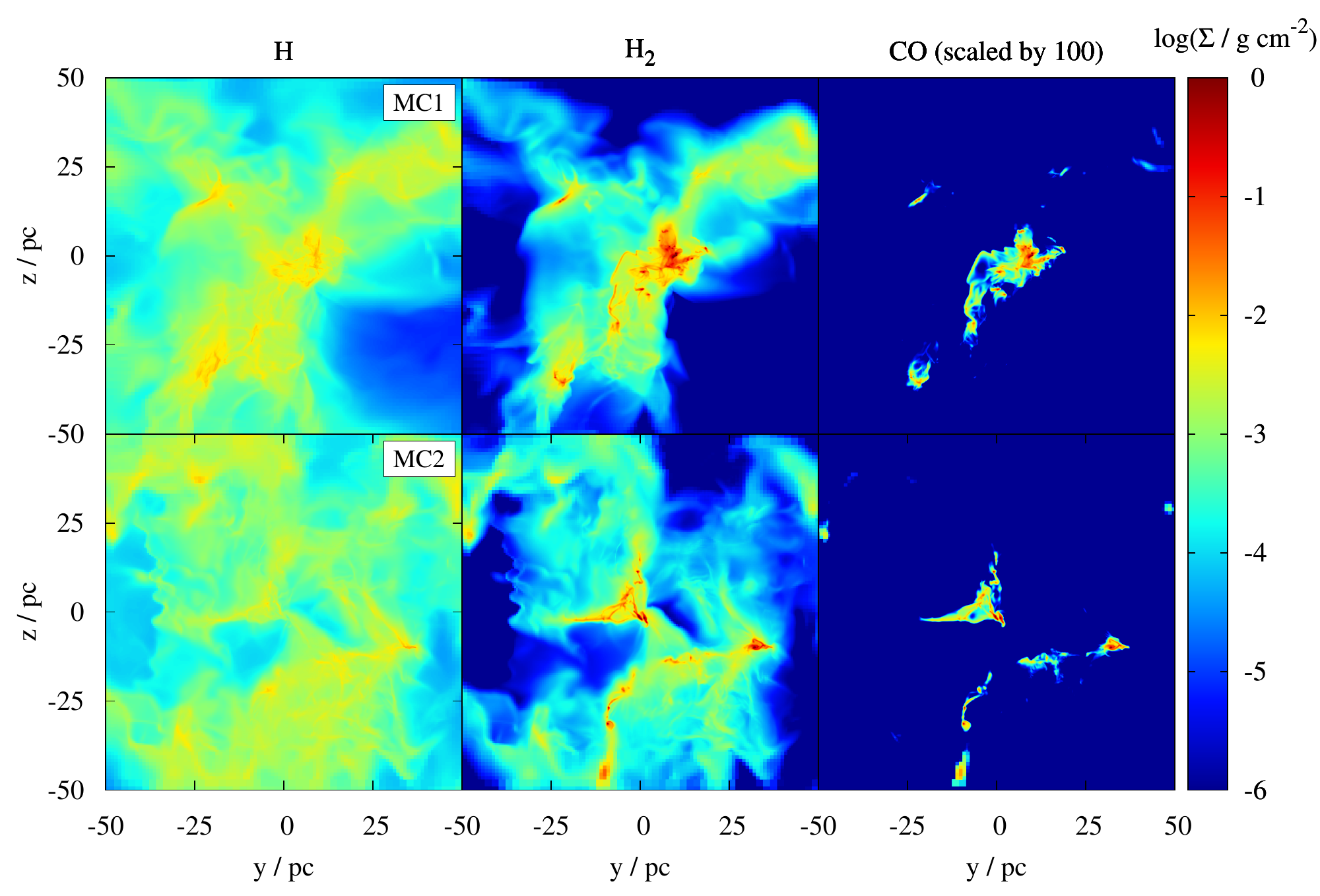}
\caption{Column density of the atomic hydrogen (left panels), H$_2$ (middle panels), and CO (right panels, scaled by 100) of run MC1\_tau-1.5\_dx-0.12 (top row) and run MC2\_tau-1.5\_dx-0.12 (bottom row) at $t_{\rm end}$. As expected, CO traces the densest regions of the MCs, which are embedded in a more extended distribution of H$_2$ and atomic hydrogen.}
\label{fig:chemistry}
\end{figure*}

\begin{figure*}
\includegraphics[width=0.85\textwidth]{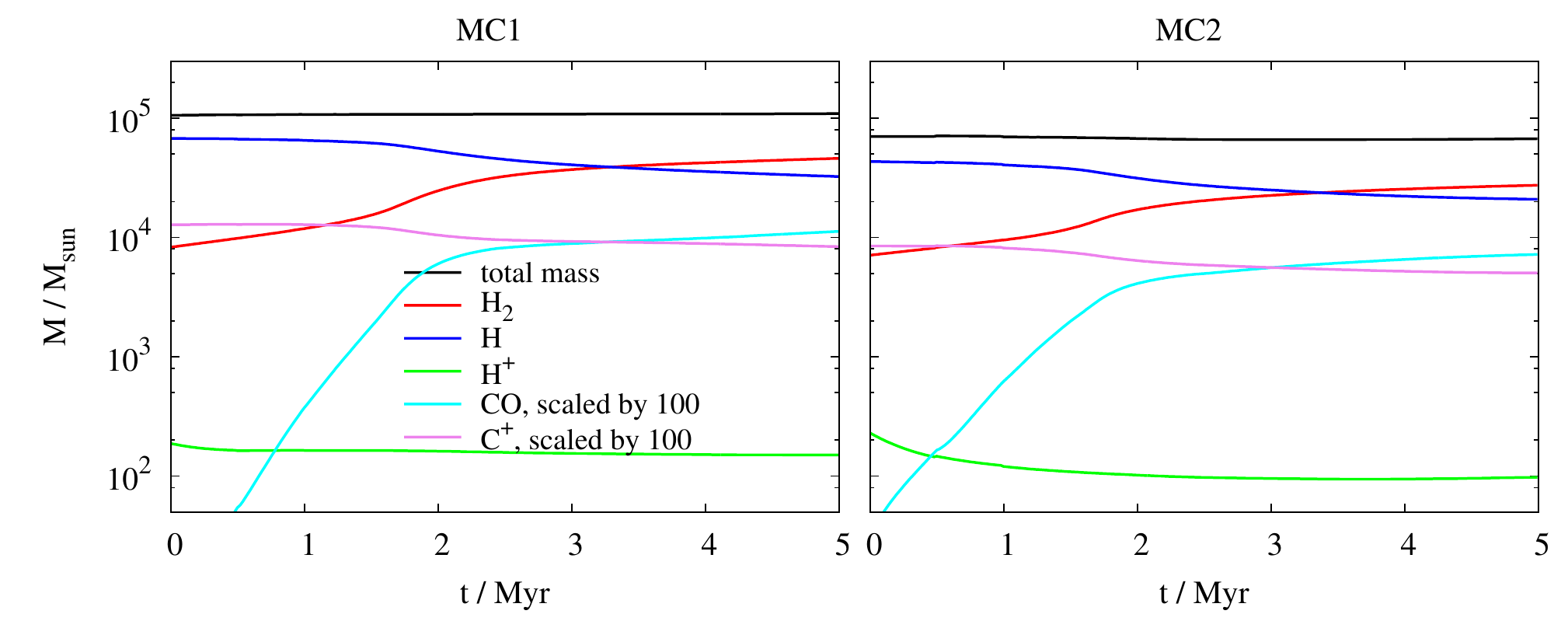}
\caption{Time evolution (relative to $t_0$) of the total mass and the mass of the various chemical species considered in this work for run MC1\_tau-1.5\_dx-0.12 (left) and MC2\_tau-1.5\_dx-0.12 (right). The considered volume is constrained to the zoom-in region. For demonstrative purposes we have scaled the mass of CO and C$^+$ by a factor of 100. H$^+$ has a negligible mass compared to that of H and H$_2$. CO and H$_2$ show a different evolution over time which causes a time-variable CO-to-H$_2$ ratio (see also Section~\ref{sec:chem}).}
\label{fig:time_chem}
\end{figure*}

\subsection{Overview}
\label{sec:overview}

In Fig.~\ref{fig:zoom} we show the edge-on view of the total gas column density $\Sigma$ of run MC1\_tau-1.5\_dx-0.12 at time $t_{\rm end}$ on different spatial scales of (from left to right) 500 pc, 100 pc, and 16 pc. The figure demonstrates the power of the zoom-in technique, which allows us to cover spatial scales over about 5 orders of magnitude to resolve the filamentary structure on pc scales and densities over about 8 orders of magnitude from the diffuse ISM to the densest structures. Further, in Fig.~\ref{fig:coldens} we show the time evolution of the column density of run MC1\_tau-1.5\_dx-0.12 (top) and run MC2\_tau-1.5\_dx-0.12 (bottom). Neither of the clouds has a simple spherical shape. MC1 has an almost ``T''-shaped structure with significant filamentary and clumpy substructure developing over time. While MC1 is dominated by one massive central clump, MC2 fragments into several clumps.
Furthermore, we find extended low-\mbox{(column-)}density regions, which are created by nearby SNe that have exploded before $t_0$.

In Fig.~\ref{fig:chemistry} we show the column density maps of atomic hydrogen, H$_2$, and CO at the end of both runs. We find that atomic hydrogen envelopes the molecular component, whereas CO is mainly present in the densest parts of the clouds \citep{Smith14b,Duarte16,Xu16}. We note that the chemical model used here, might slightly overestimate the production of CO \citep{Glover12}. However, we do not expect a more accurate chemical model to qualitatively change the aforementioned findings.

Also, we depict the evolution of the total mass and the mass in the five species which are traced with our chemical network in Fig.~\ref{fig:time_chem}. The total mass in the zoom-in region shows only minor variations over time, which indicates that we are only marginally affected by the flows through the boundaries of the zoom-in region. The H$_2$ mass in both MCs is \mbox{3 -- 4 $\times$ 10$^4$ M$_{\sun}$} towards the end. The mass in H$^+$ is negligible compared that of H and H$_2$. For both, hydrogen and carbon, the molecular component starts to dominate after about 3 Myr. However, the growth rates of the CO and H$_2$ masses are different. At $t_0$ approximately 10$^4$ M$_{\sun}$ of H$_2$ are already present while the mass of CO -- which forms at higher visual extinction, $A_\rmn{V}$, than H$_2$ \citep[e.g.][]{Roellig07,Glover10} -- is still negligible. In the following 2 Myr the H$_2$ mass shows only a moderate increase by a factor of 2 -- 3, whereas the CO mass increases by a factor of 100 once the high density regions start to form. At later stages the evolution of both species seems to be similar. We discuss the resulting time-variable CO-to-H$_2$ ratio in more detail in Section~\ref{sec:chem} and investigate its effect on the conversion factor between measured CO, $J$ = 1 -- 0 line intensity and H$_2$ mass, $X_\rmn{CO}$ \citep[see e.g.][]{Shetty11a,Shetty11b,Bolatto13} in Section~\ref{sec:XCO}.

\subsection{Molecular cloud parameters}
\label{sec:definition}

We consider the time evolution of global properties like mass, mean density, and velocity dispersion of the simulated MCs. In this context, a basic question is how to define a MC. In numerical works this has often been done by using a volume density \citep[e.g.][]{Tasker09,Ibanez15} or column density threshold \citep[e.g.][]{Dobbs08,Ward14,Bertram16}. Motivated by observations \citep[e.g.][]{Goldsmith08,Heyer09}, recent studies which include the chemical evolution of MCs also use intensity thresholds for CO line emission \citep{Clark12,Smith14b,Padoan16b,Richings16a}. In this paper, we use five simple definitions based on the 3D structure and chemistry of the cloud:
\begin{itemize}
 \item three ``volume density thresholds'', where all gas above a threshold value is counted as part of the MC. We consider the three density thresholds $\rho_{\rm thres}=$ 1.15 $\times$ 10$^{-22}$, \mbox{3.84 $\times$ 10$^{-22}$}, and \mbox{1.15 $\times$ 10$^{-21}$ g cm$^{-3}$}. Under the assumption of a mean molecular weight of 2.3, this corresponds to particle densities of $n_{\rm thres}=$ 30, 100, and 300 cm$^{-3}$. In all figure labels we refer to $n_{\rm thres}$.
 \item an "H$_2$ threshold", where we sum up the total gas mass in cells with a mass fraction of H$_2$ -- with respect to the total hydrogen mass in that cell -- that is greater or equal to 50\%, and
 \item a "CO threshold", where we sum up the total gas mass in cells with a mass fraction of carbon locked up in CO -- with respect to the total mass in carbon in that cell -- that is greater or equal to 50\%.
\end{itemize}

\begin{figure*}
\includegraphics[width=0.85\textwidth]{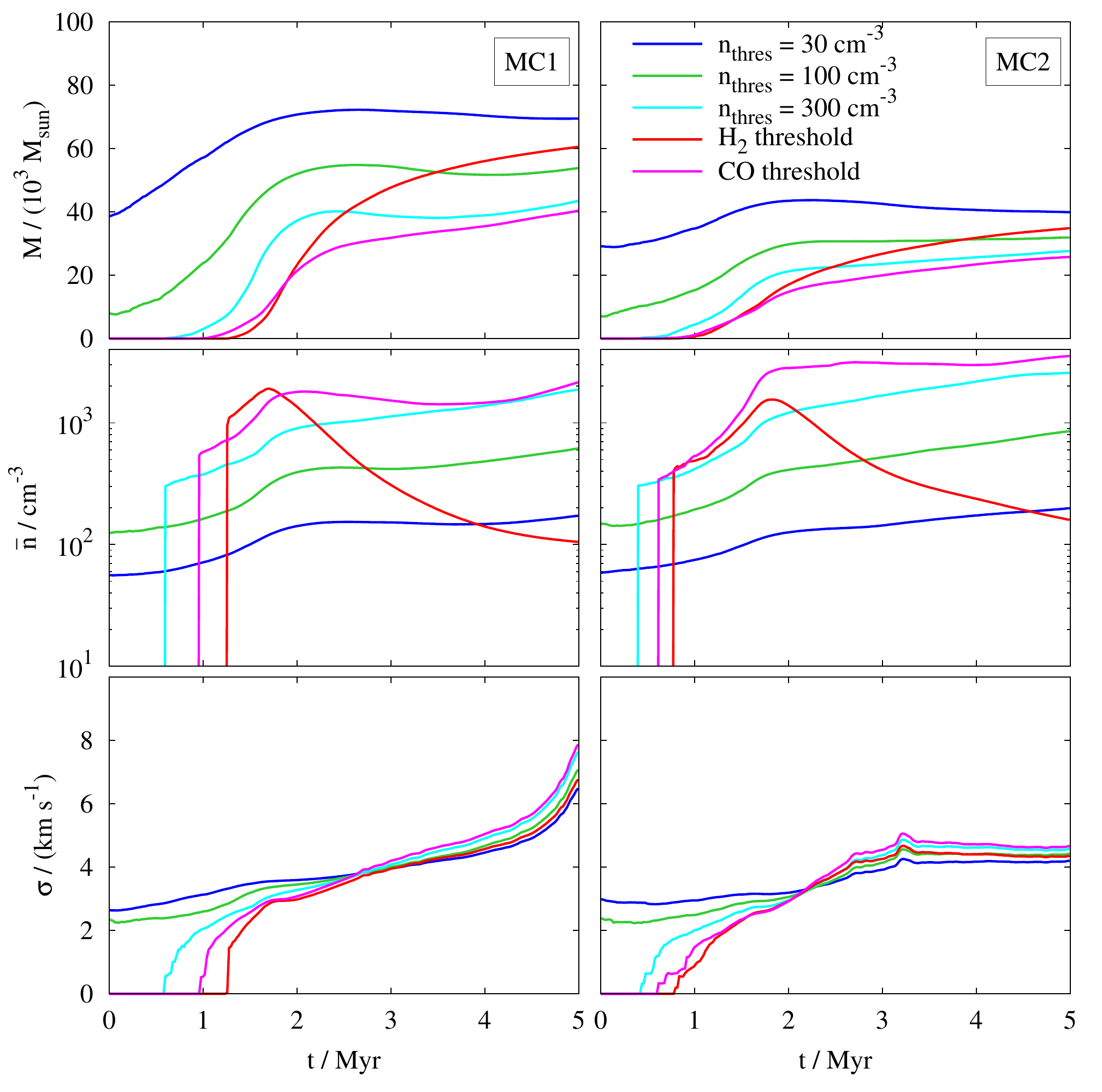}
\caption{Time evolution (relative to $t_0$) of the mass (top panels), mean volume-weighted density (middle), and velocity dispersion (bottom panels) of MC1 (left) and MC2 (right) for five different cloud definitions: gas above a particle density threshold of $n_{\rm thres}=$ 30, 100, or 300 cm$^{-3}$ (blue, green, cyan) and gas consisting to more than 50\% of H$_2$ (red) or CO (magenta). The time is given with respect to $t_0$. The mass above the H$_2$ threshold shows a more rapid increase than all other criteria, which implies that towards the end there is a significant amount of H$_2$ in gas with $n < 100$ cm$^{-3}$. For the density criteria, the mean particle density is well above the threshold values and overall increasing with time. The latter is also the case for the CO threshold. However, the mean density for the H$_2$ threshold decreases as a function of time since H$_2$ spreads out over a larger volume into the low density regions. The turbulent velocity dispersions are not very sensitive to the cloud definition apart from the general trend that denser gas also shows slightly higher dispersions.}
\label{fig:timeevol}
\end{figure*}

In Fig.~\ref{fig:timeevol}, we plot the masses (top row), mean number densities $\bar{n}$ (middle), and velocity dispersions $\sigma$ (bottom row) for MC1 (run MC1\_tau-1.5\_dx-0.12, left column) and MC2 (run MC2\_tau-1.5\_dx-0.12, right column) as a function of time. The time refers to the start of the zoom-in simulation at $t_0$. Each of these quantities is derived for the 5 different criteria. We refer to each criterion via a subscript, e.g. the masses for the three $n_{\rm thres}$ thresholds are $M_{30}$, $M_{100}$, $M_{300}$, and for H$_2$ and CO criterion $M_\rmn{H_2}$, and $M_\rmn{CO}$.

\subsection{Mass assembly and evolution of the mean density}
\label{sec:masses}

The MCs condense out of the atomic ISM within $\sim$ 2 Myr from the start of the zoom-in simulation at $t_0$. Therefore, the masses above the given density thresholds, $M_{30}$, $M_{100}$, and $M_{300}$ grow within this period of time. Depending on $n_{\rm thres}$, the final masses of MC1 are between $M_{30} \approx 7.0\times 10^4\;{\rm M}_{\odot}$ and $M_{300} \approx 4.3\times 10^4\;{\rm M}_{\odot}$. For MC2 we reach masses between $M_{30} \approx 4.0\times 10^4\;{\rm M}_{\odot}$ and $M_{300} \approx 2.8\times 10^4\;{\rm M}_{\odot}$. Overall, $M_{300}$ accounts for about 40\% of the total mass within the zoom-in region in both cases. The derived masses and sizes (see Fig.~\ref{fig:coldens}) indicate that these clouds are representative of typical MCs in the Milky Way \citep[e.g.][]{Solomon87,Elmegreen96,Heyer01,Roman10,Miville17}. 

Using the H$_2$ and CO thresholds, $M_\rmn{H_2}$ and $M_\rmn{CO}$ start to increase later than $M_{300}$ and show a different time dependence. Again, we remind the reader that $M_\rmn{H_2}$ and $M_\rmn{CO}$ are not the total mass in H$_2$ and CO (as shown in Fig.~\ref{fig:time_chem}), but those obtained by the criteria discussed before. $M_\rmn{CO}$ grows more slowly than $M_{300}$, but steadily increases with a mean rate of $5 - 7\times 10^{-3}$ M$_{\sun}$ yr$^{-1}$. Therefore, $M_\rmn{CO}$ approaches $M_{300}$ towards the end of the simulations. We note that initially $M_\rmn{H_2}$ is close to zero, although about 7 $\times$ 10$^3$ M$_{\sun}$ of H$_2$ are present within both zoom-in regions at that time (see Fig.~\ref{fig:time_chem}). This is due to the fact that at this stage hydrogen is predominantly in atomic form, and thus the diffuse H$_2$ component is not taken into account in $M_\rmn{H_2}$. Once the cloud starts to become molecular, $M_\rmn{H_2}$ shows a higher growth rate than $M_\rmn{CO}$ with $\dot{M}_\rmn{H_2}\sim1 \times 10^{-2}$ for run MC1\_tau-1.5\_dx-0.12 (and $\dot{M}_\rmn{H_2}\sim7 \times 10^{-3}$ M$_{\sun}$ yr$^{-1}$ for MC2\_tau-1.5\_dx-0.12) and thus, overtakes $M_{300}$ after $\sim$ 2.5 Myr as well as $M_{100}$ after $\sim$ 3.5 Myr. These growth rates are similar to recent observations \citep[e.g.][]{Fukui09,Kawamura09} but the different time dependence of $M_\rmn{H_2}$ and $M_\rmn{CO}$ supports the idea of a time-dependent X-factor \citep[see Fig.~\ref{fig:time_chem} and Section~\ref{sec:chem} and e.g.][]{Richings16a,Richings16b}.

The mean, volume-weighted densities $\bar{n}$ (middle row of Fig.~\ref{fig:timeevol}) inferred using the $n_{\rm thres}$ criteria show an increase over time. Given that the masses, $M_{30}$, $M_{100}$, and $M_{300}$, differ only by a factor of $\sim$ 2, the differences in $\bar{n}$ by more than one order of magnitude reflect that the occupied volume, i.e. the volume of all cells which fulfil the applied criterion, increases significantly with decreasing density threshold.

For the CO threshold, we find that $\bar{n}_{\rm CO}$ is similar to, but slightly higher than, $\bar{n}_{300}$. On the other hand, $\bar{n}_{\rm H_2}$ decreases as a function of time and even drops below $\bar{n}_{30}$ after $\sim$ 4 - 4.5 Myr. The decrease of $\bar{n}_\rmn{H_2}$ is due to the progressive extension of H$_2$ into the low-density regime. This is in agreement with our analysis so far, which implies that gas with $n<100\;{\rm cm}^{-3}$ can become predominantly molecular and shows that the volume occupied by this gas significantly increases over time.

\begin{figure}
\centering
 \includegraphics[width=\linewidth]{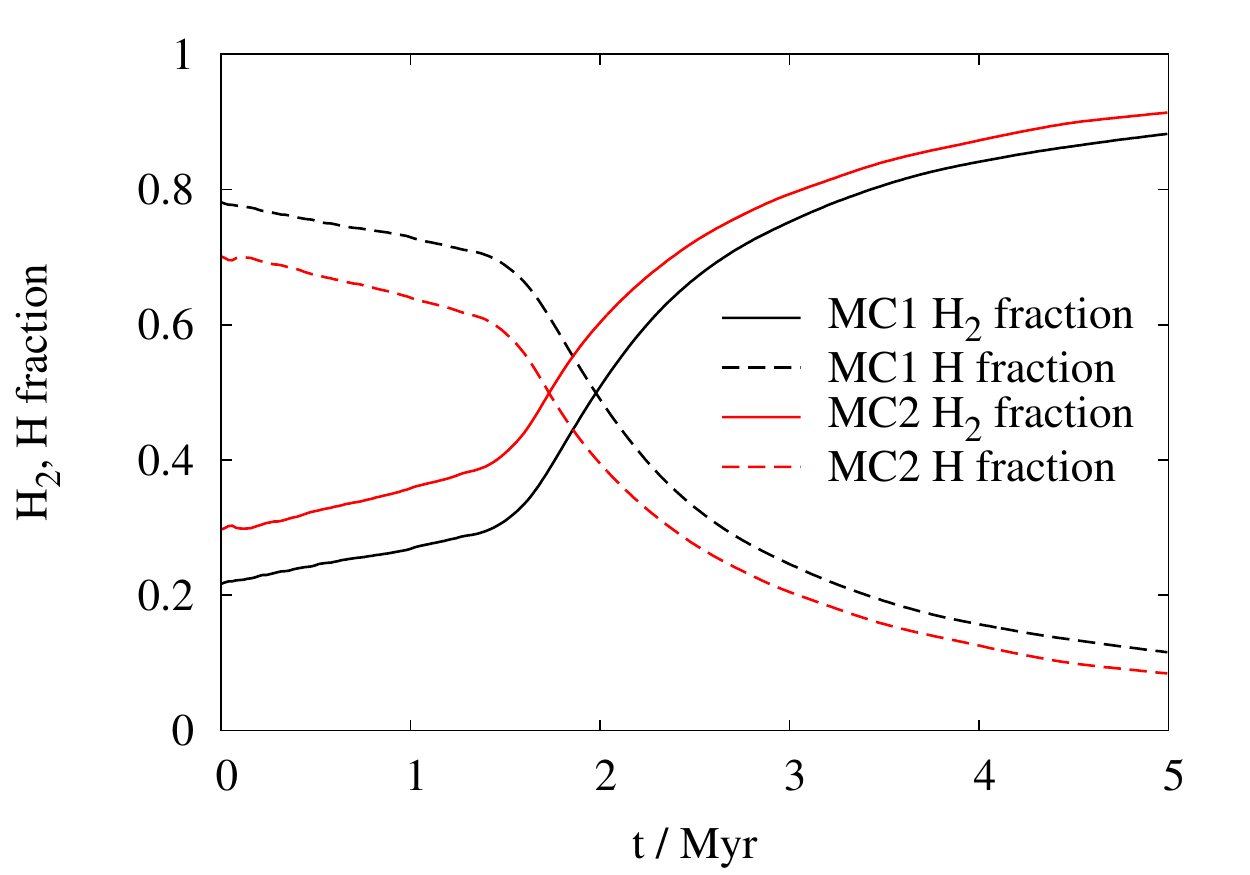}
 \caption{Fraction of H$_2$ (solid lines) and atomic hydrogen (dashed lines) in the dense gas ($n$ $>$ 100 cm$^{-3}$) of MC1 (black lines) and MC2 (red lines).}
 \label{fig:H2_fraction}
\end{figure}

\begin{figure}
\centering
 \includegraphics[width=\linewidth]{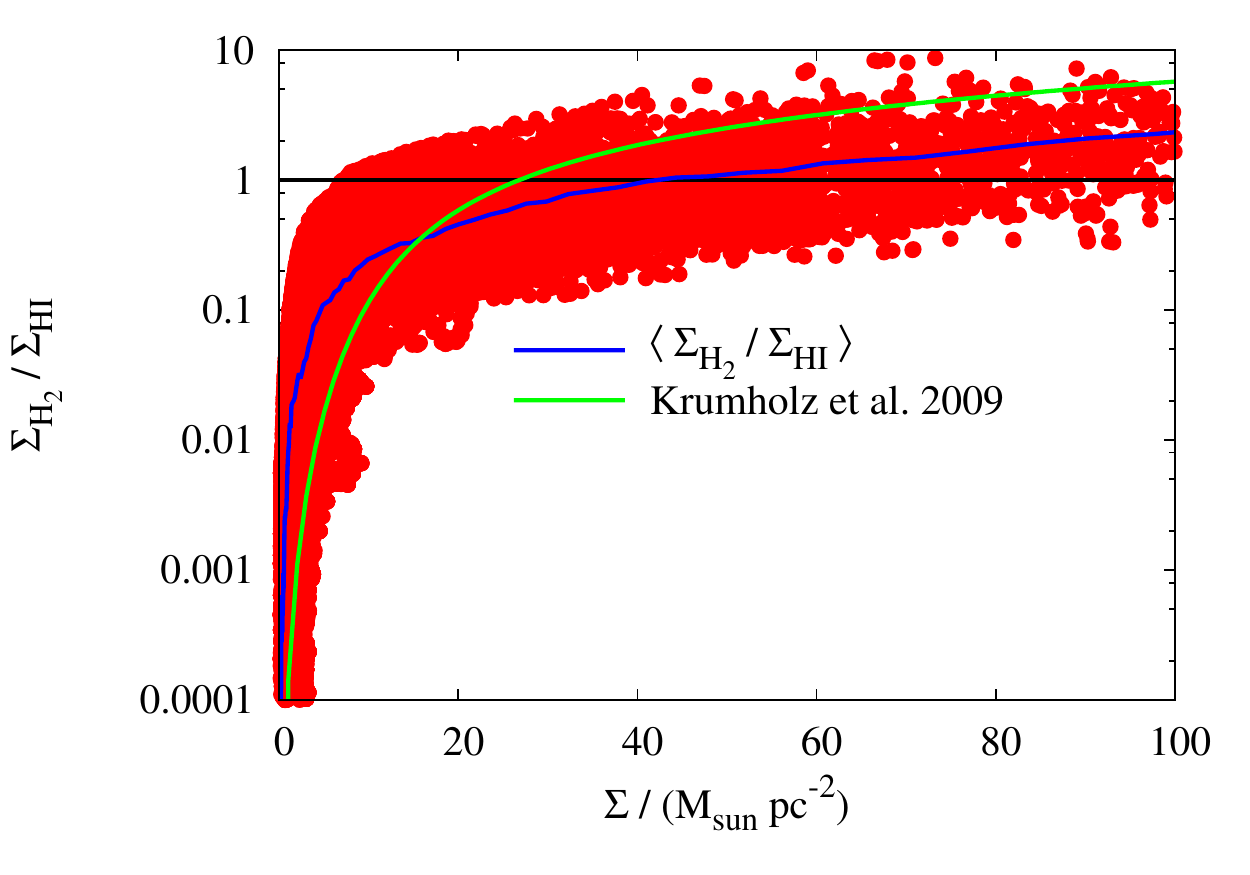}
 \caption{Ratio of the column density of H$_2$ to atomic hydrogen as a function of the total gas column density for MC1 at $t$ = $t_0$ + 5 Myr for the map shown in Fig.~\ref{fig:chemistry}. The blue line shows the mean for each column density bin. The transition to H$_2$ dominated gas occurs at a column density around 40 M$_{\sun}$ pc$^{-2}$. The green line shows the theoretical predictions by \citet{Krumholz09} for solar metallicity.}
 \label{fig:H2transition}
\end{figure}

\begin{figure}
\centering
  \includegraphics[width=\linewidth]{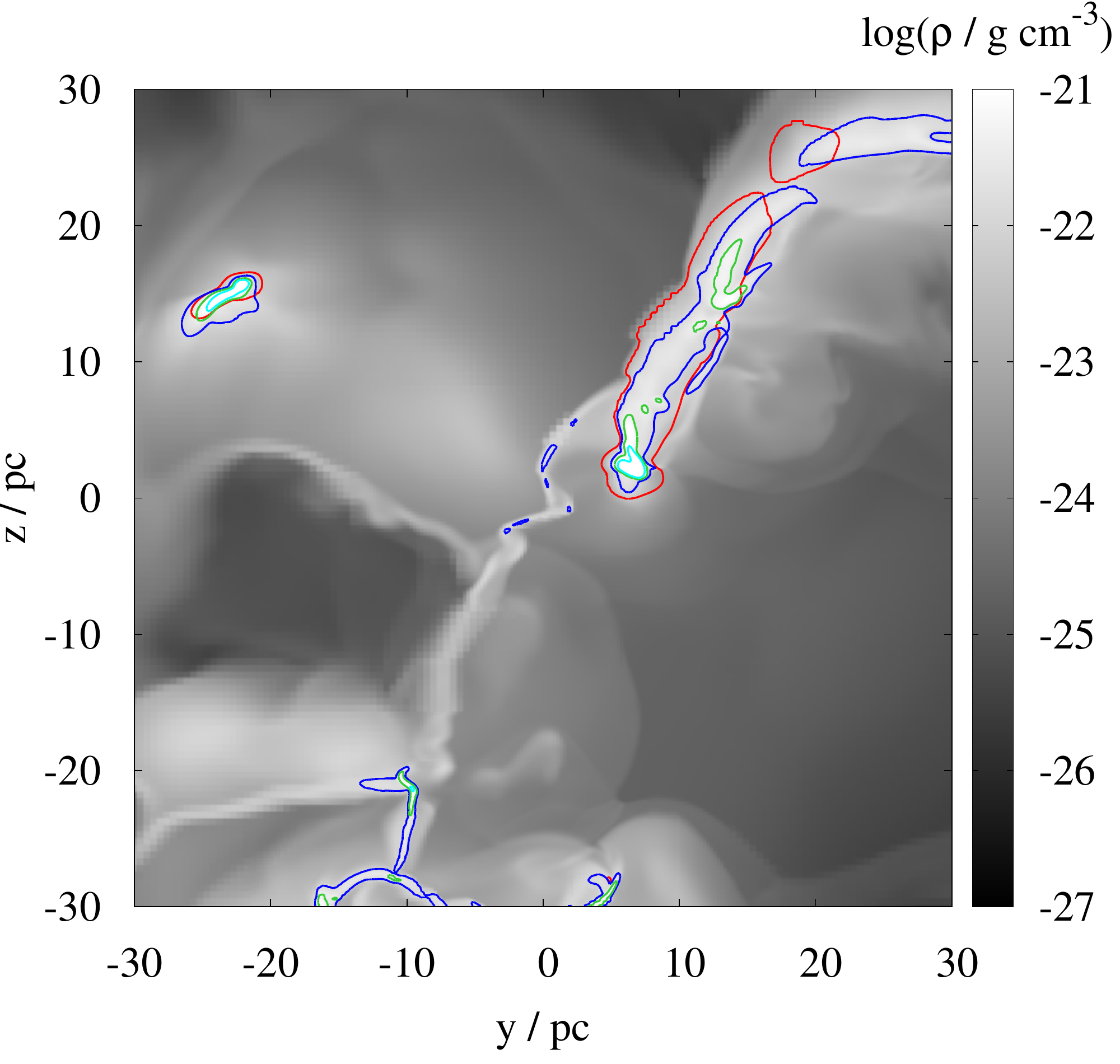}
 \caption{Density slice in the y-z-plane through the centre of the cloud in run MC1\_tau-1.5\_dx-0.12 at $t$ = 5 Myr. The blue, green, and cyan lines represent number density contours at 30, 100, and 300 cm$^{-3}$, respectively (1.15, 3.84, and 11.5 $\times$ 10$^{-22}$ g cm$^{-3}$ in corresponding mass densities). Inside the red contour the H$_2$ mass fraction is above 50\%. The cloud is no single connected region but fragmented into several dense clumps. H$_2$ is also found in low-density gas ($n \lesssim 30$ cm$^{-3}$).}
 \label{fig:slice}
\end{figure}

\subsection{Evolution of H$_2$}
\label{sec:rhomean}

In Fig.~\ref{fig:H2_fraction} we show the mass fraction of H$_2$ with respect to the total amount of hydrogen in the dense ($n$ $>$ 100 cm$^{-3}$) gas of MC1 and MC2. Initially, the gas is predominantly atomic, with an H$_2$ mass fraction of $\sim$ 0.2 -- 0.3. After about 1.5 Myr it increases more rapidly reaching an almost fully molecular state with an H$_2$ mass fraction of about 0.9 at $t$ = $t_0$ + 5 Myr. We note that for density thresholds of 30(300) cm$^{-3}$ the evolution is qualitatively very similar with 5 -- 10\% lower(higher) H$_2$ fractions. In Fig.~\ref{fig:H2transition} we plot the ratio of the H$_2$ to atomic hydrogen column density, $\Sigma_\rmn{H_2}/\Sigma_\rmn{HI}$, against the total column density for MC1 at \mbox{$t$ = $t_0$ + 5 Myr}. The transition to H$_2$ dominated gas occurs around 30 -- 50 M$_{\sun}$ pc$^{-2}$ ($\Sigma_\rmn{H_2}/\Sigma_\rmn{HI}$ $>$ 1, indicated by the blue line representing the mean value at each $\Sigma$). This is somewhat higher than the value of 10 M$_{\sun}$ pc$^{-2}$ found in the Perseus cloud \citep{Lee12,Lee15} and that of theoretical predictions for solar metallicity by \citet[][their Equation 39, green line in Fig.~\ref{fig:H2transition}]{Krumholz09}. It agrees, however, well with recent observations of W43 \citep{Bihr15}. We note that for other times and MC2 we find a very similar behaviour.

To further analyse the properties of the H$_2$ gas, we show a density slice through the centre of MC1 at \mbox{$t$ = 5 Myr} in Fig.~\ref{fig:slice}. The contours show the volume occupied by gas above $n_{\rm thres}=$ 30, 100, and 300 cm$^{-3}$ (blue, green, and cyan lines) and above the H$_2$ threshold (red line). It can be seen that the H$_2$ dominated gas extends into regions with gas densities below 30 cm$^{-3}$. The presence of H$_2$ in such low-density gas is difficult to explain with in-situ formation of the molecules in this regions: The time scale $\tau_\rmn{form}$ for H$_2$ formation in gas with \mbox{$n = 30$ cm$^{-3}$} is about \mbox{1 Gyr $\times \left( \frac{30 \, \rmn{cm}^{-3}}{1 \, \rmn{cm}^{-3}} \right)^{-1} \approx 33$ Myr} \citep{Hollenbach79,Glover07a,Glover07b,Glover10}, and thus significantly longer than the entire simulated time.

We rather argue that H$_2$ forms in dense regions and is efficiently re-distributed into the lower density environment by turbulent mixing. The turbulent mixing timescale $\tau_\rmn{mix}$, on which this is expected to happen, can be estimated by assuming that the typical distance $l_\rmn{mix}$ of H$_2$ in the low-density environment from the high density clumps is a few pc (Fig.~\ref{fig:slice}) and the typical velocity $v_\rmn{mix}$ a few km s$^{-1}$ (see bottom row of Fig.~\ref{fig:timeevol}). Hence, we obtain \mbox{$\tau_\rmn{mix} = l_\rmn{mix}/v_\rmn{mix} \simeq$ 1 Myr $<<$ $\tau_\rmn{form}$}. Furthermore, the dissociation timescale of H$_2$ due to UV radiation, which is the dominant dissociation process, is given by
\begin{equation}
 \tau_\rmn{diss} = \left(3.3 \times 10^{-11} \rmn{s}^{-1} \, \rmn{G}_0 \, f_\rmn{shield,H_2} \, e^{-3.5 A_\rmn{V}}\right)^{-1} \, ,
\end{equation}
where $f_\rmn{shield,H_2}$ is the UV shielding due to H$_2$ molecules \citep{Dishoeck88,Glover10}. In the regions with \mbox{$n$ $\gtrsim$ 30 cm$^{-3}$} we typically find \mbox{$A_\rmn{V}$ $>$ 0.3} and \mbox{$f_\rmn{shield,H_2}$ $<$ 3 $\times$ 10$^{-4}$}, which results in \mbox{$\tau_\rmn{diss}$ $\gtrsim$ 5 Myr $>$ $\tau_\rmn{mix}$}. Therefore, the relatively short time scale of a few Myr on which H$_2$ starts to dominate in gas with $n \lesssim 30\;{\rm cm}^{-3}$ implies that the mixing of H$_2$ into the low density gas is significantly more important than the in-situ formation of the molecules. This result is similar to the enhanced H$_2$ formation process reported in more idealised simulations of molecular cloud formation using turbulent periodic boxes \citep{Glover07b} and simulations of isolated clouds \citep{Valdivia16}. We note that a part of the H$_2$ found in low-density regions might also be caused by re-expansion of dense, molecular regions regions on timescales shorter than $\tau_\rmn{diss}$.

Since turbulent mixing seems to play an essential role for the formation of H$_2$, the determination of chemical abundances in a post-processing step can be inaccurate, unless the full time history of the fluid element is known (e.g. by the usage of tracer particles). A simple approach based on the current state of the cell, i.e. using $\tau_\rmn{form}$ as an approximation \cite[e.g.][]{Padoan16b}, will generally underestimate the amount of H$_2$. Moreover, this also directly affects abundance of CO: first, H$_2$ contributes to the shielding of radiation which dissociates CO \citep{Dishoeck88,Glover10}. Second, H$_2$ is involved in initiating a number of chemical pathways that lead to the formation of CO. Hence, underestimating H$_2$ will also result in a underestimation of the CO abundance. Our findings also demonstrate that for analytical estimates of atomic and molecular hydrogen column densities \citep[e.g.][]{Krumholz09,Bialy17} turbulent mixing should be taken into account.

\subsection{Velocity dispersion}
\label{sec:sigma}

Next, we consider the velocity dispersion $\sigma$ of the two MCs (bottom row of  Fig.~\ref{fig:timeevol}). We define $\sigma$ with respect to their centre-of-mass (COM) velocity, $\vec{v}_\rmn{COM}$,
\begin{equation}
 \sigma = \sqrt{ \frac{ \Sigma \, ( \textrm{d}m \times (\vec{v} - \vec{v}_\rmn{COM})^2)}{\Sigma \, \textrm{d}m} } \, ,
\end{equation}
where d$m$ and $\vec{v}$ are the mass and velocity of an individual cell, respectively, and we sum over all cells identified with the corresponding definition. Note that $\sigma$ contains contributions from all the individual fragments (see Fig.~\ref{fig:coldens}) as well as their relative motions with respect to the COM. A more detailed investigation of what dominates the value of $\sigma$ (i.e. turbulence, interclump motions, infall, or outflow motions) is postponed to a subsequent paper.

The velocity dispersion  in both  clouds reaches final values around \mbox{4 -- 8 km s$^{-1}$} in good agreement with observations \citep[e.g.][]{Solomon87,Elmegreen96,Heyer01,Roman10,Miville17}. With typical gas temperatures around 10 K, i.e. a sound speed $c_\rmn{s}$ around \mbox{0.2 km s$^{-1}$}, this corresponds to highly supersonic motions with Mach numbers above 10. This strengthens our statement in Section~\ref{sec:rhomean} that turbulent mixing is the main reason for the presence of H$_2$ in regions with $n$ $\simeq$ 30 cm$^{-3}$.

For MC1, $\sigma$ is steadily increasing with time. This could be attributed to two effects: First, turbulent motions on small scales become resolved with increasing resolution. However, since only a small fraction of energy is stored on smaller scales \citep{Kolmogorov41}, we rather attribute the increase of $\sigma$ for MC1 to a continuous gravitational collapse (see Fig.~\ref{fig:coldens}). The resulting infall velocities might be converted to turbulent motions due to the complex structure of the cloud \citep{Klessen10,Goldbaum11,Matzner15}, and thus counteract the decay of turbulent, kinetic energy, which is believed to happen within one crossing time \citep{Stone98,MacLow98,MacLow04,Elmegreen04}.

This picture is also supported by the saturation of $\sigma$ for MC2 just above 4 km s$^{-1}$ after $\sim$ 3 Myr: MC2 has fragmented into a number of spatially separated, small objects (see Fig.~\ref{fig:coldens}), which might be less gravitationally unstable. We emphasize, however, that each of the fragments is collapsing itself, which is reflected by the increase in column density and the continuously increasing value of $M_\rmn{300}$, $M_\rmn{H_2}$, and $M_\rmn{CO}$ (top right panel of Fig.~\ref{fig:timeevol}). On the other hand, MC1 seems to collapse towards a common centre (with higher infall velocities), which is reflected by the larger amount of dense gas (Fig.~\ref{fig:timeevol}). Furthermore, we find that at later times, for both MCs, $\sigma$ increases slightly with increasing density threshold. This further corroborates the picture that gravity/infall motions contribute to the overall velocity dispersion, in particular the observed increase. This interpretation is in line with a number of recent works on the origin of $\sigma$ in MCs \citep[e.g.][]{Heyer09,Ballesteros11,Kritsuk13,Ibanez15,Ibanez17}, although for the initial level of turbulence ($\sigma$  = 2 -- 3 km s$^{-1}$, Fig.~\ref{fig:timeevol}) most likely also SN feedback is important.

Finally, we note that the velocity dispersion derived for the various cloud criteria differs only little, which we attribute to the fact that it is a bulk property (i.e. its magnitude is independent of the size of the system).

\section{Convergence study}
\label{sec:convergence}

In recent years a number of authors have studied the formation of MCs in galactic environments. However, little work has been done to investigate which resolution is required to accurately model the chemical, dynamical, and structural properties of the forming MCs. Typical maximum resolutions of a number of galactic disc simulations modelling the formation of MCs are of the order of a few pc or a few M$_{\sun}$ in case of smoothed-particle hydrodynamics (SPH) (see Table~\ref{tab:resolution}). Here we present a comprehensive study to identify resolution requirements for modelling the formation of MCs in galactic discs.

In the following, we test the convergence of our results for increasing spatial resolution (see section \ref{sec:resolution}) as well as different refinement times $\tau$ (see section \ref{sec:comparison}). All runs are listed in Table~\ref{tab:runs}.

\subsection{Comparison with other work}
\label{sec:otherworks}
In order to compare our work with SPH simulations, we convert the spatial resolution to a mass resolution using the Jeans length and mass. In a grid code the Jeans length $L_\rmn{J}$ typically has to be resolved with about 10 grid cells \citep{Truelove97,Federrath11}, i.e.
\begin{equation}
 10 \, \rmn{d}x = L_\rmn{J} = c_\rmn{s} \sqrt{\frac{\pi}{G \rho}} \, ,
 \label{eq:grid}
\end{equation}
while in an SPH simulation the Jeans mass $M_\rmn{J}$ has to be resolved with a certain number $N_\rmn{neigh}$ of particles with mass $m_\rmn{SPH}$, i.e
\begin{equation}
 N_\rmn{neigh} \times m_\rmn{SPH} = M_\rmn{J} = \frac{4 \pi}{3} \, \rho \, \left(\frac{L_\rmn{J}}{2}\right)^3 =  \frac{\pi^{5/2}}{6} \frac{c_s^3}{G^{3/2} \rho^{1/2}} \, .
 \label{eq:SPH}
\end{equation}
Next, Eq.~\ref{eq:grid} can be used to eliminate $\rho$ on the right-hand side of Eq.~\ref{eq:SPH}, which allows us to solve for $m_\rmn{SPH}$:
\begin{equation}
m_\rmn{SPH} = 10 \, \rmn{d}x \, \frac{\pi^2}{6}  \frac{c_s^2}{N_\rmn{neigh} G} \, .
\end{equation}
Here, we adopt $N_\rmn{neigh} = 100$ typical for 3D simulations \citep{Bate97} and consider isothermal ($\gamma$ = 1), molecular gas with \mbox{$T$ = 10 K} and \mbox{$\mu$ = 2.3} to obtain a simple equation to convert between the effective spatial resolution of a grid code and the corresponding particle mass in an SPH code,
\begin{equation}
 \frac{m_\rmn{SPH}}{1 \rmn{M}_{\sun}} \simeq 1.4  \left(\frac{\rmn{d}x}{\rmn{1 pc}}\right) \, .
 \label{eq:conversion}
\end{equation}

In Table~\ref{tab:resolution} we list a number of recent works, which study the formation of MCs in their galactic environment. We group the papers by their spatial/mass resolution to compare them with the different spatial resolutions used in this work. Most of the works listed neither reach our highest spatial resolution nor include a chemical network. Our work is thus the first which combines high spatial resolution as well as an on-the-fly modelling of molecule formation in an extensive resolution study.
\begin{table*}
 \caption{Effective spatial resolution d$x$, corresponding particle mass $m_\rmn{SPH}$ in an SPH simulation (obtained via Eq.~\ref{eq:conversion}), and a selection of recent numerical works with that effective resolution. An italic font-style implies that no chemical network is included in the simulations.}
 \begin{tabular}{ccc}
 \hline
 d$x$ & $m_\rmn{SPH}$ & references \\
 \hline
 d$x$ $\ge$ 3.9 pc & $m_\rmn{SPH} \ge$ 5.4 M$_{\sun}$ & {\citet{Dobbs08b,Dobbs13,Duarte15}};\\
  && \textit{\citet{Hopkins13}}$^a$; {\citet{Pettitt14,Richings16a,Richings16b}}\\
 3.9 pc $>$ d$x$ $\ge$ 1 pc & 5.4 M$_{\sun} > m_\rmn{SPH}$ $\ge$ 1.4 M$_{\sun}$ & {\citet{Dobbs15,Duarte16}}; \textit{\citet{Hennebelle14}}$^a$; \\
 && \textit{\citet{Kim15b}}$^a$, {\citet{Hu16}} \\
 1 pc $>$ d$x$ $\ge$ 0.5 pc & 1.4 M$_{\sun} > m_\rmn{SPH}$ $\ge$ 0.79 M$_{\sun}$ & \textit{\citet{Ibanez15,Ibanez17}}$^a$ \\
 0.5 pc $>$ d$x$ $\ge$ 0.24 pc & 0.79 M$_{\sun} > m_\rmn{SPH}$ $\ge$ 0.33 M$_{\sun}$ & {\citet{Clark12,Smith14b}}\\
 0.24 pc $>$ d$x$ & 0.33 M$_{\sun} > m_\rmn{SPH}$ & \textit{\citet[][0.05 pc resolution]{Renaud13}}$^a$ \\
\hline
 \end{tabular}
  \\
$^a$ no chemical network included \\
 \label{tab:resolution}
\end{table*}

\subsection{Spatial resolution study}
\label{sec:resolution}

Our fiducial runs, which we discussed so far, are MC1\_tau-1.5\_dx-0.12 and MC2\_tau-1.5\_dx-0.12 with a maximum spatial resolution of d$x=0.12$ pc and a refinement time of $\tau=1.5$ Myr. \\

We repeat the simulations for both MC1 and MC2 using a different maximum refinement level. We perform runs with a lower effective spatial resolution of \mbox{d$x$ = 0.24, 0.5, 1.0, and 3.9 pc}, i.e. we stop the refinement procedure after $\tau=$1.25, 1.0, 0.5, and 0.0 Myr. These additional runs are called MC1\_tau-1.25\_dx-0.24, MC1\_tau-1.0\_dx-0.5, MC1\_tau-0.5\_dx-1.0, and MC1\_dx-3.9, as well as MC2\_tau-1.25\_dx-0.24, MC2\_tau-1.0\_dx-0.5, MC2\_tau-0.5\_dx-1.0, and MC2\_dx-3.9. In addition, we carry out two runs with a higher effective spatial resolution of d$x$ = 0.06 pc with $\tau=1.65$ Myr. These runs are called MC1\_tau-1.65\_dx-0.06 and MC2\_tau-1.65\_dx-0.06\footnote{Since these two simulations become numerically unstable, we were not able to run them for more than 2.7 Myr.}. All runs are listed in Table~\ref{tab:runs}.

\begin{figure*}
\includegraphics[width=\textwidth]{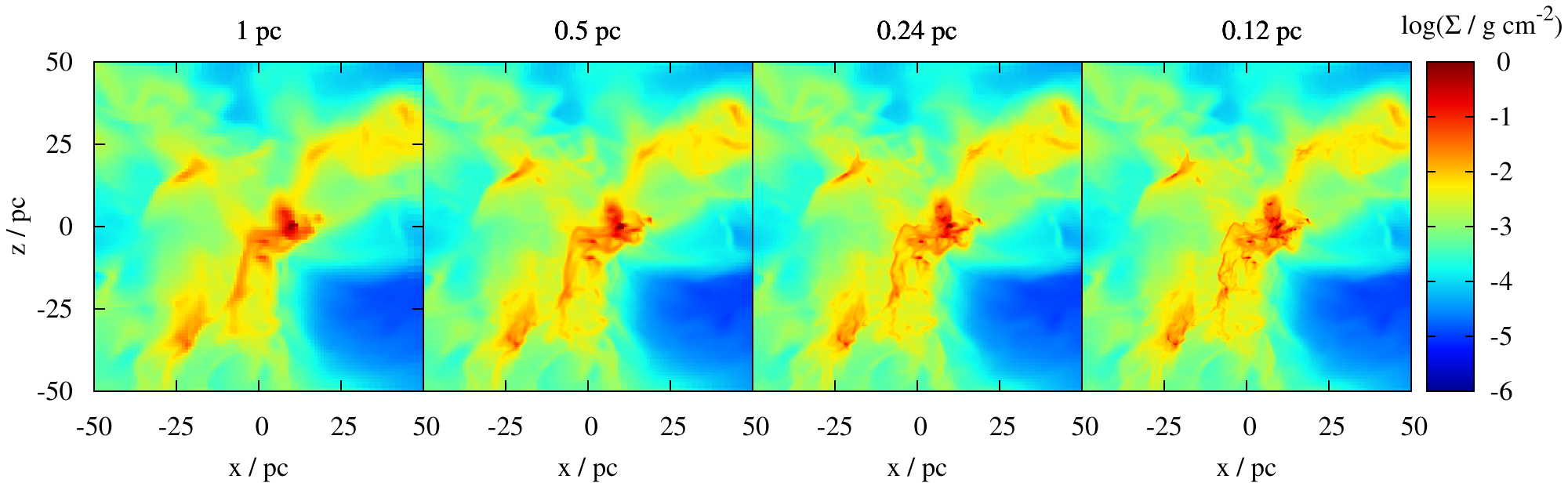}
\caption{Impact of the spatial resolution on the total gas column density of MC1 at a maximum resolution of 1 pc, 0.5 pc, 0.24 pc, and \mbox{0.12 pc} at $t$ = $t_0$ + 5 Myr (from left to right). More and more sub-structure becomes visible as the resolution increases. The filamentary structure of the clouds is clearly visible at sub-pc resolution.}
\label{fig:coldens_res}
\end{figure*}

\begin{figure*}
\includegraphics[width=0.9\textwidth]{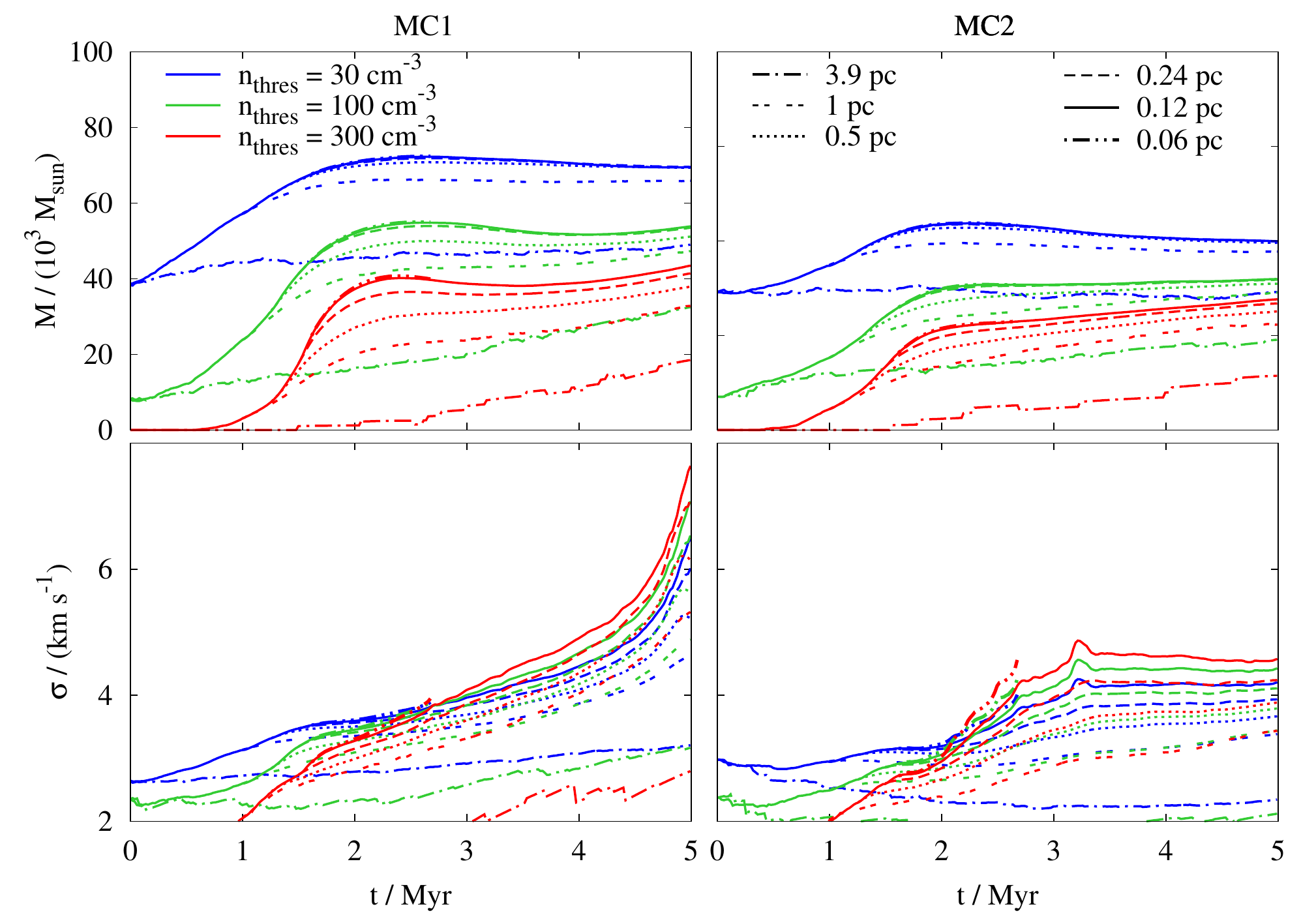}
\caption{Time evolution of the mass (top) and turbulent velocity dispersion (bottom) of MC1 (left) and MC2 (right) for different spatial resolutions, where we use the three different values of $n_{\rm thres}$ to define the clouds. There is a good agreement between the 0.06, 0.12, and 0.24 pc runs, which indicates that our simulations are converged. The runs with a resolution of 0.5 and 1 pc show systematic deviations. A resolution of 3.9 pc is clearly insufficient.}
\label{fig:res}
\end{figure*}

\subsubsection{Cloud structure, mass, and velocity dispersion}
\label{sec:structure}

In Fig.~\ref{fig:coldens_res} we show the total gas column density at $t$ = $t_0$ + 5 Myr for runs MC1\_tau-0.5\_dx-1.0, MC1\_tau-1.0\_dx-0.5, MC1\_tau-1.25\_dx-0.24, and MC1\_tau-1.5\_dx-0.12 (from left to right). While for the runs with d$x$ = 0.12 and 0.24 pc, the distribution looks relatively similar with a prominent filamentary structure, the filamentary structure seems to be only marginally resolved for d$x=$ 0.5 pc and the structure is rather clumpy for d$x$ = 1 pc. A similar change in cloud structure can be seen for the runs of MC2 (not shown here). 

In Fig.~\ref{fig:res} we show the time evolution of the masses (top panels) and velocity dispersions (bottom panels) of the two clouds as defined via the three number density thresholds $n_{\rm thres} =$ 30, 100, and 300 cm$^{-3}$. We show all runs with different spatial resolution, ranging from d$x$ = 3.9 pc (base grid resolution) to d$x$ = 0.06 pc. As expected, higher densities are increasingly well resolved with higher spatial resolution. Therefore $M_{30}$ is already well converged at d$x\le$0.5 pc, while $M_{100}$ requires $x\le$ 0.24 pc, and $M_{300}$ requires d$x\le$ 0.12 pc to meet the same time evolution. For lower resolutions, the masses are systematically underestimated. For example, for $M_{300}$, the cloud masses in the simulations with \mbox{d$x=$ 0.24 pc} are $\sim$ 5 -- 10\% below the results of the runs with \mbox{d$x\le$ 0.12 pc}. For low resolutions, d$x=$ 1 pc and d$x=$~3.9~pc, $M_{300}$ is reduced by 30\% and 60\%, respectively. Since gas with number densities greater than 100 cm$^{-3}$ can quickly become molecular, the insufficient resolution of high density gas is accompanied by an artificially reduced molecular gas fraction in the simulations (see \ref{sec:chem}). 

The velocity dispersion (bottom row of Fig.~\ref{fig:res}) systematically increases with $n_{\rm thres}$ and with resolution. However, the differences in velocity dispersion between runs with different spatial resolution hardly depend on $n_{\rm thres}$. Again we find that runs with d$x=$ 1 pc and d$x=$ 3.9 pc highly underestimate $\sigma$. For d$x=$ 0.24 pc, the velocity dispersion is a few 0.1 km s$^{-1}$ ($\sim$ 10\%) smaller than for d$x=$ 0.12 pc. Since the cloud dynamics at late times is dictated by self-gravity and turbulence, we do not expect $\sigma$ to be fully converging at \mbox{d$x\leq$ 0.12 pc} since small scales (small turbulent eddies and/or small, collapsing structures) still contribute to the kinetic energy content of the cloud.

\begin{figure*}
\includegraphics[width=\textwidth]{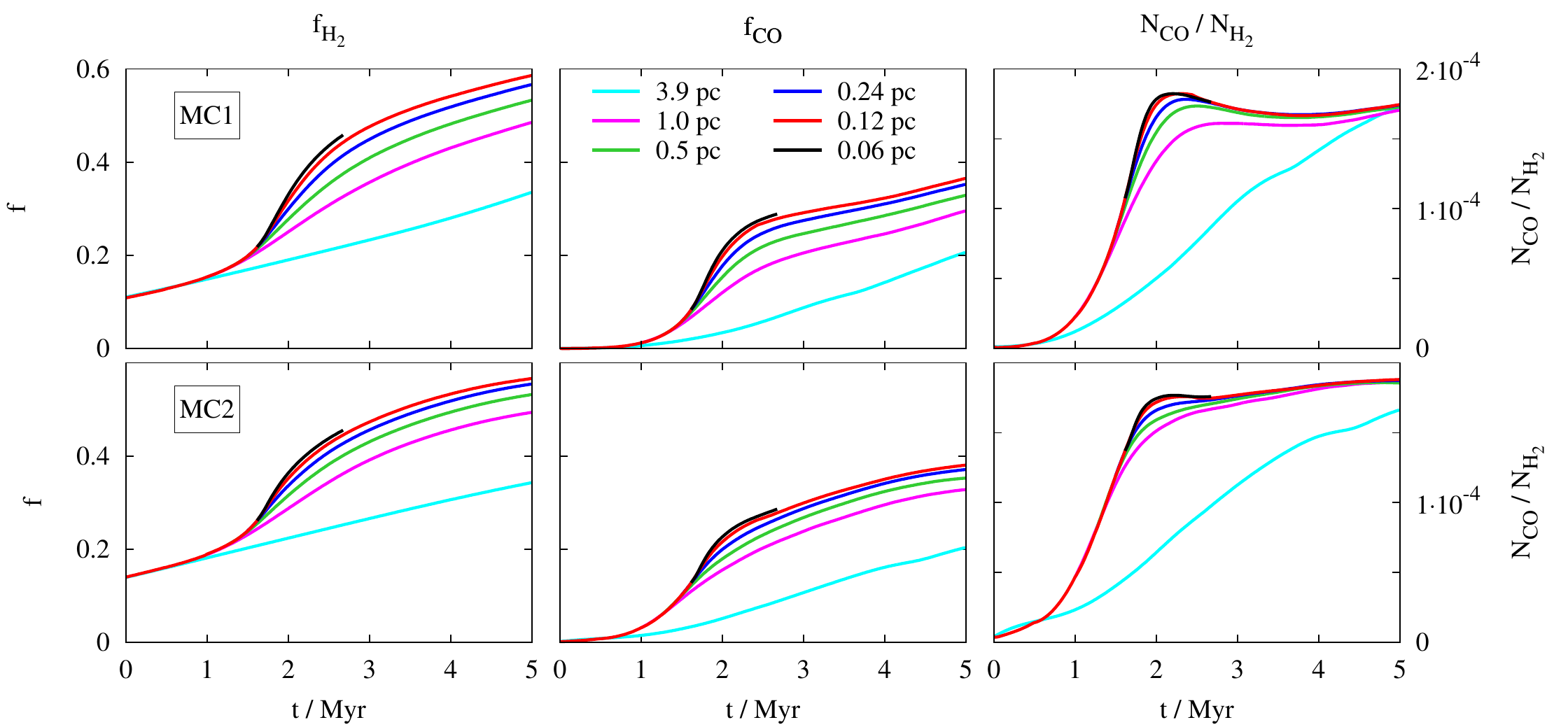}
\caption{Time evolution of the mass fraction $f$ of H$_2$ (left column) and CO (middle column) as well as the ratio of the number of CO and H$_2$ molecules (right column) within the zoom-in region for both MC1 (top row) and MC2 (bottom row) for different spatial resolutions. The lower the resolution, the lower are the molecular mass fractions. For d$x$ = 0.06, 0.12, and 0.24 pc, the relative deviations are below 5\%, which indicates that our simulations are converged. For runs with d$x$ $\geq$ 1 pc, the molecular abundances are not converged.}
\label{fig:boxres}
\end{figure*}

\begin{figure*}
\includegraphics[width=\textwidth]{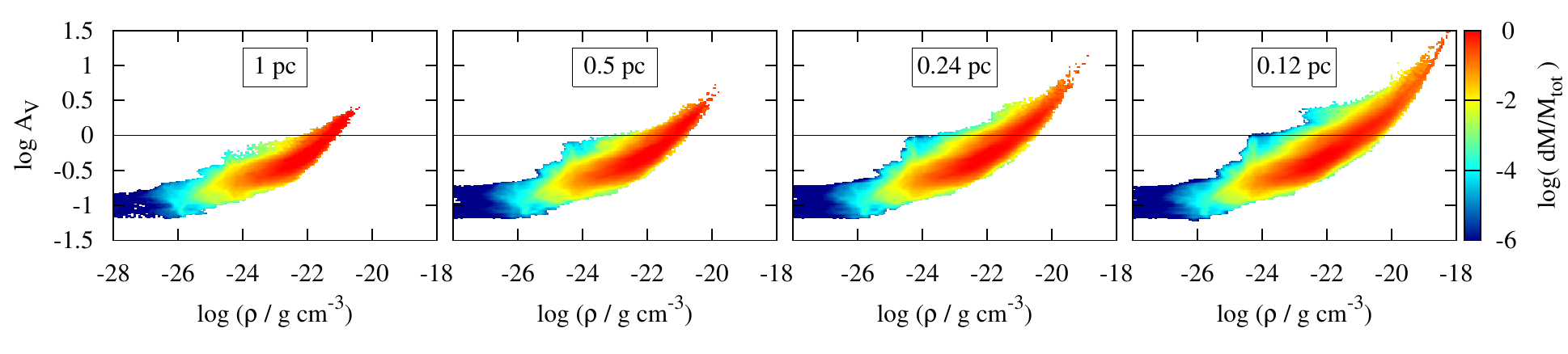}
\caption{Resolution dependence of the mass-weighted density-$A_\rmn{V}$-PDF for MC1 at $t$ = $t_0$ + 5 Myr at a maximum resolution of 1.0, 0.5, 0.24, and 0.12 pc (from left to right). The high-density/$A_\rmn{V}$ region becomes more and more populated for higher resolution runs. To guide the reader's eye, we show the $A_\rmn{V}$ = 1 line. We typically expect most carbon to be in the form of CO at $A_\rmn{V} \gtrsim 1$.}
\label{fig:dens_AV}
\end{figure*}

\subsubsection{Chemical properties}
\label{sec:chem}

In Fig.~\ref{fig:boxres} we plot the mass fractions of H$_2$ and CO, $f_{{\rm H}_2}$ (left column) and $f_{{\rm CO}}$ (middle column), in both zoom-in regions, with 
\begin{equation}
 f_\rmn{H_2} = \frac{M_\rmn{H_2,tot}}{M_\rmn{H,tot}} \; \; \; \textrm{and} \; \; \; f_\rmn{CO} = \frac{\frac{12}{28} \, M_\rmn{CO,tot} }{M_\rmn{C,tot}} \, .
\end{equation}
Here, $M_\rmn{H,tot}$ and $M_\rmn{C,tot}$ are the total mass of hydrogen and carbon in the considered volume, $M_\rmn{H_2,tot}$ and $M_\rmn{CO,tot}$ the mass of all H$_2$ and CO molecules, respectively, and the factor $\frac{12}{28}$ corrects for the mass of oxygen. We refer to Fig.~\ref{fig:time_chem} for the time evolution of the total masses $M_\rmn{H_2,tot}$ and $M_\rmn{CO,tot}$ in the zoom-in regions.

Interestingly, although MC1 and MC2 show a different evolution in the total masses (Figs.~\ref{fig:chemistry} and~\ref{fig:timeevol}), the mass fractions are remarkably similar. In general, $f_\rmn{CO}$ is always below $f_\rmn{H_2}$ because CO forms at higher visual extinction ($A_\rmn{V}$ $\gtrsim$ 1) than H$_2$ ($A_\rmn{V}$ $\gtrsim$ 0.3) \citep[e.g.][]{Roellig07,Glover10}. In agreement with the convergence of the mass evolution discussed in Fig.~\ref{fig:res}, we find that $f_\rmn{CO}$ and $f_\rmn{H_2}$ are grossly underestimated for low resolution runs and only start to converge for d$x\leq$ 0.24 pc (the difference in $f_\rmn{CO}$ and $f_\rmn{H_2}$ between runs with d$x$ = 0.12 and 0.06 pc is smaller than 5\%). Although the low resolution runs would also form significant amounts of molecular gas towards the end of the runs, the corresponding timescale would be significantly longer. This resolution dependence also affects the transition to H$_2$ dominated gas (Fig.~\ref{fig:H2transition}, blue line). With decreasing resolution, the value of 30 -- 50 M$_{\sun}$ pc$^{-2}$ for d$x$ = 0.12 pc, where $\Sigma_\rmn{H_2}/\Sigma_\rmn{HI}$ $>$ 1 , increases up to $\sim$ 100 M$_{\sun}$ pc$^{-2}$ for d$x$ = 1 pc. For d$x$ = 3.9 pc, up to this time ($t$ = $t_0$ + 5 Myr) the gas does not become H$_2$ dominated at all. In addition, the molecular gas sits in blobby configurations at low resolution, rather than in thin filamentary structures (see Fig.~\ref{fig:coldens_res}). 

Furthermore, we plot the ratio of the total number of CO and H$_2$ molecules within the zoom-in region, $N_{\rm CO}/N_{{\rm H}_2}$, in the right column of Fig.~\ref{fig:boxres}. The molecular clouds mostly form within the first $\sim$ 2 Myr of the zoom-in simulations and thus, $N_{\rm CO}/N_{{\rm H}_2}$ is rapidly increasing in this period. From this one could naively assume a highly variable $X_{\rm CO}$ during cloud formation (see Section~\ref{sec:XCO}). At later times the ratio saturates and we find that $N_{\rm CO}/N_{{\rm H}_2} \sim 1.8 -1.9\times 10^{-4}$ at $t=t_0+5$ Myr for both molecular clouds and for all but the lowest resolution run. The fractional abundance of carbon with respect to hydrogen nuclei used in the simulations is 1.4 $\times$ 10$^{-4}$. Therefore the final value of $N_{\rm CO}/N_{{\rm H}_2}$ is somewhat below the ratio of \mbox{2.8 $\times$ 10$^{-4}$} for fully molecular gas. Our resolution study indicates that d$x \lesssim$ 0.24 pc should be used to determine $X_{\rm CO}$ in simulations of molecular cloud formation.

The initial increase of $N_{\rm CO}/N_{{\rm H}_2}$ is a consequence of the rapid formation of CO in this time period. The rate-limiting reaction is the formation of hydrocarbon radicals like CH or CH$_2$ out of the (already present) reservoir of H$_2$ (left panel of Fig.~\ref{fig:boxres}) with a rate coefficient of \mbox{5 $\times$ 10$^{-16}$ cm$^{3}$ s$^{-1}$} \citep{Nelson97,Glover10}. Assuming a fractional abundance of H$_2$ of 10\%, the typical timescale for CO formation is thus 
\begin{equation}
\tau_\rmn{CO} = \frac{1}{(5 \times 10^{-16} \, \rmn{cm}^{3} \, \rmn{s}^{-1} \times 0.1 \, n)} \simeq 600 \,\rmn{Myr} \left( \frac{n}{1\;\rm{cm}^{-3}} \right)^{-1} \, ,
\end{equation}
which is slightly faster than the canonical formation time of H$_2$, $\tau_{\rm H_2} \approx 1 \;{\rm Gyr} \left( n / 1\;{\rm cm}^{-3}\right)^{-1}$, thus explaining the initial increase of $N_{\rm CO}/N_{{\rm H}_2}$.

The formation of H$_2$ and CO occurs at high column densities, which shield the forming molecules from the ambient interstellar radiation field (ISRF). A different way to analyse at which resolution we expect to converge with respect to the H$_2$ and CO mass fractions is to look at the mass-weighted probability-density-function (PDF) of visual extinction $A_\rmn{V}$ vs. gas density. Here, $A_\rmn{V}$ is the effective visual extinction of radiation in each cell in the simulation domain obtained via the TreeCol algorithm \citep{Clark12b}. In Fig.~\ref{fig:dens_AV}, we plot the resulting PDFs of MC1 at $t$ = \mbox{$t_0$ + 5 Myr} for different resolutions (from left to right: d$x$ = 1.0, 0.5, 0.24, and \mbox{0.12 pc}). We expect most hydrogen to be in the form of H$_2$ for $A_\rmn{V} \gtrsim 0.3$ \citep[see e.g.][for benchmark tests for photodissociation regions, but also \citeauthor{Glover10} \citeyear{Glover10}]{Roellig07} and most carbon to be in the form of CO for $A_\rmn{V} \gtrsim 1$. The distribution of $A_\rmn{V}$ is broad due to the different amount of shielding of different regions in the cloud, i.e. at a given density cells near the cloud surface are found at low $A_\rmn{V}$ and cells within the deeply embedded interior are found at high $A_\rmn{V}$. Most of the volume is occupied by lower density, low $A_\rmn{V}$ gas. An $A_\rmn{V}$ of 0.3 is reached for number densities between $\sim$ 1 -- 100 cm$^{-3}$, while an $A_\rmn{V}$ of 1.0 typically occurs at number densities of $\sim$ 300 cm$^{-3}$, which is barely resolved in the lower resolution simulations. 

\subsubsection{Physical interpretation}

Combining the results from the Sections~\ref{sec:structure} and~\ref{sec:chem}, we conclude that {\it an effective spatial resolution of {\rm d}$x$ $\lesssim 0.12$ pc is required to obtain a reasonably well converged result for structural, dynamical and chemical properties.}

For CO this agrees well with the fact that the mass in gas above $n_{\rm thres} = 300\;{\rm cm}^{-3}$ is well converged at d$x \lesssim$ 0.12 pc. For H$_2$, we could expect convergence at d$x \lesssim$ 0.5 pc (from the mass evolution and the $A_\rmn{V}$ distribution). However, the evolution of $f_{{\rm H}_2}$ indicates that d$x \lesssim$ 0.12 pc is also required to properly resolve the formation of molecular hydrogen. This is due to the fact that H$_2$ forms quickly in dense regions and is then efficiently mixed into the lower density environment, an effect which is not well resolved at lower resolution.

We note that the resolution requirement of \mbox{$\sim$ 0.1 pc} also agrees with the sonic scale of interstellar turbulence \citep[e.g.][and references therein]{Goodman98,Vazquez03,Federrath10}. On these scales velocity-coherent structures are believed to form, which fits with our findings that the structural properties converge around this resolution (see Fig.~\ref{fig:coldens_res}.)

This finding has implications for a number of recent studies (see Table~\ref{tab:resolution}), in which a lower spatial resolution is used. Our results suggest that there are limitations for the accuracy of the quantities derived from these simulation data. We also speculate that there are similar limitations for associated synthetic observations produced for comparison with actual observations, e.g. the linewidth-size relation.

\subsection{Impact of the refinement time}
\label{sec:comparison}

So far we have only considered simulations with a refinement time \mbox{$\tau = 1.5$ Myr} at d$x=0.12$ pc. However, one might ask why an instantaneous refinement up to the maximum refinement level, i.e. \mbox{$\tau$ = 0} should not be allowed. We discuss the results of simulations with $\tau$ = 0 in section \ref{sec:instant} and discuss why it does not work well. Furthermore, we could decrease/increase $\tau$ without going to the extreme case of instantaneous refinement. The results for runs with $\tau=$ 1.0, 2.25, and 4.5 Myr are shown in section \ref{sec:delay}, where the impact of $\tau$ on the cloud mass evolution, molecular mass fraction evolution, velocity dispersion, and cloud substructure is discussed.

\begin{figure*}
\includegraphics[width=\textwidth]{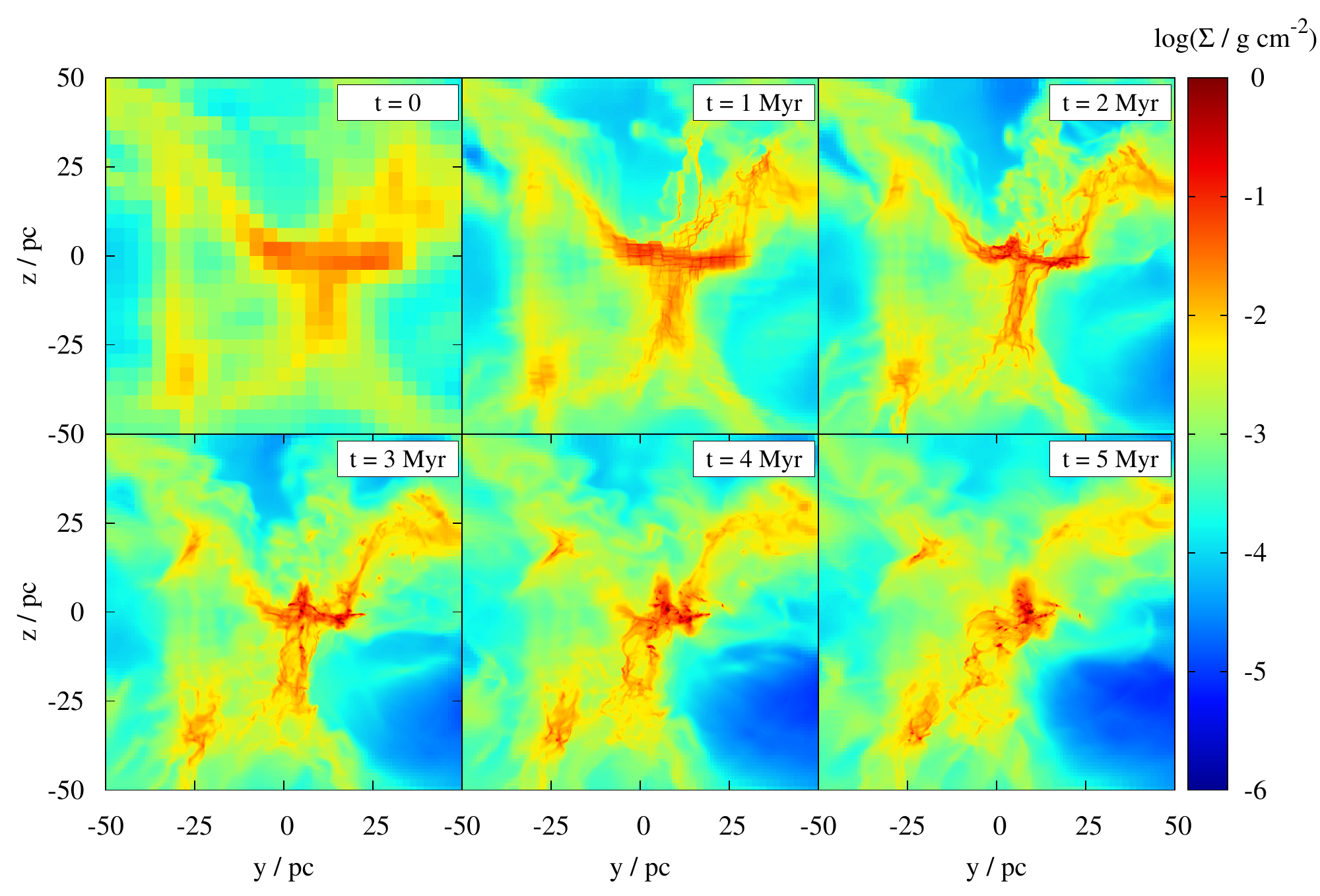}
\caption{Time evolution of the total gas column density in the zoom-in region of run MC1\_tau-0\_dx-0.12, when allowing for an instantaneous refinement. In the initial phase ($t \lesssim$ $t_0$ + 2 Myr) grid artefacts are occur, which also affect the long term evolution of the cloud resulting in an enhanced fragmentation.}
\label{fig:instant_col_dens}
\end{figure*}

\subsubsection{Instantaneous refinement $\tau=0$}
\label{sec:instant}

We consider both molecular clouds with an instantaneous refinement, i.e. the runs MC1\_tau-0\_dx-0.12 and MC2\_tau-0\_dx-0.12 (see Table~\ref{tab:runs}). In Fig.~\ref{fig:instant_col_dens} we show the column density of run MC1\_tau-0\_dx-0.12 at different times. Soon after we start the zoom-in calculation (up to 1 -- 2 Myr), the instantaneous refinement causes strong grid artefacts. The "blocky" structures are formed if a coarse (parent) cell is refined multiple times within just a few time steps and thus, all of its sub-cells still contain almost equal hydrodynamical quantities since the small-scale velocity and density fluctuations had no time to developed. This problem cannot be easily solved even by a first- or second-order prolongation scheme from coarse to fine cells since the conservation of mass and momentum has to be guaranteed. As a consequence, at the boundaries of some parent cells, gas will be pushed together and compressed, thus leading to artificial filamentary structures which were unresolved on the coarse grid. 

At later times, the grid artefacts are washed out, but the filamentary structures they seeded can survive and develop into dense filaments that cannot be distinguished from the "real" filamentary structures which form in turbulent molecular clouds. Comparing the $\tau=0$ Myr and the fiducial \mbox{$\tau$ = 1.5 Myr} runs (compare with Fig.~\ref{fig:coldens}, in particular the southern part of the cloud at $t \geq$ 3 Myr), it can be seen that the cloud substructure in run MC1\_tau-0\_dx-0.12 is different. Similar results are also found for MC2.

We see such grid artefacts for any $\tau <$ 1 Myr. Therefore, we use  $\tau=$ 1.0 Myr, which corresponds to about 200 time steps on each refinement level, as a \textit{lower} limit and do not consider runs with \mbox{$\tau < 1$ Myr} in the following section.

\begin{figure*}
\includegraphics[width=\textwidth]{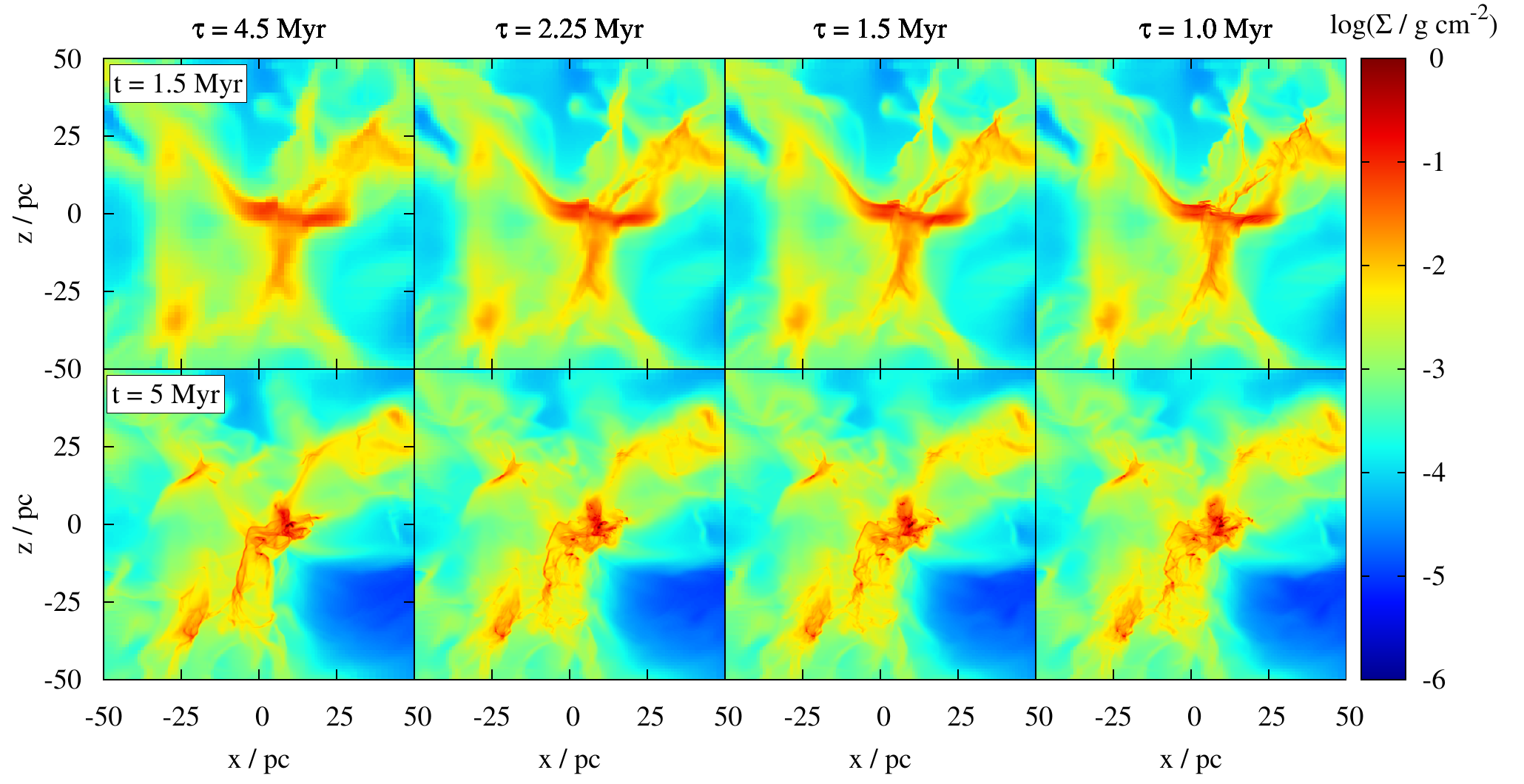}
\caption{Impact of the refinement time $\tau$ on the total gas column density of MC1 at $t$ = $t_0$ + 1.5 Myr (top row) and $t_0$ + 5 Myr (bottom row). More sub-structure becomes visible as $\tau$ decreases. Due to the numerical artefacts occurring for $\tau$ = 0, we do not include the corresponding column density map here. Recall that our fiducial case is the second to the right ($\tau$ = 1.5 Myr).}
\label{fig:coldens_tau}
\end{figure*}

\begin{figure}
 \includegraphics[width=8cm]{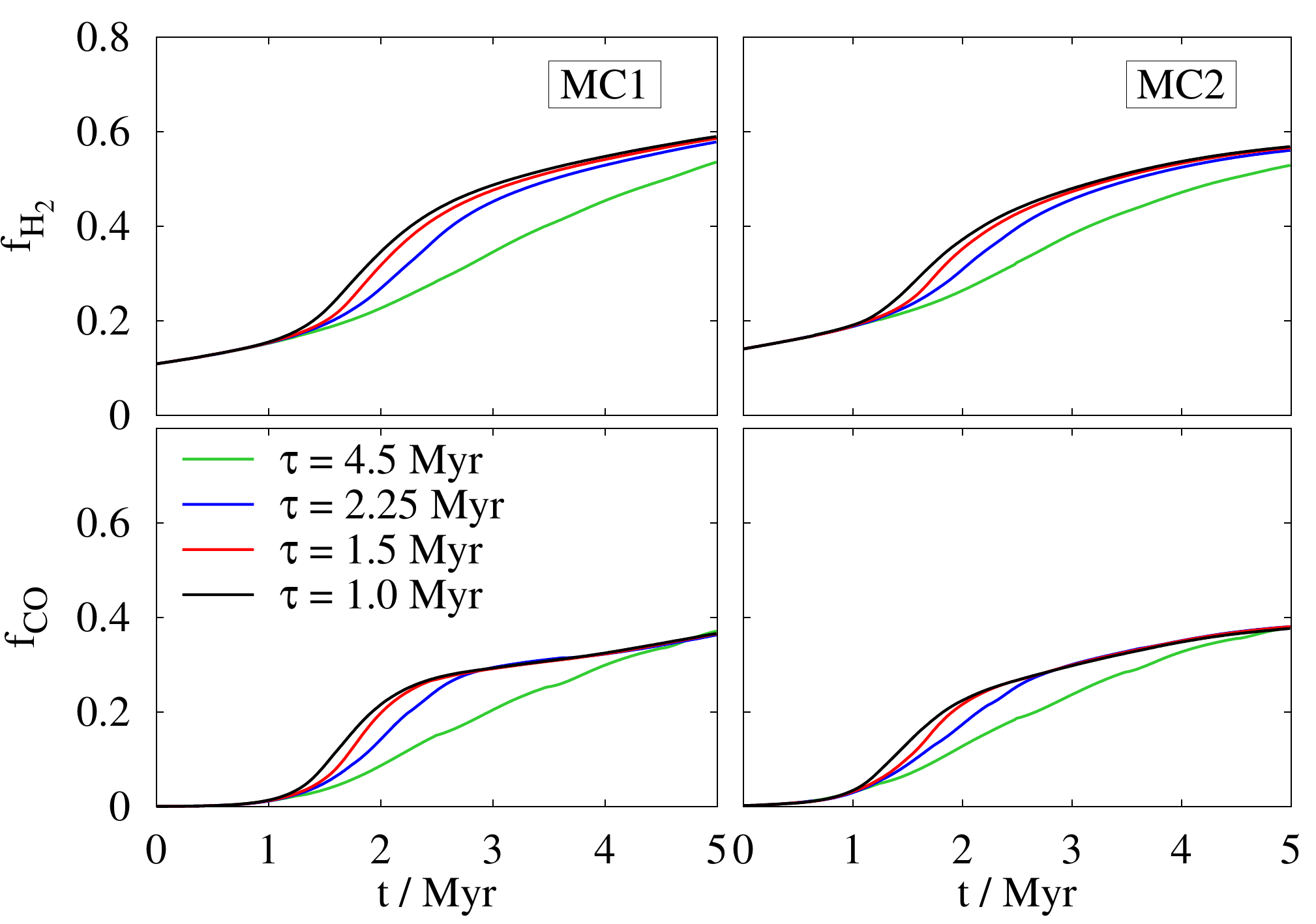}\\
  \includegraphics[width=8cm]{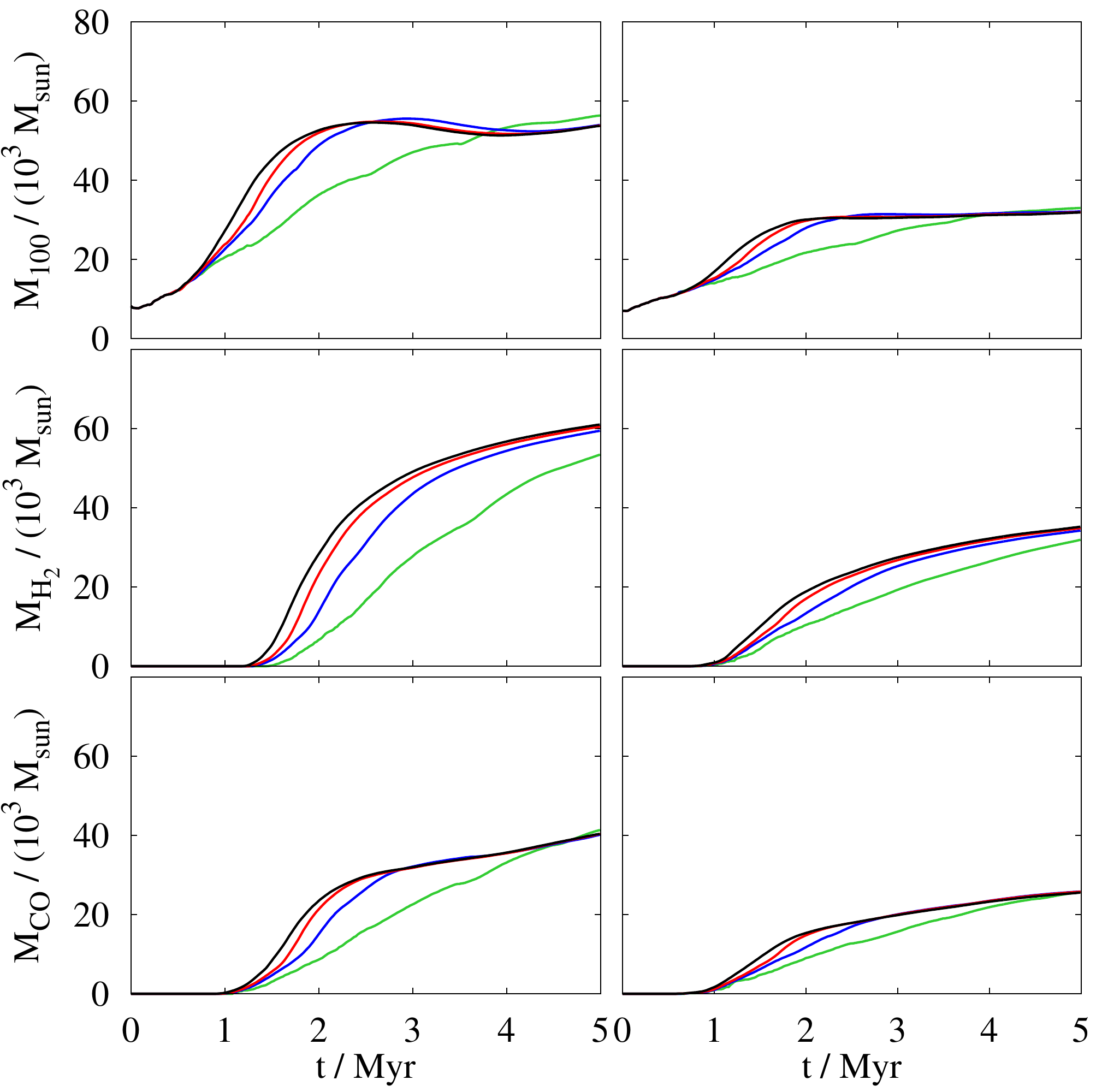} \\
  \includegraphics[width=8cm]{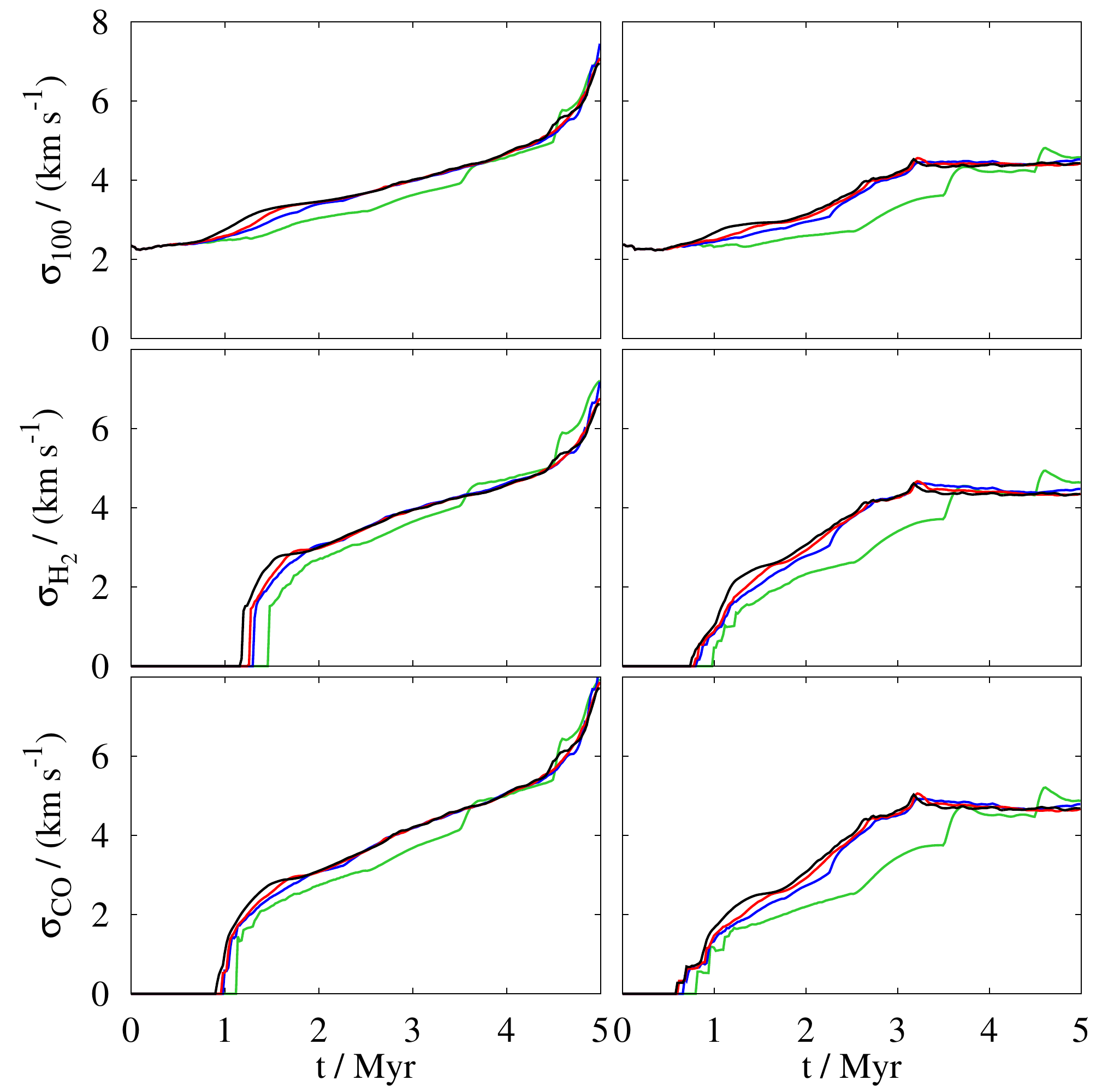}
 \caption{Impact of the refinement time $\tau$ on the evolution of MC1 (left) and MC2 (right). We show the total H$_2$ and CO mass fractions within the zoom-in region (top and 2$^{\rm nd}$ row), the gas masses above the $n_{\rm thres}$ = 100 cm$^{-3}$ (3$^{\rm rd}$ row), the H$_2$ (4$^{\rm th}$ row), and CO (5$^{\rm th}$ row) criterion, and the velocity dispersions for the three criteria. For \mbox{$\tau \leq 1.5$ Myr} no significant differences can be found.}
 \label{fig:mass_tau}
\end{figure}

\begin{figure}
\includegraphics[width=\linewidth]{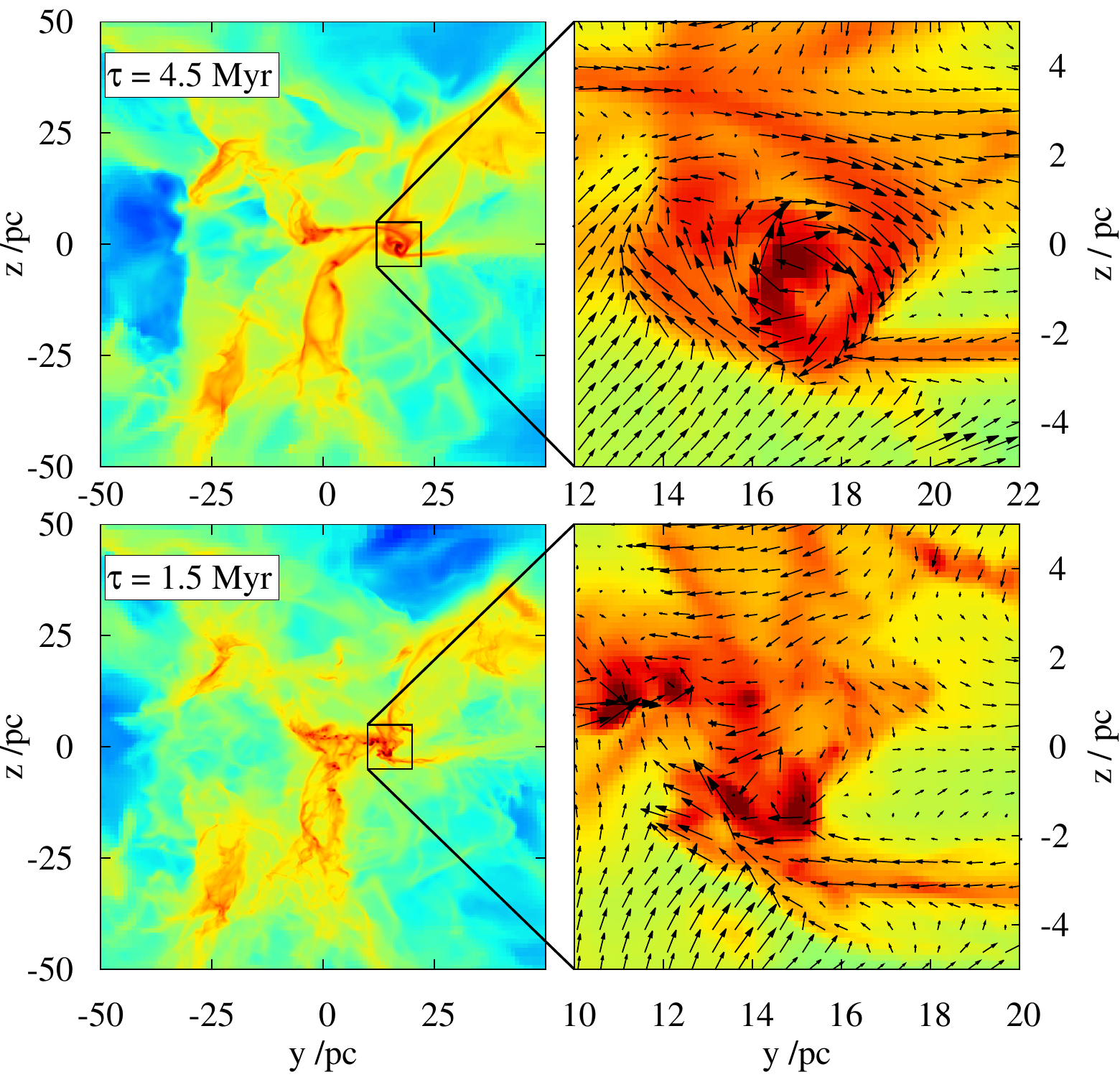}
\caption{Total gas column density of simulations of MC1 under the influence of nearby SNe performed with \mbox{$\tau$ = 4.5 Myr} (top-row) and $\tau$ = 1.5 Myr (bottom row) showing the entire MC (left) and a zoom-in onto a dense structure (right) at $t$ = $t_0$ + 4 Myr. For $\tau$ = 4.5 Myr, a large-scale, rotating, disc-like structure forms, which is possibly a consequence of unresolved, turbulent, colliding flows at an earlier stage.}
\label{fig:coldens_disk}
\end{figure}

\subsubsection{Influence of different $\tau>0$}
\label{sec:delay}

As discussed before, a refinement time \mbox{$\tau$ $<$ 1 Myr} results in undesired grid artefacts, which imposes a lower limit on $\tau$. In order to show the impact of different $\tau$, we repeat the runs for both molecular clouds with d$x=$ 0.12 pc and different $\tau$. Consequently, the time (number of time steps) which we spend on each refinement level is adapted (see Table~\ref{tab:levels}). In particular, we use one refinement time that is shorter than our fiducial value of 1.5 Myr, that is $\tau=$ 1.0 Myr, and two larger values, $\tau=$ 2.25 and 4.5 Myr. The runs are called MC1\_tau-1.0\_dx-0.12, MC1\_tau-2.25\_dx-0.12, and MC1\_tau-4.5\_dx-0.12 for MC1, and MC2\_tau-1.0\_dx-0.12, MC2\_tau-2.25\_dx-0.12, and MC2\_tau-4.5\_dx-0.12 for MC2 (see Table~\ref{tab:runs}). 

In Fig.~\ref{fig:coldens_tau} we show the column densities of MC1 at $t = t_0 + 1.5$ Myr (top row) and $t = t_0 + 5$ Myr (bottom row) for the four different $\tau$ used (from left to right). At $t = t_0 + 1.5$ Myr the runs with $\tau >$ 1.5 Myr have not yet reached their highest resolution. This affects the cloud structure at later times: As $\tau$ increases, less sub-structure is visible at $t = t_0 + 5$ Myr. While the differences between runs MC1\_tau-1.0\_dx-0.12 and MC1\_tau-1.5\_dx-0.12 are small, we see significantly less fragmentation in MC1\_tau-4.5\_dx-0.12. Again, the results for MC2 are qualitatively similar.

In Fig. ~\ref{fig:mass_tau} we compare the total mass fractions of H$_2$ and CO within the zoom-in region as well as the cloud masses and velocity dispersions obtained for three cloud definitions: a density threshold of $n_{\rm thres} = 100\;{\rm cm}^{-3}$, and the H$_2$ and CO threshold. The runs with long refinement times (in particular \mbox{$\tau=$ 4.5 Myr}) are not able to properly capture the formation of both CO and H$_2$ at early times. In particular for H$_2$ (top panels) the mass fractions remain lower than for shorter $\tau$ throughout the simulation. We attribute this to two reasons: first, the turbulent formation mechanism of H$_2$ \citep[see Section~\ref{sec:rhomean} but also][]{Glover07b} is more efficient for higher spatial resolution and thus underestimated for long $\tau$. Second, due to the lower resolution for longer $\tau$ the mass at high densities ($M_{100}$, third row of Fig.~\ref{fig:mass_tau}), in which H$_2$ is formed, is lower. For this reason, also $M_\rmn{H_2}$ and $M_\rmn{CO}$ grow more slowly for longer refinement times, in agreement with the H$_2$ and CO mass fractions. For all shorter values of $\tau \leq$ 2.25 Myr, we see only small differences within the first 2-3 Myr of evolution. In particular for $\tau$ = 1.0 Myr and 1.5 Myr the changes are minor.

Once each simulation reaches the highest refinement level, both masses and velocity dispersions seem to converge for different $\tau$, however, there are several problems which become apparent if the runs are inspected more closely.

First, the velocity dispersions do not evolve smoothly for $\tau$ = 4.5 Myr, but jump to a higher value as soon as a higher refinement level is introduced. This clearly shows that the dynamics of the cloud is not well captured if $\tau$ is too long. Second, the cloud substructure changes irrevocably if $\tau=$ 4.5 Myr, which we show for MC1\footnote{This particularly bad example occurred in a simulation of MC1 including SN explosions close to the zoom-in region between $t_0$ and $t_0$ + 5 Myr. The effect of nearby SNe will be discussed in detail in a subsequent paper. However, we note that the changes between the simulations presented here and the simulations with nearby SNe are overall minor in this particular run.} in Fig.~\ref{fig:coldens_disk}. A too long refinement time leads to a rotationally supported structure of size $\sim$ 5 -- 10 pc in the centre of the cloud (top panels). This disc-like cloud is broken up into smaller filaments and blobs if the refinement time is shorter (see $\tau$ = 1.5 Myr in the bottom panels). As turbulent motions in the molecular cloud locally create converging flows which carry some amount of net angular momentum with respect to the collision point, rotating structures on the grid scale can be formed when the flows converge under the influence of self-gravity. These structures might be unphysical if smaller-scale, turbulent motions are unresolved. Since for the run with \mbox{$\tau$ = 4.5 Myr} the resolution is lower than for \mbox{$\tau$ = 1.5 Myr} at the moment when the flows collide \mbox{($t \simeq$ $t_0$ + 3.5 Myr)}, this results in a rather large, rotating disc which does not disappear, even after the resolution is increased further. A numerical artefact like this could lead to the frequent appearance of rotating molecular clouds in larger-scale galaxy-scale simulations \citep[e.g.][]{Renaud13,Bournaud14}, while observations of such large-scale rotating molecular cloud structures are very rare \citep[e.g.][for one example]{Li16}.

\subsubsection{Physical interpretation}

Overall, we have shown that there is a lower limit ($\sim$ 1 Myr) as well as an upper limit ($\sim$ 1.5 Myr) for $\tau$. The lower limit on $\tau$ is set by a minimum number of time steps ($\gtrsim$ 200) which should be spent on each newly introduced refinement level, and is thus most likely applicable to any kind of grid code, although the exact numbers may vary.

The upper limit of $\tau$ can be explained by means of the free-fall time of the clouds. At $t$ = $t_0$, there is gas with number densities between 100 and 300 cm$^{-3}$ but no gas with $n$ $>$ 300 cm$^{-3}$ in both zoom-in regions\footnote{In the Appendix we discuss the effect when starting the entire refinement procedure earlier and show that overall we find a good agreement between the runs with different zoom-in starting times.}. The corresponding free-fall times are $\tau_{\rm ff}$ = 3.40 and 1.96 Myr, respectively. Moreover, since the clouds seem to assemble within about 2 Myr (see upper row of Fig.~\ref{fig:timeevol}), $\tau$ $<$ 2 Myr seems to be advisable. This is also indicated by the moderate differences occurring for the runs with $\tau$ = 2.25 Myr, the runs with $\tau$ = 4.5 Myr $> \tau_{\rm ff}$ lead to a clear delayed collapse of gas into dense regions (see Fig.~\ref{fig:mass_tau}). We emphasize that $\tau_{\rm ff}$ -- and thus the upper limit of $\tau$ -- naturally depends on the mean density of the simulated object.

\begin{figure*}
\includegraphics[width=0.8\textwidth]{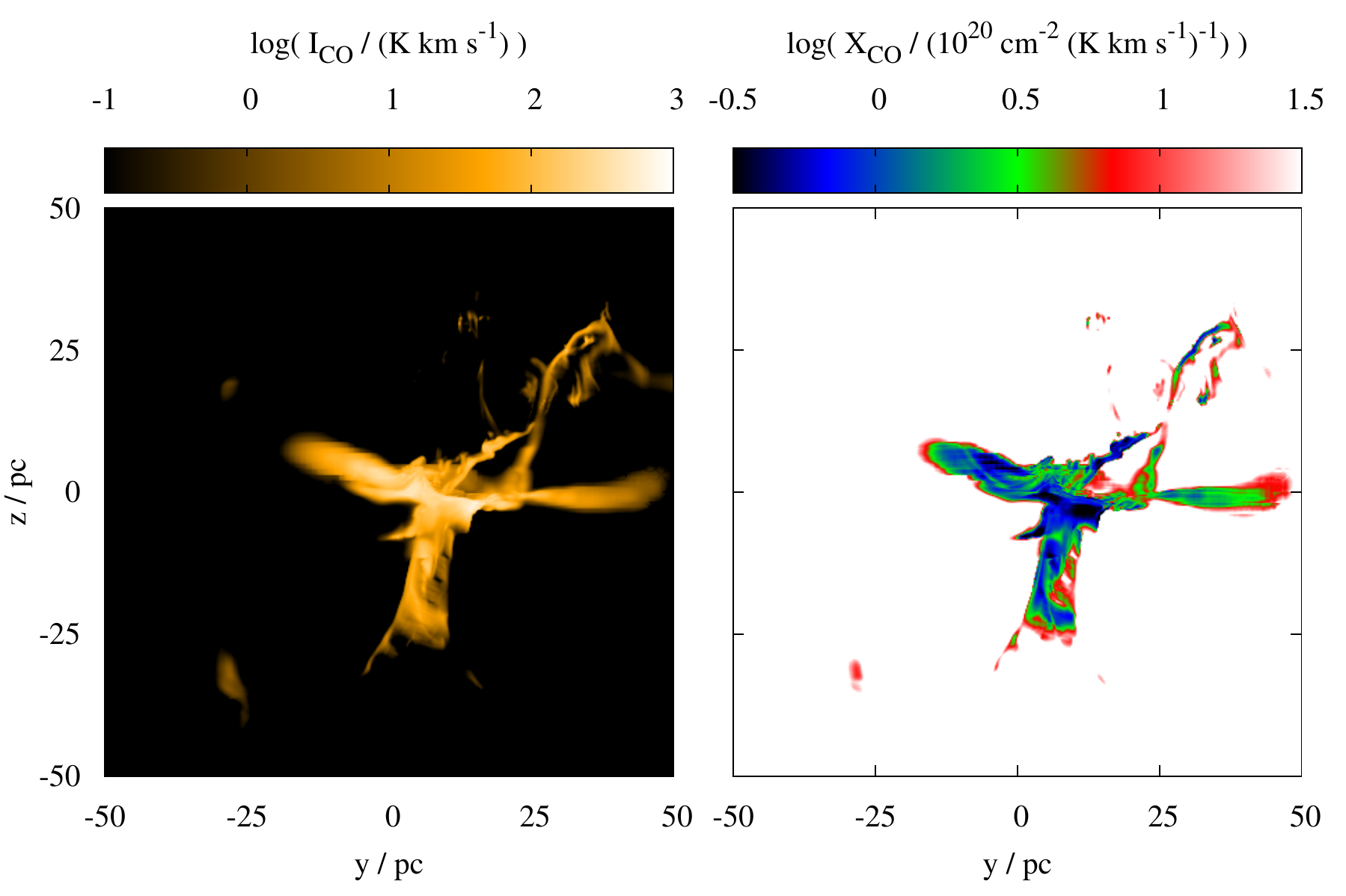}
\caption{Map of the integrated intensity (left) and the $X_\rmn{CO}$ conversion factor (right) of MC1 at $t$ = $t_0$ + 2.0 Myr for the LOS along the $x$-axis. The $X_\rmn{CO}$ factor shows a large spatial variability by more than one order of magnitude in the regions of strong emission.}
\label{fig:xco}
\end{figure*} 
 
\begin{figure}
\includegraphics[width=\linewidth]{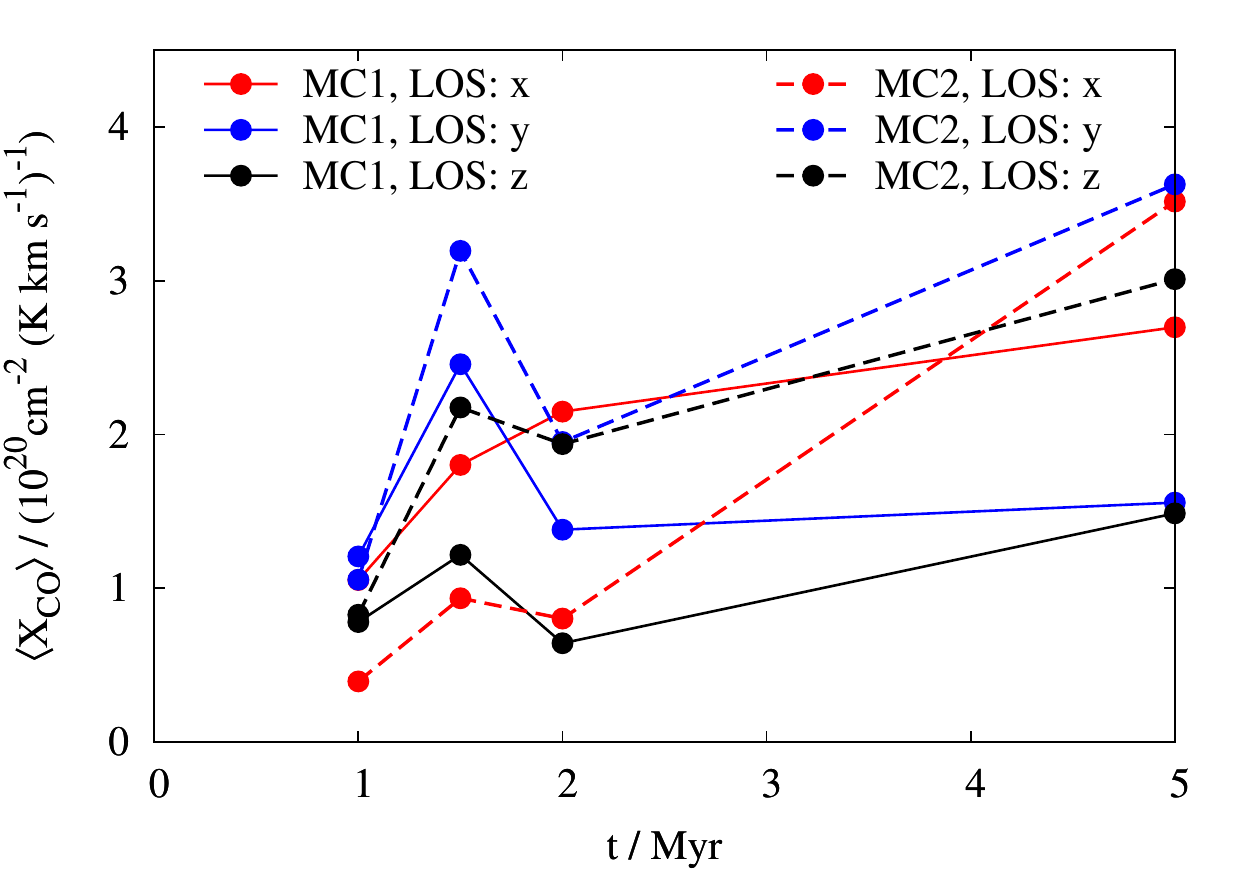}
\caption{Time evolution of the mean conversion factor $\langle X_\rmn{CO} \rangle$ calculated for the entire emission maps. The conversion factor seems to be rather robust and independent of the evolutionary stage of the considered MC.}
\label{fig:xco_time}
\end{figure}

\subsection{Impact of magnetic fields}

In the simulations presented here we have neglected magnetic fields, which are able to significantly delay the collapse time of dense objects by exerting an additional magnetic pressure, which counteracts gravity \citep[e.g.][]{Heitsch01,Vazquez11,Collins12,Federrath12,Walch15,Girichidis16,Mocz17}. This can alleviate the tight constraints on the refinement time $\tau$ discussed in Section~\ref{sec:comparison}: Since in the presence of magnetic fields the actual collapse time can become longer -- in particular longer than $\tau_\rmn{ff}$, which we take as the rough upper limit for $\tau$ -- the upper limit of $\tau$ could be increased. Furthermore, due to the slower collapse, large and well-shielded structures appear later, which in turn delays the formation of H$_2$ and CO \citep[see Fig. 10 in][]{Walch15}.  Hence, the initial formation time (see Fig.~\ref{fig:time_chem}), which has to be resolved, becomes longer, which makes it easier to accurately model the chemical evolution. We note that we plan to study the impact of magnetic fields in a follow-up paper.

\section{$X_\rmn{CO}$-factor}
\label{sec:XCO}

As demonstrated in Fig.~\ref{fig:boxres}, the ratio of CO to H$_2$ molecules increases for the first 2 Myr and then remains roughly constant. This might have some implications for the conversion factor $X_\rmn{CO}$, which is usually found to have values around 2 $\times$ 10$^{20}$ cm$^{-2}$ (K km s$^{-1}$)$^{-1}$ \citep{Bolatto13}. In order to test this, we produce $^{12}$CO, $J$ = 1 -- 0 line emission maps of our two MCs at $t$ = $t_0$ + 1.0, 1.5, 2.0, and 5.0 Myr with RADMC-3D \citep{Dullemond12} using the Large Velocity Gradient (LVG) method. The molecular data, e.g. the Einstein coefficients, are taken from the Leiden Atomic and Molecular data base \citep{Schoier05}. We use 201 velocity channels with a width of 100 m s$^{-1}$ to guarantee that all emission is captured properly. We produced emission maps for line-of-sight (LOS) directions along the $x$, $y$, and $z$ axis. We note that we do not include any observational effects like noise or resolution limitations.

In Fig.~\ref{fig:xco} we show the integrated intensity as well as the $X_\rmn{CO}$ maps of MC1 at $t$ = $t_0$ + 2.0 Myr for the LOS along the $x$-axis, i.e. the same direction as shown in Fig.~\ref{fig:coldens}. Overall, the $X_\rmn{CO}$ factor shows a strong spatial variation of more than one order of magnitude in this region, which also holds for all other maps produced. Similar results were found in numerical \citep{Clark15,Glover16,Szucs16} and observational works \citep[e.g.][]{Lee14}.

In order to quantify the time-dependence of $X_\rmn{CO}$ for both clouds and all three LOS, we calculate the mean conversion factor $\langle X_\rmn{CO} \rangle$ for all pixels in the  maps. The values of $\langle X_\rmn{CO} \rangle$ show only a weak time dependence (see Fig.~\ref{fig:xco_time}) and scatter around 1 -- 4 $\times$ 10$^{20}$ cm$^{-2}$ (K km s$^{-1}$)$^{-1}$, which is in good agreement with observations \citep[see e.g.][for a recent review]{Bolatto13}. From the CO-to-H$_2$ ratio shown in the right panel of Fig.~\ref{fig:boxres} one might expect a decreasing $\langle X_\rmn{CO} \rangle$ factor with time. However, the total CO intensity increases slower than the number of CO molecules and is roughly balanced by the increasing amount of H$_2$, which results in an almost constant $\langle X_\rmn{CO} \rangle$ factor. Overall, our results thus indicate that the $X_\rmn{CO}$ factor seems to be rather robust and independent of the evolutionary stage of the considered MC.

We attribute the slower increase of the emission of CO with time to the fact that it becomes optically thick in the densest regions. The optical depth of CO can be estimated via
\begin{equation}
 \tau_\rmn{CO, 1 - 0} = \frac{3 h c^3 A_\rmn{10}}{16 \pi k_\rmn{B} \nu_\rmn{10}^2 T_\rmn{ex} \Delta v} N_\rmn{CO} \left ( 1 - e^{-h \nu_\rmn{10}/k_\rmn{B} T_\rmn{ex}} \right ) \, ,
\end{equation}
where $h$ and $k_\rmn{B}$ are Planck and the Boltzmann constant, $A_\rmn{10}$ and $\nu_\rmn{10}$ the Einstein coefficient and frequency of the considered transition, $\Delta v$ the line-width, and $T_\rmn{ex}$ excitation temperature. This can be rewritten as
\begin{equation}
 \tau_\rmn{CO, 1 - 0} \simeq 79 \left( \frac{\textrm{km s$^{-1}$}}{\Delta v} \right) \left( \frac{10 \, \textrm{K}}{ T_\rmn{ex}} \right) \left( \frac{mag}{A_\rmn{V}} \right)\left ( 1 - e^{-5.53/T_\rmn{ex}} \right) \, .
\end{equation}
Hence, for $\Delta v$ $\simeq$ 4 km s$^{-1}$ (Fig.~\ref{fig:timeevol}) and $T_\rmn{ex}$ $\simeq$ 15 K, CO starts to become optically thick at $A_\rmn{V}$ $\sim$ 1, i.e. there is a significant amount of optically thick gas in the clouds (see Fig.~\ref{fig:dens_AV}).

There appears to be a slight trend of an increase of $\langle X_\rmn{CO} \rangle$ with time, which is in good agreement with the findings of \citet[][see their Fig. 7]{Glover16}, but in contrast to the results reported by \citet{Richings16a,Richings16b}, who find a decreasing $\langle X_\rmn{CO} \rangle$ factor over time. As we argue in Section~\ref{sec:resolution}, the result of \citet{Richings16a,Richings16b} might be an artefact of the mass resolution in their SPH simulations, which is orders of magnitude lower than the corresponding spatial resolution used in our simulations, thus leading to a significant underestimation of CO.

\section{Conclusions}
\label{sec:conclusion}

We present simulations of the self-consistent formation of two molecular clouds (MCs) in a stratified galactic disc in the context of the SILCC project \citep{Walch15,Girichidis16}. The simulations include a chemical network to model the formation of H$_2$ and CO. Furthermore, we make use of a zoom-in technique, which allows us to follow the large-scale, galactic environment and, at the same time, resolve the formation of filamentary MCs down to scales of $\sim$ 0.1 pc.

The simulated MCs typically grow from the low-density environment by accreting mass at a rate of a few times \mbox{$10^{-3}$ to 10$^{-2}$ M$_{\sun}$ yr$^{-1}$} for a couple of Myr and reach masses between \mbox{3 -- 7 $\times$ 10$^4$ M$_{\sun}$}. We tentatively argue that gravity contributes a significant part to the velocity dispersion, which reaches \mbox{4 -- 8 km s$^{-1}$}. The transition to molecular hydrogen dominated gas occurs around 30 -- 50 M$_{\sun}$ pc$^{-2}$. At later stages, the dense gas ($n \gtrsim 100$ cm$^{-3}$) is almost fully molecular with H$_2$ fractions around 0.9. We show that different cloud definitions (such as based on density thresholds, H$_2$ or CO mass fraction) result in significantly different cloud properties. Gas which contains a significant fraction of H$_2$ ($>$ 50\% H$_2$) grows differently from the dense and CO dominated gas and cannot be matched by a single density threshold. While CO follows the evolution of the denser gas with $n \geq$ 300 cm$^{-3}$, H$_2$ is found in gas with lower and lower number density (even in gas with $n \lesssim 30$ cm$^{-3}$). From estimating the relative time scales, it seems that molecular hydrogen is efficiently mixed by turbulent motions into the lower density gas rather than being formed in-situ \citep[see also][]{Glover07b,Valdivia16}. We speculate that calculating the chemical abundances from simulations without any chemical network in a post-processing step is significantly complicated by this mixing process.

The CO-to-H$_2$ ratio changes as a function of time, in particular at early times when the mass accretion rate onto the cloud is high, later on it settles to an approximately constant value of $\sim 1.8\times10^{-4}$. Since the CO(1-0) line, however, becomes quickly optically thick, the X$_\rmn{CO}$ conversion factor remains rather constant over time with values around 1 -- 4 $\times$ 10$^{20}$ cm$^{-2}$ (K km s$^{-1}$)$^{-1}$ and thus seems to be independent of the evolutionary state of a MC.

From the work presented here, we identify four requirements for numerically converged zoom-in simulations of molecular cloud formation:  
\begin{enumerate}
\item An effective resolution of \mbox{$\lesssim 0.1$ pc} is required to obtain converged mass fractions of molecular hydrogen and CO as well as realistic morphological and dynamical properties.
\item The adaptive mesh refinement (AMR) should proceed stepwise, thus allowing the hydrodynamic quantities to relax on each refinement level. Therefore, in any grid code \mbox{$\gtrsim$ 200} time steps should be spent on a level before a next higher level is introduced. Otherwise grid artefacts at the boundaries of coarse blocks might occur.
 \item On the other hand, for simulations of self-gravitating objects, the total refinement time (after which the resolution reaches its maximum) should be shorter than its free-fall time to follow the actual collapse of the dense structures (here $<$ 2 Myr). This ensures that realistic, small-scale substructures can form and avoids the formation of numerical artefacts like rotationally supported objects.
 \item The total refinement time should also be adjusted to the formation time of chemical species in turbulent environments in order to prevent the delay of H$_2$ and CO formation.
 \end{enumerate}
If fulfilled, the forming molecular clouds have converged chemical compositions, mass distributions, and substructure, and grid artefacts are minimised.

Since the paper serves as a proof-of-concept, it does by no means cover all aspects of the simulations. For future work we plan to investigate in more detail important parameters such as the origin of the velocity dispersion, the filamentary structure, the chemical evolution, the virial state as well as column density PDFs of the clouds. Moreover, since we have detailed information about the chemical composition of the gas as well as the dust temperature, we plan to produce synthetic line and continuum emission maps for a detailed comparison with current observational data.

\section*{Acknowledgements}

The authors like to thank the anonymous referee for the comments which helped to significantly improve the paper.
DS and SW acknowledge the support of the Bonn-Cologne Graduate School, which is funded through the German Excellence Initiative. SW further acknowledges support via the ERC starting grant No. 679852 "RADFEEDBACK".
PG, SW, TN, SCOG, RSK, and TP acknowledge support from the DFG Priority Program 1573 ``Physics of the Interstellar Medium''.
PG acknowledges funding from the European Research Council under ERC-CoG grant CRAGSMAN-646955.
TN acknowledges support from the DFG cluster of excellence ''Origin and Structure of the Universe''.
RW acknowledges support by the Czech Science Foundation grant 15-06012S and by the project RVO:67985815 of the Academy of Sciences of the Czech Republic.
RSK and SCOG thank the DFG for funding via the SFB 881 ``The MilkyWay System'' (subprojects B1, B2, and B8).
RSK furthermore acknowledges support from the European Research Council under the European Community's Seventh Framework Programme (FP7/2007-2013) via the ERC Advanced Grant STARLIGHT (project number 339177).
PCC acknowledges support from the Science and Technology Facilities Coun- cil (under grant ST/N00706/1) and the European Com- munitys Horizon 2020 Programme H2020-COMPET-2015, through the StarFormMapper project (number 687528).
The FLASH code used in this work was partly developed by the Flash Center for Computational Science at the University of Chicago.
The simulations were performed at SuperMUC at the Leibniz-Rechenzentrum Garching within the projects pr94du and pr45si.

\appendix

\section{Influence of the starting time}

\begin{figure*}
  \includegraphics[width=0.8\textwidth]{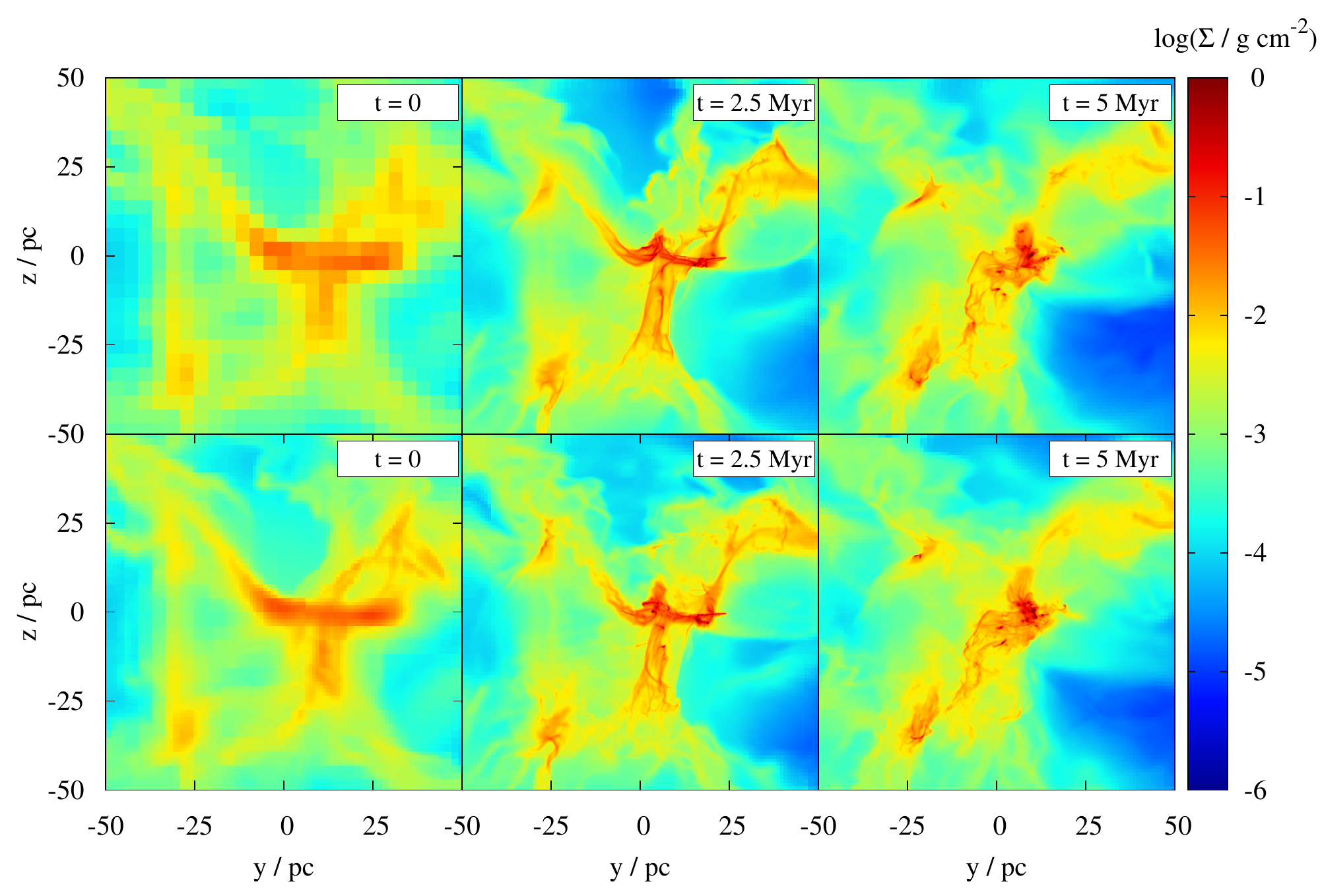}
 \caption{Time evolution of the total gas column density evolution of run fiducial run MC1 (top row) and when starting to zoom-in on MC1 at \mbox{$t$ = $t_0$ - 0.9 Myr} (bottom row). From left to right we show the same physical times $t_0$, $t_0 +2.5$ Myr, and $t_0+5$ Myr. There are small morphological differences but the overall filamentary structure and fragmentation is almost indistinguishable.}
 \label{fig:coldens_MC1_early}
\end{figure*}
\begin{figure*}
  \includegraphics[width=0.8\textwidth]{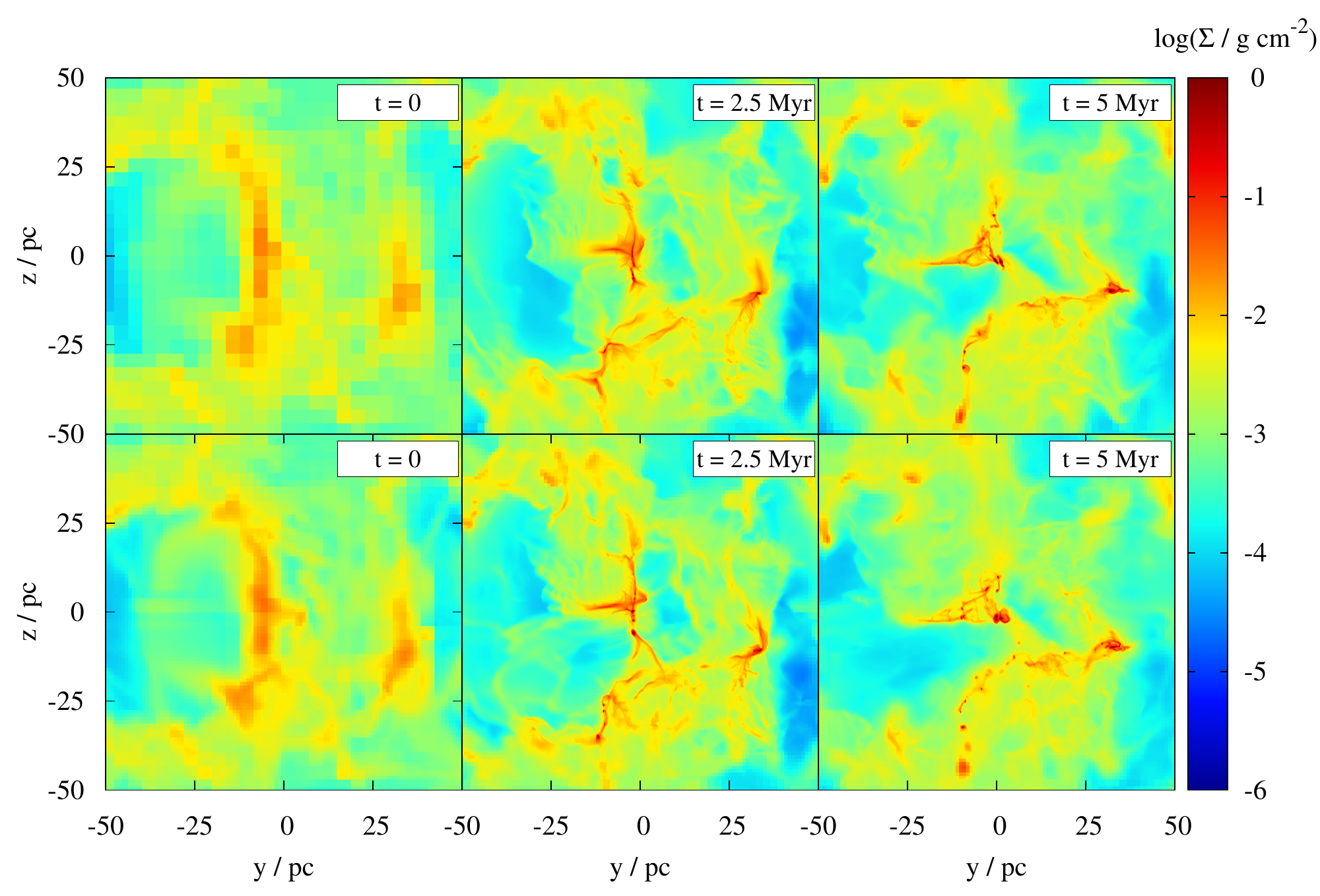}
 \caption{Same as in Fig.~\ref{fig:coldens_MC1_early} for run MC2. Again the qualitative appearance of the MCs for both starting times is similar. In particular the globally different behaviour of MC1 and MC2 is recovered independently of the starting time.}
 \label{fig:coldens_MC2_early}
\end{figure*}

\begin{figure}
 \includegraphics[width=\linewidth]{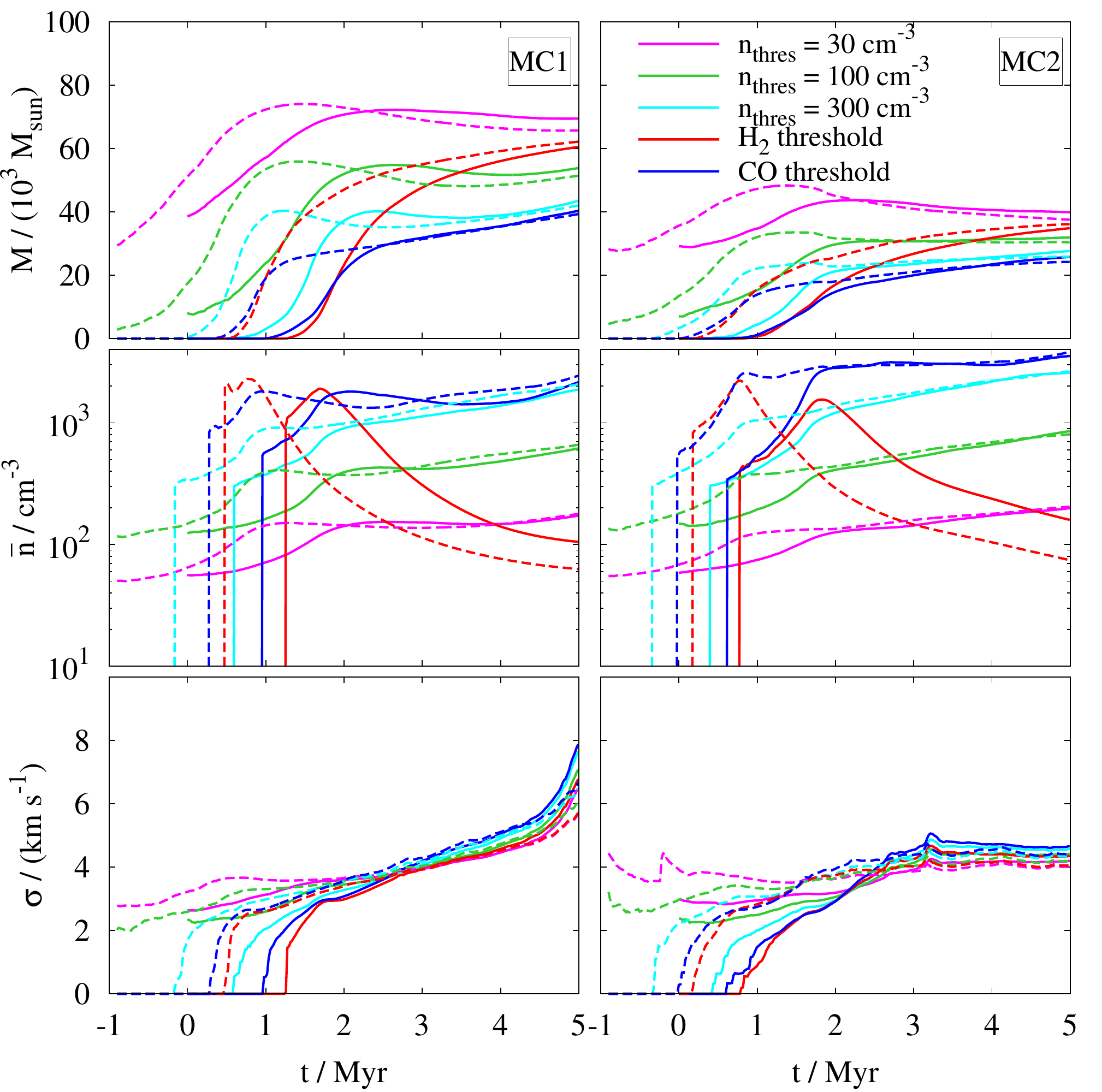}
 \caption{Time evolution of the mass (top row), mean density (middle row), and velocity dispersion (bottom row) for MC1 (left) and MC2 (right) obtained via different criteria. We compare the evolution when starting to zoom-in at $t_0$ (solid lines) and \mbox{$t$ = $t_0$ - 0.9 Myr} (dashed lines). There are differences in the evolution, in particular up to \mbox{$t \simeq t_0+2$ Myr}, as the MCs in the earlier zoom-in runs start forming earlier. However, the fiducial runs quickly catch up and the differences in all quantities towards the end of the simulations are small. }
 \label{fig:timeevol_early}
\end{figure}

\begin{figure}
 \includegraphics[width=\linewidth]{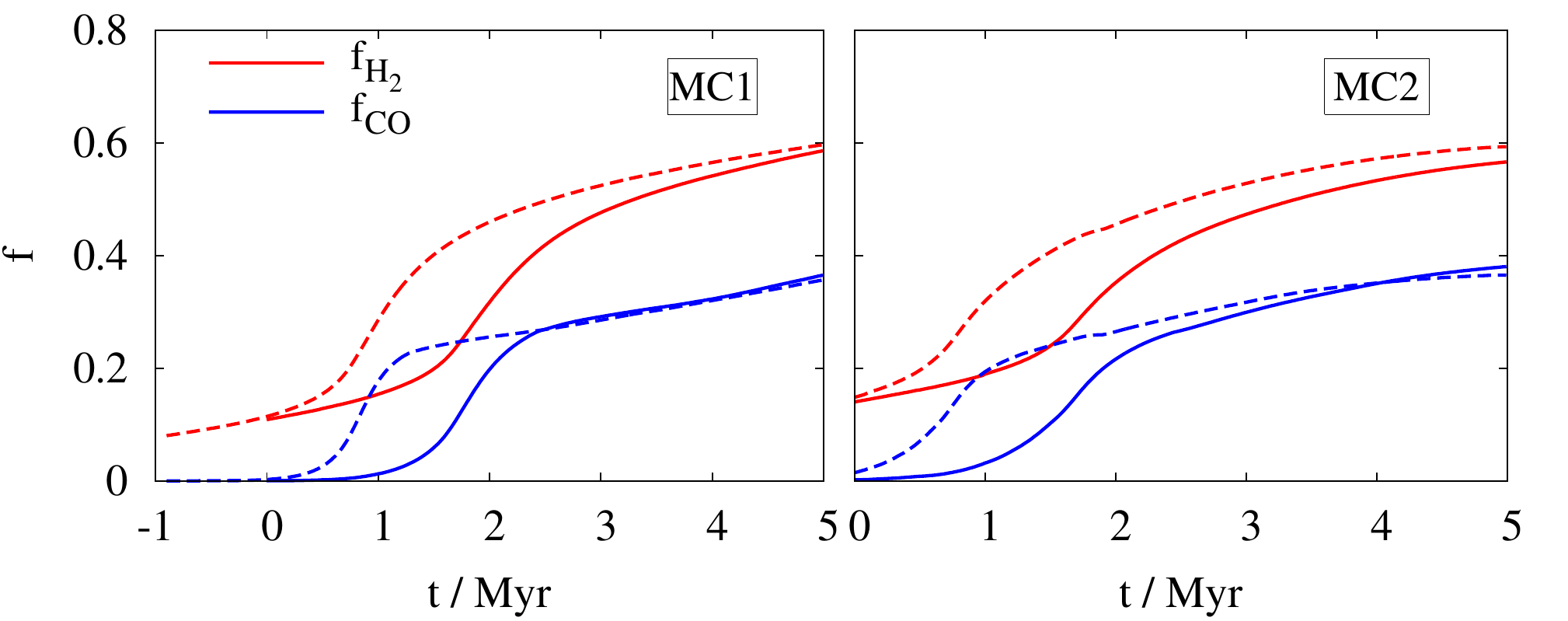}
 \caption{Time evolution of the mass fraction of H$_2$ and CO for run MC1 (left) and MC2 (right) when starting to zoom-in at $t_0$ (solid lines) and \mbox{$t$ = $t_0$ - 0.9 Myr} (dashed lines). Similar to the results in Fig.~\ref{fig:timeevol_early}, the mass fractions for the different starting times approach each other towards the end of the runs.}
\label{fig:chem_early}
\end{figure}

Here we present results of simulations of MC1 and MC2 when starting to zoom-in 0.9 Myr earlier than in our fiducial runs MC1\_tau-1.5\_dx-0.12 and  MC2\_tau-1.5\_dx-0.12, i.e. at $t$ = $t_0$ - 0.9 Myr. In Fig.~\ref{fig:coldens_MC1_early} we compare the evolution of the column density for run MC1 when zooming in earlier (bottom row) with the one obtained from the fiducial run (top row). In Fig.~\ref{fig:coldens_MC2_early} we show the same for MC2. Qualitatively, the exact time at which we start to zoom-in seems to have only little effect for both clouds. 

In order to test the dependence on the starting time more quantitatively, in Fig.~\ref{fig:timeevol_early} we compare the time evolution of the mass, mean density, and velocity dispersion obtained via the different criteria used in this paper for both MCs.

As expected, there are differences in the various quantities in particular in the beginning \mbox{($t \lesssim$ 2 Myr)}, i.e. during the refinement time. All considered quantities grow at earlier times when starting to zoom-in earlier, which reflects that in this case the gravitational collapse sets in earlier. However, towards later times the various values approach each other with typical deviations of a few 10\% or less. The only exception is $\bar{n}_\rmn{H_2}$, which is somewhat lower for the earlier zoom-in starting time. This is a consequence of the turbulent mixing of H$_2$ into the low density regions, which increases the occupied volume (compare Section~\ref{sec:rhomean}) and therefore apparently decreases the mean density of the H$_2$ dominated gas. The total mass $M_\rmn{H_2}$, however, is converge towards the end. All general trends of MC1 and MC2 are recovered, independent of the starting time.

Finally, we consider the dependence of the H$_2$ and CO mass fractions on the starting time in Fig.~\ref{fig:chem_early}. We find a similar behaviour to the mass evolution discussed above: Below $t_0$ + 2 Myr, i.e. during the zoom-in procedure, the molecular abundances are smaller for the fiducial runs where we start to zoom-in later. However, as time evolves, the differences become significantly smaller reaching relative deviations of a few percent towards the end of the runs.

In summary, for a complex, turbulent object like a molecular cloud, a perfect match between simulations with different zoom-in starting times cannot be expected. However, the overall good agreement (to within a few percent) of the structural, dynamical, and chemical properties makes us confident about the results presented in this paper.

\label{lastpage}


\begin{thebibliography}{125}
\expandafter\ifx\csname natexlab\endcsname\relax\def\natexlab#1{#1}\fi

\bibitem[{{Andr{\'e}} {et~al.}(2014){Andr{\'e}}, {Di Francesco},
  {Ward-Thompson}, {Inutsuka}, {Pudritz}, \& {Pineda}}]{Andre14}
{Andr{\'e}}, P., {Di Francesco}, J., {Ward-Thompson}, D., {et~al.} 2014,
  Protostars and Planets VI, 27

\bibitem[{{Ballesteros-Paredes} {et~al.}(2011){Ballesteros-Paredes},
  {Hartmann}, {V{\'a}zquez-Semadeni}, {Heitsch}, \&
  {Zamora-Avil{\'e}s}}]{Ballesteros11}
{Ballesteros-Paredes}, J., {Hartmann}, L.~W., {V{\'a}zquez-Semadeni}, E.,
  {Heitsch}, F., \& {Zamora-Avil{\'e}s}, M.~A. 2011, \mnras, 411, 65

\bibitem[{{Ballesteros-Paredes} {et~al.}(2007){Ballesteros-Paredes}, {Klessen},
  {Mac Low}, \& {Vazquez-Semadeni}}]{Ballesteros07}
{Ballesteros-Paredes}, J., {Klessen}, R.~S., {Mac Low}, M.-M., \&
  {Vazquez-Semadeni}, E. 2007, Protostars and Planets V, 63

\bibitem[{{Bate} \& {Burkert}(1997)}]{Bate97}
{Bate}, M.~R. \& {Burkert}, A. 1997, \mnras, 288, 1060

\bibitem[{{Bertram} {et~al.}(2016){Bertram}, {Glover}, {Clark}, {Ragan}, \&
  {Klessen}}]{Bertram16}
{Bertram}, E., {Glover}, S.~C.~O., {Clark}, P.~C., {Ragan}, S.~E., \&
  {Klessen}, R.~S. 2016, \mnras, 455, 3763

\bibitem[{{Bialy} {et~al.}(2017){Bialy}, {Burkhart}, \& {Sternberg}}]{Bialy17}
{Bialy}, S., {Burkhart}, B., \& {Sternberg}, A. 2017, \apj, 843, 92

\bibitem[{{Bihr} {et~al.}(2015){Bihr}, {Beuther}, {Ott}, {Johnston},
  {Brunthaler}, {Anderson}, {Bigiel}, {Carlhoff}, {Churchwell}, {Glover},
  {Goldsmith}, {Heitsch}, {Henning}, {Heyer}, {Hill}, {Hughes}, {Klessen},
  {Linz}, {Longmore}, {McClure-Griffiths}, {Menten}, {Motte}, {Nguyen-Luong},
  {Plume}, {Ragan}, {Roy}, {Schilke}, {Schneider}, {Smith}, {Stil}, {Urquhart},
  {Walsh}, \& {Walter}}]{Bihr15}
{Bihr}, S., {Beuther}, H., {Ott}, J., {et~al.} 2015, \aap, 580, A112

\bibitem[{{Bolatto} {et~al.}(2013){Bolatto}, {Wolfire}, \& {Leroy}}]{Bolatto13}
{Bolatto}, A.~D., {Wolfire}, M., \& {Leroy}, A.~K. 2013, \araa, 51, 207

\bibitem[{Bouchut {et~al.}(2007)Bouchut, Klingenberg, \& Waagan}]{Bouchut07}
Bouchut, F., Klingenberg, C., \& Waagan, K. 2007, Numerische Mathematik, 108,
  7, 10.1007/s00211-007-0108-8

\bibitem[{Bouchut {et~al.}(2010)Bouchut, Klingenberg, \& Waagan}]{Bouchut10}
Bouchut, F., Klingenberg, C., \& Waagan, K. 2010, Numerische Mathematik, 115,
  647, 10.1007/s00211-010-0289-4

\bibitem[{{Bournaud} {et~al.}(2014){Bournaud}, {Perret}, {Renaud}, {Dekel},
  {Elmegreen}, {Elmegreen}, {Teyssier}, {Amram}, {Daddi}, {Duc}, {Elbaz},
  {Epinat}, {Gabor}, {Juneau}, {Kraljic}, \& {Le Floch'}}]{Bournaud14}
{Bournaud}, F., {Perret}, V., {Renaud}, F., {et~al.} 2014, \apj, 780, 57

\bibitem[{{Brunt} {et~al.}(2009){Brunt}, {Heyer}, \& {Mac Low}}]{Brunt09}
{Brunt}, C.~M., {Heyer}, M.~H., \& {Mac Low}, M.-M. 2009, \aap, 504, 883

\bibitem[{{Chabrier}(2001)}]{Chabrier01}
{Chabrier}, G. 2001, \apj, 554, 1274

\bibitem[{{Clark} \& {Glover}(2015)}]{Clark15}
{Clark}, P.~C. \& {Glover}, S.~C.~O. 2015, \mnras, 452, 2057

\bibitem[{{Clark} {et~al.}(2012{\natexlab{a}}){Clark}, {Glover}, \&
  {Klessen}}]{Clark12b}
{Clark}, P.~C., {Glover}, S.~C.~O., \& {Klessen}, R.~S. 2012{\natexlab{a}},
  \mnras, 420, 745

\bibitem[{{Clark} {et~al.}(2012{\natexlab{b}}){Clark}, {Glover}, {Klessen}, \&
  {Bonnell}}]{Clark12}
{Clark}, P.~C., {Glover}, S.~C.~O., {Klessen}, R.~S., \& {Bonnell}, I.~A.
  2012{\natexlab{b}}, \mnras, 424, 2599

\bibitem[{{Collins} {et~al.}(2012){Collins}, {Kritsuk}, {Padoan}, {Li}, {Xu},
  {Ustyugov}, \& {Norman}}]{Collins12}
{Collins}, D.~C., {Kritsuk}, A.~G., {Padoan}, P., {et~al.} 2012, \apj, 750, 13

\bibitem[{{Dobbs}(2008)}]{Dobbs08}
{Dobbs}, C.~L. 2008, \mnras, 391, 844

\bibitem[{{Dobbs}(2015)}]{Dobbs15}
{Dobbs}, C.~L. 2015, \mnras, 447, 3390

\bibitem[{{Dobbs} {et~al.}(2008){Dobbs}, {Glover}, {Clark}, \&
  {Klessen}}]{Dobbs08b}
{Dobbs}, C.~L., {Glover}, S.~C.~O., {Clark}, P.~C., \& {Klessen}, R.~S. 2008,
  \mnras, 389, 1097

\bibitem[{{Dobbs} {et~al.}(2014){Dobbs}, {Krumholz}, {Ballesteros-Paredes},
  {Bolatto}, {Fukui}, {Heyer}, {Low}, {Ostriker}, \&
  {V{\'a}zquez-Semadeni}}]{Dobbs14}
{Dobbs}, C.~L., {Krumholz}, M.~R., {Ballesteros-Paredes}, J., {et~al.} 2014,
  Protostars and Planets VI, 3

\bibitem[{{Dobbs} \& {Pringle}(2013)}]{Dobbs13}
{Dobbs}, C.~L. \& {Pringle}, J.~E. 2013, \mnras, 432, 653

\bibitem[{{Draine}(1978)}]{Draine78}
{Draine}, B.~T. 1978, \apjs, 36, 595

\bibitem[{{Duarte-Cabral} {et~al.}(2015){Duarte-Cabral}, {Acreman}, {Dobbs},
  {Mottram}, {Gibson}, {Brunt}, \& {Douglas}}]{Duarte15}
{Duarte-Cabral}, A., {Acreman}, D.~M., {Dobbs}, C.~L., {et~al.} 2015, \mnras,
  447, 2144

\bibitem[{{Duarte-Cabral} \& {Dobbs}(2016)}]{Duarte16}
{Duarte-Cabral}, A. \& {Dobbs}, C.~L. 2016, \mnras, 458, 3667

\bibitem[{{Dubey} {et~al.}(2008){Dubey}, {Fisher}, {Graziani}, {Jordan},
  {Lamb}, {Reid}, {Rich}, {Sheeler}, {Townsley}, \& {Weide}}]{Dubey08}
{Dubey}, A., {Fisher}, R., {Graziani}, C., {et~al.} 2008, in Astronomical
  Society of the Pacific Conference Series, Vol. 385, Numerical Modeling of
  Space Plasma Flows, ed. N.~V. {Pogorelov}, E.~{Audit}, \& G.~P. {Zank}, 145

\bibitem[{{Dullemond}(2012)}]{Dullemond12}
{Dullemond}, C.~P. 2012, {RADMC-3D: A multi-purpose radiative transfer tool},
  astrophysics Source Code Library

\bibitem[{{Elmegreen} \& {Falgarone}(1996)}]{Elmegreen96}
{Elmegreen}, B.~G. \& {Falgarone}, E. 1996, \apj, 471, 816

\bibitem[{{Elmegreen} \& {Scalo}(2004)}]{Elmegreen04}
{Elmegreen}, B.~G. \& {Scalo}, J. 2004, \araa, 42, 211

\bibitem[{{Federrath} \& {Klessen}(2012)}]{Federrath12}
{Federrath}, C. \& {Klessen}, R.~S. 2012, \apj, 761, 156

\bibitem[{{Federrath} {et~al.}(2010){Federrath}, {Roman-Duval}, {Klessen},
  {Schmidt}, \& {Mac Low}}]{Federrath10}
{Federrath}, C., {Roman-Duval}, J., {Klessen}, R.~S., {Schmidt}, W., \& {Mac
  Low}, M.-M. 2010, \aap, 512, A81

\bibitem[{{Federrath} {et~al.}(2011){Federrath}, {Sur}, {Schleicher},
  {Banerjee}, \& {Klessen}}]{Federrath11}
{Federrath}, C., {Sur}, S., {Schleicher}, D.~R.~G., {Banerjee}, R., \&
  {Klessen}, R.~S. 2011, \apj, 731, 62

\bibitem[{{Fryxell} {et~al.}(2000){Fryxell}, {Olson}, {Ricker}, {Timmes},
  {Zingale}, {Lamb}, {MacNeice}, {Rosner}, {Truran}, \& {Tufo}}]{Fryxell00}
{Fryxell}, B., {Olson}, K., {Ricker}, P., {et~al.} 2000, \apjs, 131, 273

\bibitem[{{Fukui} {et~al.}(2009){Fukui}, {Kawamura}, {Wong}, {Murai},
  {Iritani}, {Mizuno}, {Mizuno}, {Onishi}, {Hughes}, {Ott}, {Muller},
  {Staveley-Smith}, \& {Kim}}]{Fukui09}
{Fukui}, Y., {Kawamura}, A., {Wong}, T., {et~al.} 2009, \apj, 705, 144

\bibitem[{{Gatto} {et~al.}(2015){Gatto}, {Walch}, {Mac Low}, {Naab},
  {Girichidis}, {Glover}, {W{\"u}nsch}, {Klessen}, {Clark}, {Baczynski},
  {Peters}, {Ostriker}, {Ib{\'a}{\~n}ez-Mej{\'{\i}}a}, \& {Haid}}]{Gatto15}
{Gatto}, A., {Walch}, S., {Mac Low}, M.-M., {et~al.} 2015, \mnras, 449, 1057

\bibitem[{{Girichidis} {et~al.}(2016){Girichidis}, {Walch}, {Naab}, {Gatto},
  {W{\"u}nsch}, {Glover}, {Klessen}, {Clark}, {Peters}, {Derigs}, \&
  {Baczynski}}]{Girichidis16}
{Girichidis}, P., {Walch}, S., {Naab}, T., {et~al.} 2016, \mnras, 456, 3432

\bibitem[{{Glover} \& {Clark}(2012)}]{Glover12}
{Glover}, S.~C.~O. \& {Clark}, P.~C. 2012, \mnras, 421, 116

\bibitem[{{Glover} \& {Clark}(2016)}]{Glover16}
{Glover}, S.~C.~O. \& {Clark}, P.~C. 2016, \mnras, 456, 3596

\bibitem[{{Glover} {et~al.}(2010){Glover}, {Federrath}, {Mac Low}, \&
  {Klessen}}]{Glover10}
{Glover}, S.~C.~O., {Federrath}, C., {Mac Low}, M.-M., \& {Klessen}, R.~S.
  2010, \mnras, 404, 2

\bibitem[{{Glover} \& {Mac Low}(2007{\natexlab{a}})}]{Glover07a}
{Glover}, S.~C.~O. \& {Mac Low}, M.-M. 2007{\natexlab{a}}, \apjs, 169, 239

\bibitem[{{Glover} \& {Mac Low}(2007{\natexlab{b}})}]{Glover07b}
{Glover}, S.~C.~O. \& {Mac Low}, M.-M. 2007{\natexlab{b}}, \apj, 659, 1317

\bibitem[{{Goldbaum} {et~al.}(2011){Goldbaum}, {Krumholz}, {Matzner}, \&
  {McKee}}]{Goldbaum11}
{Goldbaum}, N.~J., {Krumholz}, M.~R., {Matzner}, C.~D., \& {McKee}, C.~F. 2011,
  \apj, 738, 101

\bibitem[{{Goldsmith} {et~al.}(2008){Goldsmith}, {Heyer}, {Narayanan}, {Snell},
  {Li}, \& {Brunt}}]{Goldsmith08}
{Goldsmith}, P.~F., {Heyer}, M., {Narayanan}, G., {et~al.} 2008, \apj, 680, 428

\bibitem[{{Goodman} {et~al.}(2014){Goodman}, {Alves}, {Beaumont}, {Benjamin},
  {Borkin}, {Burkert}, {Dame}, {Jackson}, {Kauffmann}, {Robitaille}, \&
  {Smith}}]{Goodman14}
{Goodman}, A.~A., {Alves}, J., {Beaumont}, C.~N., {et~al.} 2014, \apj, 797, 53

\bibitem[{{Goodman} {et~al.}(1998){Goodman}, {Barranco}, {Wilner}, \&
  {Heyer}}]{Goodman98}
{Goodman}, A.~A., {Barranco}, J.~A., {Wilner}, D.~J., \& {Heyer}, M.~H. 1998,
  \apj, 504, 223

\bibitem[{{Habing}(1968)}]{Habing68}
{Habing}, H.~J. 1968, \bain, 19, 421

\bibitem[{{Haid} {et~al.}(2016){Haid}, {Walch}, {Naab}, {Seifried}, {Mackey},
  \& {Gatto}}]{Haid16}
{Haid}, S., {Walch}, S., {Naab}, T., {et~al.} 2016, \mnras, 460, 2962

\bibitem[{{Heitsch} {et~al.}(2001){Heitsch}, {Mac Low}, \&
  {Klessen}}]{Heitsch01}
{Heitsch}, F., {Mac Low}, M.-M., \& {Klessen}, R.~S. 2001, \apj, 547, 280

\bibitem[{{Hennebelle} \& {Iffrig}(2014)}]{Hennebelle14}
{Hennebelle}, P. \& {Iffrig}, O. 2014, \aap, 570, A81

\bibitem[{{Heyer} {et~al.}(2009){Heyer}, {Krawczyk}, {Duval}, \&
  {Jackson}}]{Heyer09}
{Heyer}, M., {Krawczyk}, C., {Duval}, J., \& {Jackson}, J.~M. 2009, \apj, 699,
  1092

\bibitem[{{Heyer} {et~al.}(2001){Heyer}, {Carpenter}, \& {Snell}}]{Heyer01}
{Heyer}, M.~H., {Carpenter}, J.~M., \& {Snell}, R.~L. 2001, \apj, 551, 852

\bibitem[{{Hollenbach} \& {McKee}(1979)}]{Hollenbach79}
{Hollenbach}, D. \& {McKee}, C.~F. 1979, \apjs, 41, 555

\bibitem[{{Hopkins} {et~al.}(2013){Hopkins}, {Narayanan}, \&
  {Murray}}]{Hopkins13}
{Hopkins}, P.~F., {Narayanan}, D., \& {Murray}, N. 2013, \mnras, 432, 2647

\bibitem[{{Hu} {et~al.}(2017){Hu}, {Naab}, {Glover}, {Walch}, \&
  {Clark}}]{Hu17}
{Hu}, C.-Y., {Naab}, T., {Glover}, S.~C.~O., {Walch}, S., \& {Clark}, P.~C.
  2017, \mnras, 471, 2151

\bibitem[{{Hu} {et~al.}(2016){Hu}, {Naab}, {Walch}, {Glover}, \&
  {Clark}}]{Hu16}
{Hu}, C.-Y., {Naab}, T., {Walch}, S., {Glover}, S.~C.~O., \& {Clark}, P.~C.
  2016, \mnras, 458, 3528

\bibitem[{{Ib{\'a}{\~n}ez-Mej{\'{\i}}a}
  {et~al.}(2016){Ib{\'a}{\~n}ez-Mej{\'{\i}}a}, {Mac Low}, {Klessen}, \&
  {Baczynski}}]{Ibanez15}
{Ib{\'a}{\~n}ez-Mej{\'{\i}}a}, J.~C., {Mac Low}, M.-M., {Klessen}, R.~S., \&
  {Baczynski}, C. 2016, \apj, 824, 41

\bibitem[{{Ib{\'a}{\~n}ez-Mej{\'{\i}}a}
  {et~al.}(2017){Ib{\'a}{\~n}ez-Mej{\'{\i}}a}, {Mac Low}, {Klessen}, \&
  {Baczynski}}]{Ibanez17}
{Ib{\'a}{\~n}ez-Mej{\'{\i}}a}, J.~C., {Mac Low}, M.-M., {Klessen}, R.~S., \&
  {Baczynski}, C. 2017, ArXiv e-prints, 1705.01779

\bibitem[{{Inoue} \& {Inutsuka}(2009)}]{Inoue09}
{Inoue}, T. \& {Inutsuka}, S.-i. 2009, \apj, 704, 161

\bibitem[{{Inoue} \& {Inutsuka}(2012)}]{Inoue12}
{Inoue}, T. \& {Inutsuka}, S.-i. 2012, \apj, 759, 35

\bibitem[{{Inutsuka} {et~al.}(2015){Inutsuka}, {Inoue}, {Iwasaki}, \&
  {Hosokawa}}]{Inutsuka15}
{Inutsuka}, S.-i., {Inoue}, T., {Iwasaki}, K., \& {Hosokawa}, T. 2015, \aap,
  580, A49

\bibitem[{{Kawamura} {et~al.}(2009){Kawamura}, {Mizuno}, {Minamidani},
  {Filipovi{\'c}}, {Staveley-Smith}, {Kim}, {Mizuno}, {Onishi}, {Mizuno}, \&
  {Fukui}}]{Kawamura09}
{Kawamura}, A., {Mizuno}, Y., {Minamidani}, T., {et~al.} 2009, \apjs, 184, 1

\bibitem[{{Kennicutt}(1998)}]{Kennicutt98}
{Kennicutt}, Jr., R.~C. 1998, \apj, 498, 541

\bibitem[{{Kim} \& {Ostriker}(2015)}]{Kim15b}
{Kim}, C.-G. \& {Ostriker}, E.~C. 2015, \apj, 815, 67

\bibitem[{{Kim} {et~al.}(2003){Kim}, {Ostriker}, \& {Stone}}]{Kim03}
{Kim}, W.-T., {Ostriker}, E.~C., \& {Stone}, J.~M. 2003, \apj, 599, 1157

\bibitem[{{Klessen} \& {Glover}(2016)}]{Klessen16}
{Klessen}, R.~S. \& {Glover}, S.~C.~O. 2016, Star Formation in Galaxy
  Evolution: Connecting Numerical Models to Reality, Saas-Fee Advanced Course,
  Volume 43.~ISBN 978-3-662-47889-9.~Springer-Verlag Berlin Heidelberg, 2016,
  p.~85, 43, 85

\bibitem[{{Klessen} \& {Hennebelle}(2010)}]{Klessen10}
{Klessen}, R.~S. \& {Hennebelle}, P. 2010, \aap, 520, A17

\bibitem[{{Kolmogorov}(1941)}]{Kolmogorov41}
{Kolmogorov}, A. 1941, Akademiia Nauk SSSR Doklady, 30, 301

\bibitem[{{Koyama} \& {Inutsuka}(2000)}]{Koyama00}
{Koyama}, H. \& {Inutsuka}, S.-I. 2000, \apj, 532, 980

\bibitem[{{Kritsuk} {et~al.}(2013){Kritsuk}, {Lee}, \& {Norman}}]{Kritsuk13}
{Kritsuk}, A.~G., {Lee}, C.~T., \& {Norman}, M.~L. 2013, \mnras, 436, 3247

\bibitem[{{Krumholz} \& {Burkhart}(2016)}]{Krumholz16}
{Krumholz}, M.~R. \& {Burkhart}, B. 2016, \mnras, 458, 1671

\bibitem[{{Krumholz} {et~al.}(2009){Krumholz}, {McKee}, \&
  {Tumlinson}}]{Krumholz09}
{Krumholz}, M.~R., {McKee}, C.~F., \& {Tumlinson}, J. 2009, \apj, 693, 216

\bibitem[{{Kuffmeier} {et~al.}(2016){Kuffmeier}, {Haugboelle}, \&
  {Nordlund}}]{Kuffmeier16}
{Kuffmeier}, M., {Haugboelle}, T., \& {Nordlund}, {\AA}. 2016, ArXiv e-prints,
  1611.10360

\bibitem[{{Larson}(1981)}]{Larson81}
{Larson}, R.~B. 1981, \mnras, 194, 809

\bibitem[{{Lee} {et~al.}(2012){Lee}, {Stanimirovi{\'c}}, {Douglas}, {Knee}, {Di
  Francesco}, {Gibson}, {Begum}, {Grcevich}, {Heiles}, {Korpela}, {Leroy},
  {Peek}, {Pingel}, {Putman}, \& {Saul}}]{Lee12}
{Lee}, M.-Y., {Stanimirovi{\'c}}, S., {Douglas}, K.~A., {et~al.} 2012, \apj,
  748, 75

\bibitem[{{Lee} {et~al.}(2015){Lee}, {Stanimirovi{\'c}}, {Murray}, {Heiles}, \&
  {Miller}}]{Lee15}
{Lee}, M.-Y., {Stanimirovi{\'c}}, S., {Murray}, C.~E., {Heiles}, C., \&
  {Miller}, J. 2015, \apj, 809, 56

\bibitem[{{Lee} {et~al.}(2014){Lee}, {Stanimirovi{\'c}}, {Wolfire}, {Shetty},
  {Glover}, {Molina}, \& {Klessen}}]{Lee14}
{Lee}, M.-Y., {Stanimirovi{\'c}}, S., {Wolfire}, M.~G., {et~al.} 2014, \apj,
  784, 80

\bibitem[{{Li} {et~al.}(2017){Li}, {Wyrowski}, \& {Menten}}]{Li16}
{Li}, G.-X., {Wyrowski}, F., \& {Menten}, K. 2017, \aap, 598, A96

\bibitem[{{Li} {et~al.}(2015){Li}, {Ostriker}, {Cen}, {Bryan}, \&
  {Naab}}]{Li15}
{Li}, M., {Ostriker}, J.~P., {Cen}, R., {Bryan}, G.~L., \& {Naab}, T. 2015,
  \apj, 814, 4

\bibitem[{{Lohner}(1987)}]{Lohner87}
{Lohner}, R. 1987, Computer Methods in Applied Mechanics and Engineering, 61,
  323

\bibitem[{{Mac Low} \& {Klessen}(2004)}]{MacLow04}
{Mac Low}, M.-M. \& {Klessen}, R.~S. 2004, Reviews of Modern Physics, 76, 125

\bibitem[{{Mac Low} {et~al.}(1998){Mac Low}, {Klessen}, {Burkert}, \&
  {Smith}}]{MacLow98}
{Mac Low}, M.-M., {Klessen}, R.~S., {Burkert}, A., \& {Smith}, M.~D. 1998,
  Physical Review Letters, 80, 2754

\bibitem[{MacNeice {et~al.}(2000)MacNeice, Olson, Mobarry, de~Fainchtein, \&
  Packer}]{MacNeice00}
MacNeice, P., Olson, K.~M., Mobarry, C., de~Fainchtein, R., \& Packer, C. 2000,
  Computer Physics Communications, 126, 330

\bibitem[{{Matzner} \& {Jumper}(2015)}]{Matzner15}
{Matzner}, C.~D. \& {Jumper}, P.~H. 2015, \apj, 815, 68

\bibitem[{{Micic} {et~al.}(2012){Micic}, {Glover}, {Federrath}, \&
  {Klessen}}]{Micic12}
{Micic}, M., {Glover}, S.~C.~O., {Federrath}, C., \& {Klessen}, R.~S. 2012,
  \mnras, 421, 2531

\bibitem[{{Miville-Desch{\^e}nes} {et~al.}(2017){Miville-Desch{\^e}nes},
  {Murray}, \& {Lee}}]{Miville17}
{Miville-Desch{\^e}nes}, M.-A., {Murray}, N., \& {Lee}, E.~J. 2017, \apj, 834,
  57

\bibitem[{{Mocz} {et~al.}(2017){Mocz}, {Burkhart}, {Hernquist}, {McKee}, \&
  {Springel}}]{Mocz17}
{Mocz}, P., {Burkhart}, B., {Hernquist}, L., {McKee}, C.~F., \& {Springel}, V.
  2017, \apj, 838, 40

\bibitem[{{Naab} \& {Ostriker}(2016)}]{Naab16}
{Naab}, T. \& {Ostriker}, J.~P. 2016, ArXiv e-prints, 1612.06891

\bibitem[{{Nelson} \& {Langer}(1997)}]{Nelson97}
{Nelson}, R.~P. \& {Langer}, W.~D. 1997, \apj, 482, 796

\bibitem[{{Padoan} {et~al.}(2016{\natexlab{a}}){Padoan}, {Juvela}, {Pan},
  {Haugb{\o}lle}, \& {Nordlund}}]{Padoan16b}
{Padoan}, P., {Juvela}, M., {Pan}, L., {Haugb{\o}lle}, T., \& {Nordlund},
  {\AA}. 2016{\natexlab{a}}, \apj, 826, 140

\bibitem[{{Padoan} {et~al.}(2016{\natexlab{b}}){Padoan}, {Pan}, {Haugb{\o}lle},
  \& {Nordlund}}]{Padoan16}
{Padoan}, P., {Pan}, L., {Haugb{\o}lle}, T., \& {Nordlund}, {\AA}.
  2016{\natexlab{b}}, \apj, 822, 11

\bibitem[{{Pettitt} {et~al.}(2014){Pettitt}, {Dobbs}, {Acreman}, \&
  {Price}}]{Pettitt14}
{Pettitt}, A.~R., {Dobbs}, C.~L., {Acreman}, D.~M., \& {Price}, D.~J. 2014,
  \mnras, 444, 919

\bibitem[{{Ragan} {et~al.}(2014){Ragan}, {Henning}, {Tackenberg}, {Beuther},
  {Johnston}, {Kainulainen}, \& {Linz}}]{Ragan14}
{Ragan}, S.~E., {Henning}, T., {Tackenberg}, J., {et~al.} 2014, \aap, 568, A73

\bibitem[{{Renaud} {et~al.}(2013){Renaud}, {Bournaud}, {Emsellem}, {Elmegreen},
  {Teyssier}, {Alves}, {Chapon}, {Combes}, {Dekel}, {Gabor}, {Hennebelle}, \&
  {Kraljic}}]{Renaud13}
{Renaud}, F., {Bournaud}, F., {Emsellem}, E., {et~al.} 2013, \mnras, 436, 1836

\bibitem[{{Rey-Raposo} {et~al.}(2015){Rey-Raposo}, {Dobbs}, \&
  {Duarte-Cabral}}]{Rey15}
{Rey-Raposo}, R., {Dobbs}, C., \& {Duarte-Cabral}, A. 2015, \mnras, 446, L46

\bibitem[{{Richings} \& {Schaye}(2016{\natexlab{a}})}]{Richings16a}
{Richings}, A.~J. \& {Schaye}, J. 2016{\natexlab{a}}, \mnras, 460, 2297

\bibitem[{{Richings} \& {Schaye}(2016{\natexlab{b}})}]{Richings16b}
{Richings}, A.~J. \& {Schaye}, J. 2016{\natexlab{b}}, \mnras, 458, 270

\bibitem[{{R{\"o}llig} {et~al.}(2007){R{\"o}llig}, {Abel}, {Bell}, {Bensch},
  {Black}, {Ferland}, {Jonkheid}, {Kamp}, {Kaufman}, {Le Bourlot}, {Le Petit},
  {Meijerink}, {Morata}, {Ossenkopf}, {Roueff}, {Shaw}, {Spaans}, {Sternberg},
  {Stutzki}, {Thi}, {van Dishoeck}, {van Hoof}, {Viti}, \&
  {Wolfire}}]{Roellig07}
{R{\"o}llig}, M., {Abel}, N.~P., {Bell}, T., {et~al.} 2007, \aap, 467, 187

\bibitem[{{Roman-Duval} {et~al.}(2010){Roman-Duval}, {Jackson}, {Heyer},
  {Rathborne}, \& {Simon}}]{Roman10}
{Roman-Duval}, J., {Jackson}, J.~M., {Heyer}, M., {Rathborne}, J., \& {Simon},
  R. 2010, \apj, 723, 492

\bibitem[{{Schmidt}(1959)}]{Schmidt59}
{Schmidt}, M. 1959, \apj, 129, 243

\bibitem[{{Sch{\"o}ier} {et~al.}(2005){Sch{\"o}ier}, {van der Tak}, {van
  Dishoeck}, \& {Black}}]{Schoier05}
{Sch{\"o}ier}, F.~L., {van der Tak}, F.~F.~S., {van Dishoeck}, E.~F., \&
  {Black}, J.~H. 2005, \aap, 432, 369

\bibitem[{{Seifried} \& {Walch}(2016)}]{Seifried16}
{Seifried}, D. \& {Walch}, S. 2016, \mnras, 459, L11

\bibitem[{{Sembach} {et~al.}(2000){Sembach}, {Howk}, {Ryans}, \&
  {Keenan}}]{Sembach00}
{Sembach}, K.~R., {Howk}, J.~C., {Ryans}, R.~S.~I., \& {Keenan}, F.~P. 2000,
  \apj, 528, 310

\bibitem[{{Shetty} {et~al.}(2011{\natexlab{a}}){Shetty}, {Glover}, {Dullemond},
  \& {Klessen}}]{Shetty11a}
{Shetty}, R., {Glover}, S.~C., {Dullemond}, C.~P., \& {Klessen}, R.~S.
  2011{\natexlab{a}}, \mnras, 412, 1686

\bibitem[{{Shetty} {et~al.}(2011{\natexlab{b}}){Shetty}, {Glover}, {Dullemond},
  {Ostriker}, {Harris}, \& {Klessen}}]{Shetty11b}
{Shetty}, R., {Glover}, S.~C., {Dullemond}, C.~P., {et~al.} 2011{\natexlab{b}},
  \mnras, 415, 3253

\bibitem[{{Smith} {et~al.}(2014){Smith}, {Glover}, {Clark}, {Klessen}, \&
  {Springel}}]{Smith14b}
{Smith}, R.~J., {Glover}, S.~C.~O., {Clark}, P.~C., {Klessen}, R.~S., \&
  {Springel}, V. 2014, \mnras, 441, 1628

\bibitem[{{Solomon} {et~al.}(1987){Solomon}, {Rivolo}, {Barrett}, \&
  {Yahil}}]{Solomon87}
{Solomon}, P.~M., {Rivolo}, A.~R., {Barrett}, J., \& {Yahil}, A. 1987, \apj,
  319, 730

\bibitem[{{Spitzer}(1942)}]{Spitzer42}
{Spitzer}, Jr., L. 1942, \apj, 95, 329

\bibitem[{{Stone} {et~al.}(1998){Stone}, {Ostriker}, \& {Gammie}}]{Stone98}
{Stone}, J.~M., {Ostriker}, E.~C., \& {Gammie}, C.~F. 1998, \apjl, 508, L99

\bibitem[{{Sz{\H u}cs} {et~al.}(2016){Sz{\H u}cs}, {Glover}, \&
  {Klessen}}]{Szucs16}
{Sz{\H u}cs}, L., {Glover}, S.~C.~O., \& {Klessen}, R.~S. 2016, \mnras, 460, 82

\bibitem[{{Tasker} \& {Tan}(2009)}]{Tasker09}
{Tasker}, E.~J. \& {Tan}, J.~C. 2009, \apj, 700, 358

\bibitem[{{Tielens} \& {Hollenbach}(1985)}]{Tielens85}
{Tielens}, A.~G.~G.~M. \& {Hollenbach}, D. 1985, \apj, 291, 722

\bibitem[{{Truelove} {et~al.}(1997){Truelove}, {Klein}, {McKee}, {Holliman},
  {Howell}, \& {Greenough}}]{Truelove97}
{Truelove}, J.~K., {Klein}, R.~I., {McKee}, C.~F., {et~al.} 1997, \apjl, 489,
  L179

\bibitem[{{Valdivia} {et~al.}(2016){Valdivia}, {Hennebelle}, {G{\'e}rin}, \&
  {Lesaffre}}]{Valdivia16}
{Valdivia}, V., {Hennebelle}, P., {G{\'e}rin}, M., \& {Lesaffre}, P. 2016,
  \aap, 587, A76

\bibitem[{{van Dishoeck} \& {Black}(1988)}]{Dishoeck88}
{van Dishoeck}, E.~F. \& {Black}, J.~H. 1988, \apj, 334, 771

\bibitem[{{V{\'a}zquez-Semadeni} {et~al.}(2003){V{\'a}zquez-Semadeni},
  {Ballesteros-Paredes}, \& {Klessen}}]{Vazquez03}
{V{\'a}zquez-Semadeni}, E., {Ballesteros-Paredes}, J., \& {Klessen}, R.~S.
  2003, \apjl, 585, L131

\bibitem[{{V{\'a}zquez-Semadeni} {et~al.}(2011){V{\'a}zquez-Semadeni},
  {Banerjee}, {G{\'o}mez}, {Hennebelle}, {Duffin}, \& {Klessen}}]{Vazquez11}
{V{\'a}zquez-Semadeni}, E., {Banerjee}, R., {G{\'o}mez}, G.~C., {et~al.} 2011,
  \mnras, 414, 2511

\bibitem[{{V{\'a}zquez-Semadeni} {et~al.}(2000){V{\'a}zquez-Semadeni}, {Gazol},
  \& {Scalo}}]{Vazquez00}
{V{\'a}zquez-Semadeni}, E., {Gazol}, A., \& {Scalo}, J. 2000, \apj, 540, 271

\bibitem[{{V{\'a}zquez-Semadeni} {et~al.}(2008){V{\'a}zquez-Semadeni},
  {Gonz{\'a}lez}, {Ballesteros-Paredes}, {Gazol}, \& {Kim}}]{Vazquez08}
{V{\'a}zquez-Semadeni}, E., {Gonz{\'a}lez}, R.~F., {Ballesteros-Paredes}, J.,
  {Gazol}, A., \& {Kim}, J. 2008, \mnras, 390, 769

\bibitem[{Waagan(2009)}]{Waagan09}
Waagan, K. 2009, Journal of Computational Physics, 228, 8609

\bibitem[{{Waagan} {et~al.}(2011){Waagan}, {Federrath}, \&
  {Klingenberg}}]{Waagan11}
{Waagan}, K., {Federrath}, C., \& {Klingenberg}, C. 2011, Journal of
  Computational Physics, 230, 3331

\bibitem[{{Walch} {et~al.}(2015){Walch}, {Girichidis}, {Naab}, {Gatto},
  {Glover}, {W{\"u}nsch}, {Klessen}, {Clark}, {Peters}, {Derigs}, \&
  {Baczynski}}]{Walch15}
{Walch}, S., {Girichidis}, P., {Naab}, T., {et~al.} 2015, \mnras, 454, 238

\bibitem[{{Walch} \& {Naab}(2015)}]{Walch15b}
{Walch}, S. \& {Naab}, T. 2015, \mnras, 451, 2757

\bibitem[{{Ward} {et~al.}(2014){Ward}, {Wadsley}, \& {Sills}}]{Ward14}
{Ward}, R.~L., {Wadsley}, J., \& {Sills}, A. 2014, \mnras, 439, 651

\bibitem[{{W{\"u}nsch} {et~al.}(2017){W{\"u}nsch}, {Walch}, {Whitworth}, \&
  {Dinnbier}}]{Wunsch17}
{W{\"u}nsch}, R., {Walch}, S., {Whitworth}, A.~P., \& {Dinnbier}, F. 2017,
  ArXiv e-prints, 1708.06142

\bibitem[{{Xu} {et~al.}(2016){Xu}, {Li}, {Yue}, \& {Goldsmith}}]{Xu16}
{Xu}, D., {Li}, D., {Yue}, N., \& {Goldsmith}, P.~F. 2016, \apj, 819, 22

\end{thebibliography}
\end{document}